%% file: main.tex
\documentclass[onecolumn]{aastex63}

\usepackage{natbib}
\usepackage{multirow}
\usepackage{booktabs}
\usepackage{epsfig}
\usepackage{color}
\usepackage{ulem}
\usepackage{amsmath}
\usepackage{rotating}
\usepackage{hyperref}
\usepackage{graphicx}
\usepackage{comment}
\usepackage{verbatim}
\usepackage{longtable}

\usepackage{lineno}

\usepackage[colorinlistoftodos]{todonotes}

\usepackage{savesym}
\savesymbol{tablenum}
\usepackage{siunitx}
\restoresymbol{SIX}{tablenum}

\usepackage[caption=false]{subfig}

\hypersetup{urlcolor=blue, colorlinks=true, linkcolor=blue, citecolor=blue} 
\definecolor{amethyst}{rgb}{0.6, 0.4, 0.8}
\definecolor{darkred}{RGB}{148,35,35}
\definecolor{purp}{RGB}{154,0,255}
\definecolor{orangered}{rgb}{1.0, 0.27, 0.0}

\newcommand{\humna}[1]{\textcolor{black}{#1}}

\newcommand{\rachel}[1]{{\color{black}#1}}
\newcommand{\husni}[1]{{\color{black}#1}}
\newcommand{\EG}[1]{{\color{black}#1}}
\newcommand{\ml}[1]{{\color{black}#1}}
\newcommand{\PhG}[1]{{\color{black} #1}}
\newcommand{\suyu}[1]{\textcolor{black}{#1}}
\newcommand{\anguita}[1]{\textcolor{black}{#1}}
\newcommand{\ds}[1]{\textcolor{black}{#1}}
\newcommand{\tanjap}[1]{\textcolor{black}{#1}}

\newcommand{\humnaNew}[1]{\textcolor{black}{#1}}
\newcommand{\rachelNew}[1]{{\color{black}#1}}
\newcommand{\husniNew}[1]{{\color{black}#1}}

\newcommand{\mlNew}[1]{{\color{black}#1}}

\newcommand{\suyuNew}[1]{\textcolor{black}{#1}}

\newcommand{\huberNew}[1]{\textcolor{black}{#1}}

\newcommand{\ttt}[1]{\texttt{#1}}

\newcommand{\snx}{$x_0$}

\newcommand{\sncolor}{$c$}
\newcommand{\daymax}{$T_0$}

\newcommand{\zfaint}{$z_{\mathrm{faint}}$}
\newcommand{\nsnfaint}{$N_{z<z_{\mathrm{faint}}}\ $}
\newcommand{\zmed}{$z_{\mathrm{med}}$}
\newcommand{\nsnmed}{$N_{z<z_{\mathrm{med}}}\ $}

\newcommand{\sne}{{SNe Ia}}
\newcommand\plasticc{PLAsTiCC}
\newcommand\tides{TiDES}

\newcommand{\snstretch}{$x_1$}

\newcommand{\zlim}{$z_{\mathrm{lim}}$}
\newcommand{\nsnlim}{$N_{z<z_{\mathrm{lim}}}$}

\newcommand{\authormacro}[1]{}

\graphicspath{{figures/}}

\begin{document}

\title{The Impact of Observing Strategy on Cosmological Constraints with LSST}

\input{authors}

\begin{abstract}

The generation-defining Vera C. Rubin Observatory will make state-of-the-art measurements of both the static and transient universe through its Legacy Survey for Space and Time (LSST). With such capabilities, it is immensely challenging to optimize the LSST observing strategy across the survey's wide range of science drivers. Many aspects of the LSST observing strategy relevant to the LSST Dark Energy Science Collaboration, such as survey footprint definition, single visit exposure time and the cadence of repeat visits in different filters, are yet to be finalized. Here, we present metrics used to assess the impact of observing strategy on the cosmological probes considered most sensitive to survey design; these are large-scale structure, weak lensing, type Ia supernovae, kilonovae and strong lens systems (as well as photometric redshifts, which enable many of these probes). We evaluate these metrics for over 100 different simulated potential survey designs. Our results show that multiple observing strategy decisions can profoundly impact cosmological constraints with LSST; these include adjusting the survey footprint, ensuring repeat nightly visits are taken in different filters and enforcing regular cadence. We provide public code for our metrics, which makes them readily available for evaluating further modifications to the survey design. We conclude with a set of recommendations and highlight observing strategy factors that require further research.
\end{abstract}

\section{Introduction}
\label{sec:introduction}
The Vera C. Rubin Observatory Legacy Survey of Space and Time (LSST), with its ability to make rapid, deep observations over a wide sky area, will deliver unprecedented advances in a diverse set of science cases (\citealp{abell2009}, hereafter The LSST Science Book).  

LSST has an ambitious range of science goals that span the Universe: solar system studies, mapping the Milky Way, astrophysical transients, and cosmology; 
these
are all to be achieved 
with a single ten-year survey. 
Around 80\% of LSST's observing time will be dedicated to the main or WFD survey, which will cover at least 18,000 deg$^2$. The remainder of the time will be dedicated to ``mini-surveys'' (for instance, a dedicated Galactic plane survey), ``deep drilling fields'' (DDFs) and potentially, ``targets of opportunity'' (ToOs).

Because LSST has such broad science goals, the choice of observing strategy is a difficult but critical problem. Important early groundwork was laid in the community-driven paper on LSST observing strategy optimization \citep[][hereafter COSEP]{COSEP}. To further address this challenge, in 2018, the LSST Project Science Team and the LSST Science Advisory Committee released a call for community white papers proposing observing strategies for the LSST WFD survey, as well as the DDFs and mini-surveys \citep{Ivezic2018}. In response to this call, the DESC Observing Strategy 
Working Group (hereafter DESC OSWG) performed a detailed investigation of the impact of observing strategy on cosmology with LSST. The DESC OSWG made an initial set of recommendations for both the WFD \citep{Lochner2018} and DDF \citep{Scolnic2018} surveys, with proposals for an observing strategy that will optimize cosmological measurements with LSST. \ds{Following this call, many new survey strategies have been simulated to answer the ideas in various white papers submitted; these strategies are discussed in \cite{lynne_jones}.  Furthermore, a Survey Cadence Optimization Committee (SCOC) was formed\footnote{\url{https://www.lsst.org/content/charge-survey-cadence-optimization-committee-scoc}} with the charge of guiding the convergence of survey strategy decisions across the multiple LSST collaborations.  
 The SCOC released a series of top-level survey strategy questions\footnote{\url{https://docushare.lsst.org/docushare/dsweb/Get/Document-36755}}, where answers can be supported using analyses of the simulations in \cite{lynne_jones}.  In this paper, we evaluate a number of simulated LSST observing strategies to support decisions on the survey strategy.} 

A review of the dark energy analyses planned by the DESC (which is a subset of the fundamental cosmological physics that will be probed by LSST) is given in the DESC Science Requirements Document (\citealp{DESCSRD2018}; hereafter DESC SRD).  Each analysis working group \mlNew{(weak lensing, large-scale structure, galaxy clusters, type Ia supernovae and strong lensing)} within DESC provided a forecast of the constraints on dark energy expected from their probe, given a baseline observing strategy.  As a metric, the DESC SRD used the Dark Energy Task Force Figure of Merit (DETF FoM), defined as the reciprocal of the area of the contour enclosing 68\% of the credible interval constraining the dark energy parameters, $w_0$ and $w_a$, after marginalizing over other parameters \citep{Albrecht2006}. Once statistical constraints were quantified, each group determined the control of systematic uncertainties needed to reach the goals for a Stage IV dark energy mission.  

Three ways to increase the likelihood of \mlNew{achieving the goals set out in the DESC SRD} are to (a) improve the statistical precision of each probe, (b) reduce each probe's sensitivity to systematic uncertainties, or (c) reduce the total uncertainty when combining multiple probes. Changes in observing strategy have the potential to \mlNew{affect each of these}. For instance, observing strategies that yield more uniform coverage across the survey footprint, or strategies with improved cadence, can have a strong impact on both the statistical precision and the systematic control.  

\ds{DESC encompasses multiple cosmological probes  and it is the ultimate goal of the DESC OSWG
to be able to compute the combined DETF FoM using all probes, as well as other combined metrics, for each proposed observing strategy.} However, at the level of LSST precision, careful treatment of systematic effects is required, and work is still ongoing to include these in the forecasting analysis tools. In addition, a full cosmological analysis is computationally intensive and not feasible to test on hundreds of simulated LSST observing strategies. Thus, while we include an emulated DETF FoM for certain dark-energy probes, we also introduce a suite of metrics that are anticipated to correlate with cosmological constraints but 
that 
are faster to run and simpler to interpret. Most of the metrics in this paper are focused on the WFD survey, but we make use of many of the same metrics (particularly for supernova cosmology) for the DDFs.
\mlNew{It should be noted that one of the cosmological probes mentioned, galaxy clusters, is not explicitly included in our analysis. This is because it is expected that clusters will have identical requirements to large-scale structure and so should already be accommodated.}

While the metrics we have developed focus entirely on the extragalactic part of the WFD survey, there is one cosmological probe that \mlNew{relates to observations near the Galactic plane}: the study of dark matter with microlensing. \ml{Microlensing is the light magnification due to the transit of a massive compact object (lens) close enough to the line of sight of a background star \citep{Paczynski_1986}. The search for the dark matter component of intermediate mass black holes within the Milky Way through microlensing involves several months timescale events, which can be efficiently detected only with long time-scale surveys such as LSST \citep{Mirhosseini_2018}.} This search will not be sensitive to the details of observing strategy, as long as gaps larger than a few months are avoided. Thus, for this work we focus only on extragalactic probes.

\ml{Although all the metrics described in this paper are useful for understanding the impact of observing strategy on cosmological measurements with LSST, some are more closely related to the primary cosmology goals as outlined in the DESC SRD than others. The 3x2pt correlation function, which measures structure growth, and supernovae, which probe the expansion history of the Universe, have the most constraining power. However, novel probes such as strong lensing and kilonovae can be complementary and offer unique tests of cosmology beyond the DETF FoM. Our recommendations and conclusions are generally guided by the priorities outlined in the DESC SRD, but we \mlNew{attempt to quantify performance of observing strategies in terms of the scientific opportunities offered by more novel probes as well}.}

We structure the paper as follows: \autoref{sec:obs_strat} outlines the factors that affect LSST observing strategy, the simulator used and resulting sets of simulations of different observing strategies. We split the metrics descriptions as follows: general static science metrics (\autoref{sec:generalstatic}); static science-driven metrics  (\autoref{sec:static}), which include WL, LSS and photometric redshifts; and transient science-driven metrics (\autoref{sec:transients}), which include supernovae, strong lensing of supernovae/quasars, and kilonovae. We draw together the results of the analysis of our metrics on various simulated observing strategies in \autoref{sec:discussion} and~\autoref{sec:conclusions}.
In addition to describing the analysis supporting the proposal for various observing strategy choices, we provide metrics, recommendations and conclusions in this paper that are meant to be of more  general use to future large-scale cosmology surveys.

\section{LSST Observing Strategy}
\label{sec:obs_strat}
\mlNew{In this section, we 
describe the Rubin Observatory,
the baseline LSST observing strategy, the software used to generate realistic LSST observing schedules that we make use of in this work, and the metrics framework used to evaluate different strategies.}
\subsection{LSST Overview}
\ds{
An overview of the Vera C. Rubin Observatory telescope specifications can be found in \cite{LSST2013}; we summarize here the specifications that impact observing strategy.  The Rubin Observatory is \mlNew{under construction} in the Southern Hemisphere, in Cerro Pach\'on in Northern Chile, \mlNew{and will undertake the Legacy Survey of Space and Time (LSST); a 10-year survey expected to start in 2023.} The system has an 8.4m (6.7m effective) diameter primary mirror, a 9.6 deg$^2$ field of view, and a 3.2 Gigapixel camera.  \mlNew{The integrated filter exchange system can hold up five filters at a time}, and there are six filters available: $ugrizy$ that cover a wavelength range of 320--1050 nm.  Typical \suyuNew{$5\sigma$} depth (i.e., the apparent brightness in magnitudes at which a point source is detected at $5\sigma$ significance) of 30\,s exposures in $ugrizy$ are 
$23.9,25.0,24.7,24.0,23.3,22.1$ mag and co-added over the full survey will reach approximately $26.1,27.4,27.5,26.8,26.1,24.9$ mag.  \mlNew{Several performance specifications influence the survey cadence\footnote{The cadence is defined as the median inter-night gap over a season.}}: filter change time (120\,s), closed optics loop delay or slews where altitude changes by more than 9 degrees (36\,s), read time (2\,s), shutter time (1\,s), median slew time between fields (4.94\,s)\footnote{These times are estimated from the expected performance from the various components of the telescope and camera and are what is used in the scheduling simulators}. The estimated fraction of photometric time is 53\% of the 10-year survey.  Standard `visit' exposures are typically 30 seconds long (referred to as $1\times30$\,s). An alternative exposure strategy, to mitigate cosmic-ray and satellite trail artifacts, are two successive short exposures called ``snaps'' (this is referred to as $2\times15$\,s). \mlNew{The decision between these exposure strategies has not yet been made. Throughout this paper, all simulations use $1\times30$\,s exposures, however we will return to this point in \autoref{sec:discussion} to explicitly examine the impact of using $2\times15$\,s exposures instead}.}

\newpage
\subsection{LSST Observing Strategy Requirements}

The observing strategy of LSST is impacted by several factors, and its optimization is a complex challenge. 
The LSST Science Requirements Document (\citealp{LSST2013}, hereafter LSST SRD)  defines top-level specifications for the survey such that:
\begin{itemize}
\item The sky area uniformly covered by the main survey will include at least 15,000 deg$^2$, and 18,000 deg$^2$ by design.
\item  The sum over all bands of the median number of visits in each band across the sky area
 will not be smaller than 750, with a design goal of 825.
 \item At least 1,000 deg$^2$ of the sky, with a design goal of 2,000 deg$^2$, will have multiple observations separated by nearly uniformly sampled time scales ranging from 40 sec to 30 min.
\end{itemize}
There are other additional requirements on PSF ellipticity correlations and parallax constraints, the former of which will be indirectly analyzed in this paper.

\mlNew{Given that these requirements only constrain a few aspects of observing strategy, many remaining factors can still be optimized to maximize scientific return}.

\subsection{Baseline Strategy}

Here we summarize the baseline observing strategy \citep{lynne_jones}, \mlNew{which has evolved significantly over the years. The strategy described here is} considered the current nominal observing strategy plan pending further modifications:

\begin{itemize}
\item Visits are always $1\times30$\,s long (not $2\times15$\,s)\footnote{\ml{We note that the choice between $1\times30$\,s and $2\times15$\,s exposures will not be finalized until the commissioning phase of the LSST survey.}}. The baseline simulation achieves about 2.2M visits over 10 years.
\item Pairs of visits in each night are in two filters as follows: $u-g$, $u-r$, $g-r$, $r-i$, $i-z$, $z-y$ or $y-y$. Pairs are scheduled for approximately 22 minutes separation. Almost every visit in $gr$ or $i$ has another visit within 50 minutes. \ml{These visit pairs assist with asteroid identification.}
\item \ds{ The survey footprint is the standard baseline footprint, with 18,000 deg$^2$ in the WFD survey spanning declinations from -62 to +2 degrees (excluding the Galactic equator), and additional coverage for the North Ecliptic Spur (NES), the Galactic Plane (GP) and South Celestial Pole (SCP). The baseline footprint includes WFD, NES ($griz$ only), GP, and SCP.  WFD is $\sim82\%$ of the total time.}

\item \ds{Five DDFs are included\footnote{Details of the four selected DDFs can be found here: \url{https://www.lsst.org/scientists/survey-design/ddf}}, with the fifth field being composed of two pointings covering the Euclid Deep Field - South (EDF-S)\footnote{\ds{This field has not been officially confirmed as part of the LSST survey.  Information on the Euclid field can be found here: \url{https://www.cosmos.esa.int/web/euclid/euclid-survey}}}, devoting 5\% of the total survey time to DD fields.}
\item The standard balance of visits between filters is 6\% in $u$, 9\% in $g$, 22\% in $r$, 22\% in $i$, 20\% in $z$, and 21\% in $y$.

\item \ds{Owing to the limitation of five installed filters in the camera filter exchange system, if at the start of the night the moon is 40\% illuminated or more (corresponding to approximately full moon +/- 6 nights), the $z$-band filter is installed; otherwise the $u$-band filter is installed.}

\item \ds{The camera is rotationally dithered nightly between -80 and 80 degrees.  At the beginning of each night, the rotation is randomly selected.  The \mlNew{camera is rotated to cancel field rotation during an exposure} then reverts back to the chosen rotation angle for the next exposure.  }
\item \ds{Twilight observations are done in $r$,$i$,$z$,$y$, and are determined by a `greedy' algorithm which takes the maximum of the reward function at each observation. }

\item Non-rolling cadence: the nominal baseline strategy observes the entire footprint each observing season. A \emph{rolling cadence} would prioritize sections of the footprint at different times (e.g., observing half the footprint for one year and changing to the other half the next) to improve cadence in that section.

\end{itemize}

\subsection{Survey Simulators}
\label{sec:obs_strat_simulators}

\mlNew{The simulations analyzed here are created using the Feature-Based Scheduler (FBS, \citealp{Naghib2018}), which uses a modified Markov Decision Process to decide \mlNew{the next observing direction and filter selection}, allowing a flexible approach to scheduling, including the ability to compute a detailed reward function throughout a night. FBS is the new default scheduler for the LSST, replacing the LSST Operations Simulator \citep[OpSim][]{Ridgway2010,Delgado2014}.}

We note that, besides the LSST default schedulers (OpSim and FBS), there is an alternate scheduler, AltSched, presented in \citet{Rothchild2019}. AltSched is a simple, deterministic scheduler, 
which ensures
that telescope
observations take place as close to the meridian as possible, alternating between sky regions North and South of the observatory latitude on alternate nights, while cycling through the filter set and changing filters after observing blocks.  We do not include simulations from this scheduler; however we \mlNew{note that its approach does produce encouraging results.}

\subsection{Observing Strategy Simulations}
\label{sec:obs_strat_sims}

 Sets of simulations have been periodically released \mlNew{for use by the community}.  \autoref{tab:simfam}, we summarize the families of simulations used, number of simulations in each family, and versions\footnote{Simulations can be downloaded at \url{http://astro-lsst-01.astro.washington.edu:8081/}}. 
\mlNew{New versions of simulated strategies are released regularly with improvements to the scheduler, weather simulation and changes to the baseline strategy.} In this paper, we mostly focus on version 1.5 simulations, but select v1.6 and v1.7 simulations are included in certain plots \ml{\citep[see][for details of the simulations]{lynne_jones}}. \mlNew{Certain simulations are excluded from specific plots because they are unrealistic or differ significantly from the baseline (for instance, with a dramatically different footprint or visit allocation in WFD).} It is very important to note that for each version, the baseline changes somewhat. \mlNew{In particular, the default choice of exposure strategy has changed from $2\times15$\,s in older versions, to $1\times30$\,s in v1.5 and v1.6 and back to $2\times15$\,s in v1.7, which has a large impact on overall efficiency and hence metric performance.} All figures in this paper are for relative improvements compared to the baseline \ml{strategy corresponding to that simulation's version}. \ml{\autoref{sec:simulations_appendix} captures in detail exactly which simulations are used for which plot and the corresponding baseline simulation. \autoref{tab:sims_short} is a lookup table for the short, simpler names for the simulations used in some figures in this paper.}

\begin{table*}
\begin{centering}
\label{tab:simfam}
\caption{Description of FBS simulation families used including a descriptive name of each family, the FBS version, the number of simulations, and a brief explanation of the family.}
 \centering
 \begin{tabular}{lllp{8.5cm}}
 
 \hline
 Name & FBS Version & \# of simulations & Description \\
 \hline
Baseline & 1.4/1.5/1.6 & 3 & Baseline as described, with choice of 2 versus 1 snap, and mixed filters or not\\
U\_pairs & 1.4 & 12 & Varies how $u$-band visits are paired with other filters, how many visits occur in $u$-band and when $u$-band is loaded in or out of filter\\
Third\_obs & 1.5 & 6 & \ds{Adds third visit to some fields night, where total amount of time dedicated to these visits is 15--120 minutes} per night\\
Wfd\_depth & 1.5 & 16 & Amount spent on WFD compared to other areas changes from 60--99\%\\
Footprints & 1.5 & 12 & Changes in WFD footprint, in North/South, Galactic coverage, Large/Small Magellanic Clouds \\
Bulge & 1.5 & 6 & Different strategies for observing the Galactic bulge \\
Filter\_dist & 1.5 & 8 & Varying the ratios of time spent in $u,g,r,i,z,y$ filters \\
Alt\_roll\_dust & 1.5 & 3 & Dust-limited WFD footprint with the alt-scheduler scheduling algorithm, with and without rolling, \ds{where a rolling strategy only observes a set fraction of the survey footprint each year}  \\
DDFs & 1.5 & 3 & Different strategies for the DDFs, ranging from 3--5.5\% of the total survey time\\
Goodseeing & 1.5 &5 & Aims to acquire at least one good-seeing visit at each field each year, varies which filters it's needed for \\
Twilight\_neo & 1.5 & 4 & Adds a mini-survey during twilight to search for \ds{Near-Earth Objects (NEOs)} \\
Short exposures & 1.5 & 5 & Adds short exposures in all filters, from 1--5 seconds, 2--5 exposures per year \\
U60 & 1.5 & 1 & Swaps 60 second $u$-band visit instead of 30 s \\
Var\_expt & 1.5 & 1 & Changes exposure time so that the single image depth is constant \\
DCR & 1.5 & 6 & Adds high airmass observations in different combinations of filters 1 or 2 times per year \\
Even\_filters & 1.6 & 4 &  Bluer filters are observed in moon bright time \\
Greedy footprint & 1.5 & 1 & \ds{A greedy survey not run on ecliptic, \mlNew{where} a portion of the sky that has the highest reward function is observed two times over a given time span (about 20--40 minutes).}\\

Potential Schedulers & 1.6 & 17 & Multiple variations at once for a particular science goal. \\

 \hline 
 \end{tabular}
 \end{centering}
\end{table*}

\subsection{Proxy Metrics and Metrics Analysis Framework }
\label{sec:maf}
As stated in \autoref{sec:introduction}, the ultimate goal of the DESC OSWG is to compute cosmology figures of merit to evaluate observing strategies. However, this is difficult and computationally intensive, \mlNew{making such an approach impractical for evaluating many simulations}. We thus largely focus on proxy metrics, which can be quickly computed on any simulation. We make use of and incorporate our metrics into the Metrics Analysis Framework \citep[MAF,][]{jones2014}. MAF is a python framework designed to easily evaluate aspects of the simulated strategies. It can compute simple quantities, such as the total co-added depth or number of visits, but it can also be used to construct more complex metrics that can evaluate the \mlNew{expected performance of a simulation for a given science case.} Here, we use a combination of independent codes (which are too slow to run as part of MAF) and custom MAF \mlNew{metrics to evaluate the simulated observing strategies. MAF metrics and external metrics created for this paper are linked when described.} \mlNew{Unless otherwise specified, each metric is run on a simulation of the full ten year survey.}

\ml{We note that in order to compare metrics directly, they must be transformed to be able to be interpreted as ``larger is better'' and placed on the same scale. \autoref{sec:metrics_appendix} includes a table that describes how all metrics are transformed. In all plots, to put the metrics on the same scale, they are normalized using their values for the baseline simulation \mlNew{(which is different for each FBS version)} as:

\begin{equation}
    x_{\rm{normed}} = \frac{x - x_{\rm{baseline}}}{x_{\rm{baseline}}},
\end{equation}
where $x$ is the metric in question, \mlNew{ $x_{\rm{baseline}}$ is the value of that metric for the corresponding baseline simulation and  $x_{\rm{normed}}$ is the normalized metric.}

A final point to note before introducing the metrics is that the focus of this paper is the optimization of the WFD survey; thus all metrics are evaluated only on the WFD observations of each simulation. However, we include some discussion of DDF optimisation in \autoref{sec:future}, which is particularly important for supernovae and also photometric redshift calibration.}

\begin{table*}
\centering
\caption{A lookup table for the shorter, simpler names used in the figures in  \autoref{sec:static} and \autoref{sec:transients}.}
\label{tab:sims_short}
\begin{tabular}{ll}
\hline
\input{sims_short.tbl}
\end{tabular}

\end{table*}

\section{General Static Science Metrics}
\label{sec:generalstatic}
In this section, we introduce metrics \mlNew{relevant to static science topics} that will be useful to multiple cosmological probes.  This \mlNew{includes} metrics related to general WFD characteristics as well as to photometric redshift 
(photo-$z$) 
characteristics from the WFD galaxy sample.

\subsection{WFD Depth, Uniformity and Area}
\label{sec:static_metrics}
\authormacro{H. Awan, E. Gawiser}

\begin{figure}

\centering
\includegraphics[width=1\linewidth]{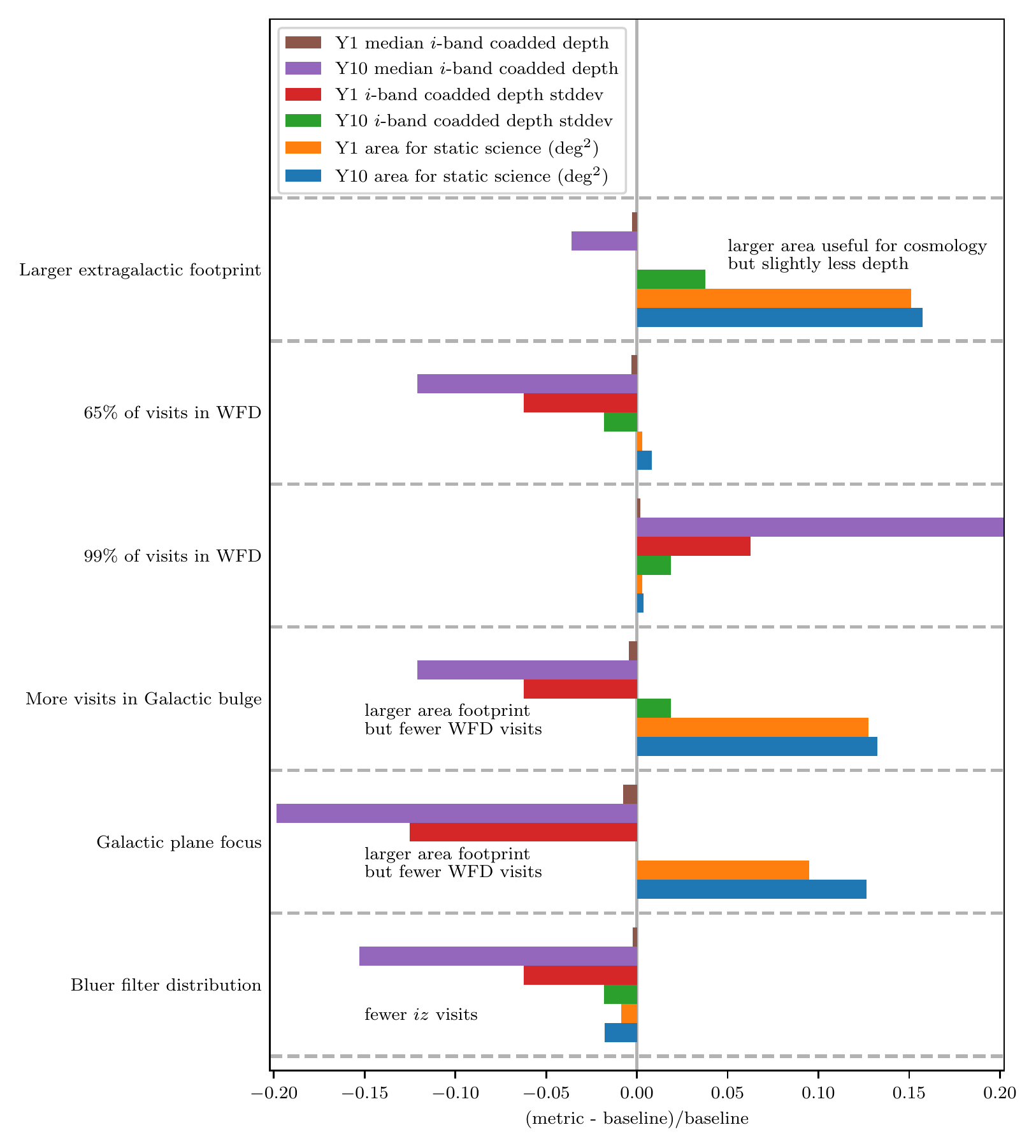}
\caption{Static science metrics as a function of selected observing strategies. \ml{\autoref{tab:sims_short} contains the exact simulation names corresponding to the short names used here. Note that as for all plots in this paper, metrics are transformed using the equations in \autoref{tab:metric_transforms} and are taken relative to their values at baseline in order to be directly comparable,} \EG{with larger values always being better.}}
\label{fig:static}
\end{figure}

In this subsection, we introduce three sets of metrics, where each set includes information after years Y1, Y3, Y6 and Y10 respectively \ml{(where Y1 refers to the data collected after the first year, Y3 after the first three years, etc.)}.  \autoref{fig:static} shows the results from these metrics.

\begin{enumerate}
\item Y1/Y3/Y6/Y10 area for static science (deg$^2$): The effective survey area for static science after Y1/Y3/Y6/Y10\humnaNew{; this area is limited by depth and extinction, and requires coverage in all 6 bands, and is described in more detail below. Note that this area is also referred to as ``the extragalactic area" in later discussion}.
\item Y1/Y3/Y6/Y10 median $i$-band coadded depth:  The median $i$-band \humnaNew{5$\sigma$} coadded depth in \humnaNew{the effective survey area 
for static science} after Y1/Y3/Y6/Y10.
\item Y1/Y3/Y6/Y10 $i$-band coadded depth stddev:  The standard deviation of \humnaNew{the} $i$-band \humnaNew{5$\sigma$} coadded depth \humnaNew{distribution }in \humnaNew{the }effective survey area for static science after \humna{Y1/Y3/Y6/Y10, quantifying the non-uniformity in depth coverage; smaller values of this metric indicate a more uniform survey.}
\end{enumerate}

We follow the LSST Science Book in using $i$-band to track the brightness \EG{limit} of galaxies that can be detected in the survey, motivated by the fact that 
\EG{almost all galaxies are brighter in $i$ than in $g$ or $r$,}
\mlNew{while the co-added depths are similar for these three filters}.
We note that this could be misleading when comparing observing strategies that vary the relative time spent observing in $i$ versus other bands by a significant factor.

\humna{The metrics above are calculated using \ttt{HEALPix}\footnote{\url{http://healpix.sourceforge.net/}} \citep{healpix} maps, with pixel resolution of 13.74 arcmin (achieved using the \ttt{HEALPix} resolution parameter $N_\mathrm{side}$ = 256)}. For extragalactic static science using high S/N measurements, we must restrict our analysis to a footprint that provides the deep, high S/N galaxy samples  needed for our science. To achieve this, we implement an extinction cut and a depth cut, retaining only the area with E(B-V)$<$0.2 (where E(B-V) is the dust reddening in magnitudes) with limiting $i$-band coadded 5$\sigma$ point source detection depths of \humnaNew{24.65 for Y1, 25.25 for Y3, 25.62 for Y6, and 25.9 for Y10}; the E(B-V) cut ensures that we consider the area with small dust uncertainties\footnote{As noted by e.g., \citet{leike+2019}, the behavior of Galactic dust becomes more uncertain as the amount of dust increases, and  \citet{schlafly+2011} identify E(B-V)=0.2 as a threshold where the dust properties change.}  while the depth cut ensures that we have high S/N galaxies, with the Y10 cut fixed by the LSST SRD goal of yielding a ``gold sample" of galaxies with \humnaNew{$i$-band coadded 5$\sigma$ (extended source detection depth), $i<25.3$} after Y10. This is achieved using a the \ttt{MAF Metric} object, \ttt{egFootprintMetric}\footnote{ \url{https://github.com/humnaawan/sims_maf_contrib/blob/master/mafContrib/lssmetrics/egFootprintMetric.py}}. 

\subsubsection{Uniformity and dithering}
\label{sec:uniformity}

\ml{Survey uniformity, as measured by our $i$-band coadded depth stddev metric, is critical for all static science probes. Non-uniformity can be introduced by spending more observing time or having better atmospheric conditions in certain parts of the sky,  or when a survey is tiled and the overlaps in fields introduce an artificial structure to the survey. The latter effect can be effectively mitigated using dithering: small offsets in the pointing of the telescope when it returns to a field \citep[see, e.g.,][]{Awan+2016}. Dithering can be translational or rotational, both of which are useful for reducing different types of systematics. \mlNew{\autoref{fig:static} shows that the stddev metric varies by less than 5\% across the simulations}, meaning that the current observing strategy proposals are implementing effective translational dithering strategies. While most metrics in this paper focus on the performance of the full ten-year LSST survey, we note that it is important that survey uniformity is achieved at specific release intervals, such as Y1, Y3, Y6 and Y10, in order to enable periodic analyses of datasets suitable for cosmology. The current baseline achieves this by default, but it will be important to consider if a strategy is chosen whereby only parts of the footprint are observed each season (so-called rolling cadence).}

\subsubsection{General Conclusions from Static Science}

Static science systematics can be reduced by increasing survey uniformity via frequent translational and rotational dithers; the impacts of these can be probe-specific, as noted below.  We note that Y1 is especially sensitive to the specific cadence, and while the different kinds of cadences/footprints converge for Y3-Y10 area, very few simulations yield close to the desired 18,000 deg$^2$ WFD area for extragalactic science. 

We also note that spectroscopic observations in the extragalactic part of the survey will be essential to calibrate LSST photometric redshifts. \mlNew{For this purpose, overlap with  upcoming spectroscopic surveys is quite critical and is further discussed in \autoref{sec:discussion}.}

\subsection{Photometric Redshifts}
\authormacro{M.~L.~Graham, H.~Awan}
\label{sec:pz}
\mlNew{While photometric redshifts impact multiple probes, including transients such as supernovae, they are in turn not affected by time-sensitive aspects of observing strategy. We thus generally include photo-$z$ metrics with the static science metrics.} 

\begin{figure}
    \centering
    \includegraphics[width=1\linewidth]{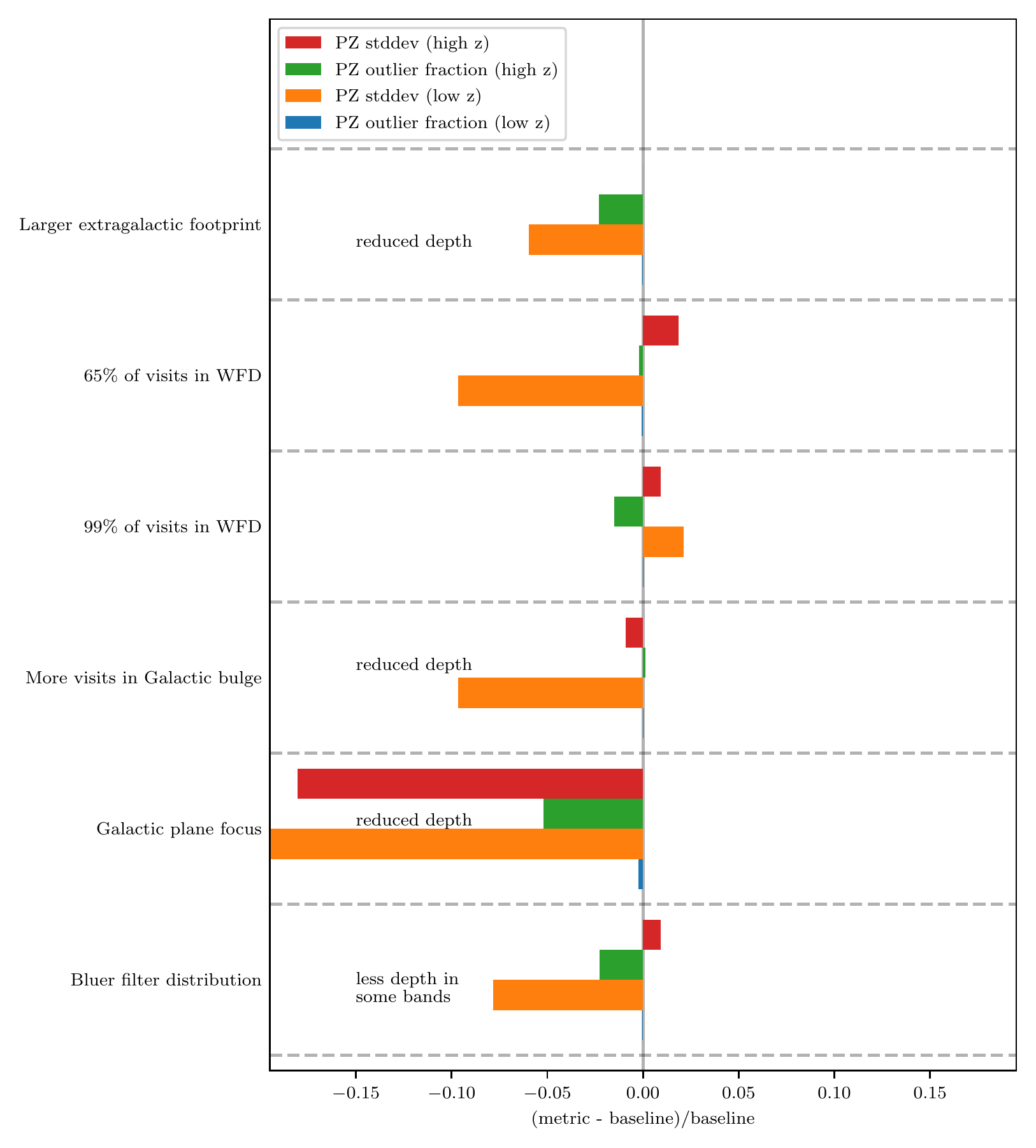}
    \caption{Photometric redshift metrics as a function of selected observing strategies. \ml{\autoref{tab:sims_short} contains the exact simulation names corresponding to the short names used here. As described in \autoref{sec:pz}, the low $z$ bin is (0.6--1.0) and the high $z$ bin is (1.8--2.2).}}
    \label{fig:pz}
\end{figure}

We introduce four metrics for the quality of photo-$z$ determination in the WFD. A summary of the results of these metrics can be seen in \autoref{fig:pz}.
\begin{enumerate}
\item Photo-$z$ standard deviation at high (1.8--2.2) and at low (0.6--1.0) redshift.
\item Outlier fraction at high and at low redshift.
\end{enumerate}
 
We evaluate the relative quality of simulated photometric redshift estimates for each simulation by determining the average coadded depth in extragalactic fields, and using those depths to simulate observed apparent magnitudes and photo-$z$ for a mock galaxy catalog using the color-matched nearest neighbors (CMNN) photometric redshift estimator \cite{2018AJ....155....1G}\footnote{A demonstration of the CMNN photo-$z$ estimator is available on {\tt GitHub} at \url{https://github.com/dirac-institute/CMNN_Photoz_Estimator}.}.
The CMNN estimator does not produce the ``best" or ``official" LSST photo-$z$, but does produce results for which the {\it relative quality} of the photo-$z$ is directly related to the input photometric quality, and thus is an appropriate photo-$z$ estimator for evaluating the bulk impact on photo-$z$ results due to changes in the photometric depth of a survey.

First, we determine the 5$\sigma$ point-source limiting magnitudes of the 10-year coadded images from the WFD program in sky regions ($\sim$220 arcmin wide) for all simulations.
We consider extragalactic fields as appropriate for cosmological studies if their Galactic dust extinction is E(B-V)$<$0.2 mag, and if they receive at least 5 visits per year in all six filters (i.e., the 5 visits define a minimum coadded depth).
The median 10-year 6-filter depths across all such appropriate extragalactic fields for each simulation are then input to the CMNN photo-$z$ estimator.
The CMNN estimator uses the depths to to synthesize apparent observed magnitudes and errors for a simulated galaxy catalog; then it splits the catalog into test and training sets, and uses the training set to estimate photo-$z$ for the test set. 
The 5$\sigma$ depths are the only input; thus, only observing strategies which result in photometric depths that differ significantly from the baseline cadence will result in significantly different photometric redshift results.

We used the same mock galaxy catalog as described in \cite{2018AJ....155....1G} and \cite{2020AJ....159..258G}, which is based on the Millennium simulation \citep{2005Natur.435..629S} and the galaxy formation models of \citet{2014MNRAS.439..264G}, and was fabricated using the lightcone construction techniques described by \citet{2013MNRAS.429..556M}\footnote{Documentation for this catalog can be found at \url{http://galaxy-catalogue.dur.ac.uk}}.
To both the test and training sets we apply cuts on the observed apparent magnitudes of 25.0, 26.0, 26.0, 25.0, 24.8, and 24.0 mag in filters {\it ugrizy}, and enforce that all galaxies are detected in all six filters.
These cuts are all about half a magnitude brighter than the brightest 5$\sigma$ depth of any given simulation we considered.
This cut is applied because it imposes a galaxy detection limit across all simulations that is independent of the depth (i.e., independent of the photometric quality).
If such a cut is not imposed, the default setting is for the CMNN estimator to apply a cut equal to the 5$\sigma$ limiting magnitude.
This default setting results in more fainter galaxies being included in the test and training sets for simulations with deeper coadds.
Although this default setting is realistic -- fainter galaxies are included in real galaxy catalogs made from deeper coadds -- due to the fact that fainter galaxies generally have poorer-quality photo-$z$ estimates, this also results in some simulations with deeper coadded depths appearing to produce {\it worse} photo-$z$ estimates.
These bright magnitude cuts ensure an ``apples-to-apples" comparison across all simulations. 

All other CMNN estimator input parameters are left at their default values, except for the number of test (\humna{50,000}) and training (\humna{200,000}) set galaxies, and the minimum number of colors which is set to 5 (from a default of 3) to only include galaxies that are detected in all six filters.
The other CMNN parameters\footnote{Other CMNN parameters include, e.g., the minimum number of CMNN training-set galaxies, the mode of determining the photo-$z$ from the CMNN subset, and the percent-point function value which is used to generate the CMNN subset.} impact the final absolute photo-$z$ quality, and so it is important to keep in mind that the results of the CMNN estimator should not be interpreted as absolute predictions for the photo-$z$ quality, but as {\it relative} predictions for the photo-$z$ quality produced by different observing strategies (i.e., different 10-year coadded depths).
It is important to note that, because the test and training sets are drawn from the same population, they have the same apparent magnitude and redshift distributions.
This contrived scenario in which the training set is perfectly representative of the test set does not produce photo-$z$ results with realistic systematic effects or biases.
Additionally, we use input parameters for the CMNN estimator that produce photo-$z$s with a very good absolute quality.
The combination of nearly perfectly matched test and training sets, optimized input parameters, and bright magnitude cuts all result in very small bias values (where bias is the average of $|z_{true}-z_{phot}|$ over all test galaxies) for our simulations, which is why the photo-$z$ bias is not being used as one of the metrics (described below) for evaluating the simulations.
In future photo-$z$ simulations, variations in the test and training sets could be established which correspond to different simulations (e.g., building a deep training set from the DDFs) -- but we consider this out of the scope of the present analysis.

We evaluate the photo-$z$ results with two statistics: precision and inlier fraction.
To calculate the precision, we first reject catastrophic outliers with $|z_{\rm true}-z_{\rm phot}|>1.5$ (this is a non-standard definition, chosen for this simulated data set, and used also in \citealt{2020AJ....159..258G}).
Then we calculate the robust standard deviation in the photo-$z$ error, $\Delta z _{1+z} = (z_{\rm true}-z_{\rm phot})/(1+z_{\rm phot})$, as the full-width at half maximum (FWHM) of the interquartile range (IQR) divided by $1.349$, and use this as the precision.
The inlier fraction is one minus the outlier fraction, which is the number of galaxies with a photo-$z$ error greater than three times the standard deviation or greater than three times 0.06, whichever is larger (this definition matches that used for photo-$z$ outliers by the LSST SRD. 
We calculate these two statistics for a low-$z_{\rm phot}$ bin (0.6-1.0) and a high-$z_{\rm phot}$ bin (1.8-2.2), for each of the simulations.

The results are shown in \autoref{fig:pz}.
As mentioned above, the results of the CMNN estimator should be considered as {\it relative} predictions for the photo-$z$ quality.
Thus in \autoref{fig:pz} we show the results as {\it fractional} changes from the results for the baseline simulation.  

\subsubsection{General Conclusions from Photometric Redshifts} 

We find that the photo-$z$ quality is optimized by observing strategies that lead to deeper limiting magnitudes, which is as expected.
As there is a trade-off in any survey between depth and area, and because areal coverage is required by a variety of LSST science goals, we recognize that the observing strategies which optimize photo-$z$ quality might not be optimal for cosmological studies. 
Different science cases have different photo-$z$ needs, and a single photo-$z$ metric that captures the science impact is very challenging to define.
Full end-to-end simulations of scientific results that incorporate photo-$z$ quality would be the correct approach, but are beyond the scope of this work.

\section{Weak Lensing and Large-Scale Structure}
\label{sec:static}

\subsection{Weak Lensing}
\authormacro{R. Mandelbaum, T. Eifler, H. Almoubayyed}
\label{sec:wl}
Weak gravitational lensing \rachelNew{(WL)} is the deflection of light from distant sources due to the gravitational influence of matter along the line-of-sight.  In practice, the coherent distortions of background galaxy shapes, or `shear' (measured in different redshift ranges), reveal the clustering of matter as a function of time, including both luminous and dark matter.  The evolution of matter clustering is affected by the expansion history of the Universe, which means that \rachelNew{WL} is also sensitive to the accelerated expansion rate of the Universe caused by dark energy \citep[for a review, see][]{2015RPPh...78h6901K}.  

\begin{figure}[htb]
    \centering
    \includegraphics[width=1.0\linewidth]{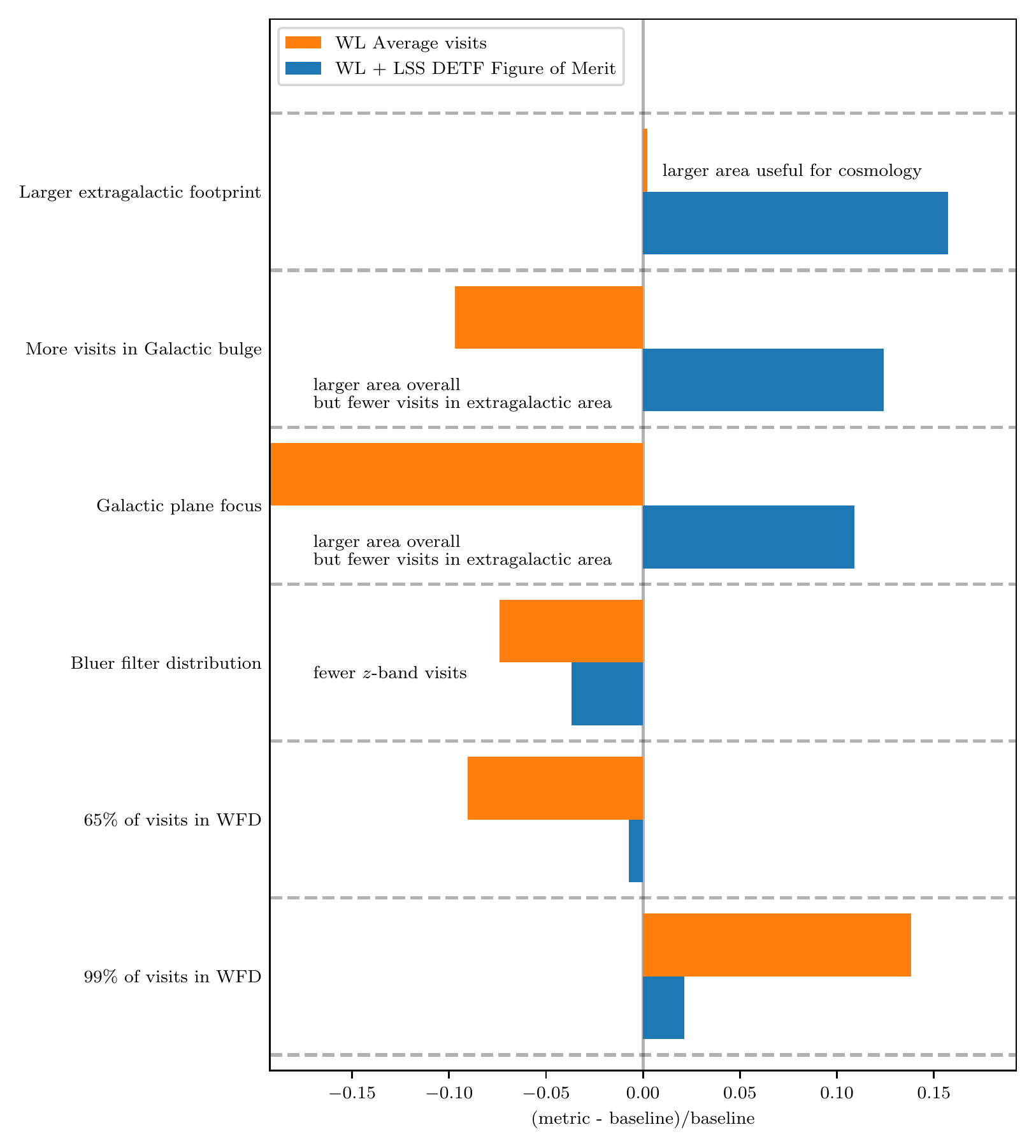}
    \caption{WL metrics as a function of selected observing strategies. \ml{\autoref{tab:sims_short} contains the exact simulation names corresponding to the short names used here.} \rachel{As shown, the metrics that describe statistical constraining power (`WL + LSS DETF Figure of Merit') and reduction in systematic uncertainties in \rachelNew{WL} shear (`WL Average visits') do not necessarily correlate, as the former tends to prefer increased area while the latter tends to prefer an increased number of visits at each point in the WFD survey on average.  Thus, observing strategy changes such as increasing the fraction of time in the WFD survey can increase both metrics, while observing strategy changes that increase area at the expense of decreasing the \husniNew{average number of visits across the survey area} will lead to opposing changes in the metrics.}}
    \label{fig:wl}
\end{figure}

We introduce two new metrics associated with \rachelNew{WL}.  A summary of the results from the WL metrics can be found in \autoref{fig:wl}.


\begin{enumerate}
    \item WL+LSS 
    Figure of Merit (\autoref{sec:static_fom})- \mlNew{DETF} Figure of Merit for \rachel{cosmological \rachelNew{WL} and large-scale structure measurement}; larger numbers correspond to larger statistical constraining power.
    \item WL Average visits (\autoref{sec:wl_sys})- Average visits metric in $r$, $i$, and $z$-bands; higher numbers are better for WL shear systematics mitigation.
\end{enumerate}

\subsubsection{WL+LSS Figure of Merit}
\label{sec:static_fom}

\husniNew{The 3x2pt (DETF) Figure of Merit (FoM) is the inverse of the area of the 68\%  confidence interval in the space of the dark energy equation of state parameters $w_0, w_a$}, and is an indicator of the statistical constraining power of the survey's \husniNew{static probes}. In this section, we follow the convention taken by ongoing \rachelNew{WL} surveys that the canonical analysis is a joint measurement of \rachelNew{WL} and large-scale structure.  \rachelNew{Since forecasting the cosmological constraining power for such a measurement is quite resource-intensive, our approach is to carry out the calculation for a limited number of survey strategies, and use that to build an emulator of the 3x2pt FoM for arbitrary scenarios based on interpolation.} This section briefly describes  \rachelNew{the emulation process for the 3x2pt FoM}; for more detail, see Eifler \& Motka {\em in prep}.\footnote{\url{https://github.com/hsnee/sims_maf/blob/master/python/lsst/sims/maf/metrics/summaryMetrics.py}}  

The emulator is defined in a six-dimensional parameter space, where the dimensions are: effective survey area, survey median depth in $i$-band, systematic uncertainty in the \rachelNew{WL} shear calibration, photometric redshift scatter, and the size of priors on photometric redshift bias and scatter.  A total of 36 points in this parameter space were selected to build the emulator, following a Latin hypercube design (often used for efficient sampling of high-dimensional parameter spaces in cosmological emulators - e.g., \citealt{2021MNRAS.502.1401M}).  For each selected point in this parameter space, the galaxy redshift distributions, and the observable quantities for joint \rachelNew{WL} and large-scale structure measurement, and their covariance matrix are calculated using \textsc{CosmoLike} \citep{2017MNRAS.470.2100K}. \rachel{This joint \rachelNew{WL} and large-scale structure measurement is often referred to as `3x2pt', as it involves the combination of three two-point correlation functions: shear-shear, galaxy-shear, and galaxy-galaxy correlations.}   The 3x2pt FoM is then calculated using MCMC constraints on cosmological parameters (marginalizing over key sources of astrophysical systematic uncertainty, such as galaxy bias, intrinsic alignments, and baryonic physics) in a simulated likelihood analysis based on the observables and their covariances. \rachelNew{The reason to marginalize over those systematic uncertainties when computing the FoM is that this is the process used in a real WL analysis to propagate the aforementioned systematic uncertainties into uncertainties on cosmological parameters.} The emulator was then built from those 36 FoM values in the six-dimensional space using a Gaussian process regression.

Area and median depth are calculated after making an extinction cut of $E (B-V) < 0.2$, to exclude high extinction areas, along with minimum depth cuts  of 24.5 and 25.9 for Y1 and Y10 respectively, to ensure that the survey depth is relatively homogeneous; as well as a cut that guarantees at least some coverage in all 6 bands to ensure photo-$z$ quality. \husniNew{These cuts are consistent with the extragalactic cuts applied in Section \ref{sec:static_metrics}}.
See the DESC SRD for more details about sample definition and methods of estimating redshift distributions and number counts, though the figures of merit in that document were calculated with slightly different choices of redshift binning and modeling of systematic uncertainty.

Note that the plots in this paper only use the area and depth dimensions of the emulator, with the remaining parameters fixed to their fiducial values from the DESC SRD. \husni{We used the emulator with marginalization over default values for the photometric redshift systematics parameters; therefore the only effects it captures are the varying area and median depth of the different survey strategies. The effect of area and median depth changes are the dominant factors for the emulator, \husniNew{with changes in photometric redshift systematics being subdominant}. For example, a 15\% change in photometric redshift variance (maximum change for the strategies considered in this work) would cause a $\sim$ 2\% change in the 3x2pt FoM (Eifler \& Motka {\em in prep}), while a similar change in area or median depth changes the 3x2pt FoM by $\sim$ 10\%. Changes due to the priors on the photometric redshift variance and bias have negligible effects for the type of strategies considered (\husniNew{specifically, the considered changes of overlap with spectroscopic surveys used for calibrating photo-$z$ errors}), and are likewise not included in this work.}  As a rule, the 3x2pt FoM prefers greater median depths and larger survey areas. The 3x2pt FoM metric described here is an improved version of the associated metric presented in \cite{Lochner2018}; while the improved version includes more sources of systematic uncertainty, its trends with survey area and depth are similar to those in \cite{Lochner2018}.

\subsubsection{WL Average visits and WL Shear Systematics} 
\label{sec:wl_sys}

The statistical constraining power for \rachelNew{WL} was covered in the previous subsection.  For this reason, the text below summarizes observing strategy considerations related to \rachelNew{WL} shear systematics in the WFD survey, for which a metric has been introduced in MAF\footnote{\url{https://github.com/lsst/sims_maf/blob/master/python/lsst/sims/maf/metrics/weakLensingSystematicsMetric.py}}.  See \cite{2020MNRAS.499.1140A} for further detail.

\rachelNew{WL}
analysis 
typically involves measuring coherent patterns in galaxy shapes due to WL shear.  For this reason, any effects that are not associated with \rachelNew{WL} but that cause apparent galaxy shape distortions with any spatial coherence must be well understood and controlled to avoid the measurement being systematics-dominated.  
The LSST provides a new opportunity to control WL systematics using the observing strategy. This opportunity was not as feasible in previous surveys because the LSST will be the first survey to dither at large scales (relative to the field of view) with a very large number of exposures.   This large number of exposures means that a source of systematic error with a particular spatial direction in one exposure may contribute with a different direction in other exposures for a given object, thus reducing the amount of systematic error that must be controlled in the image analysis process. Similar studies for systematics associated with differential chromatic refraction and CCD fixed frame distortions were conducted in the COSEP. and were found to be minimized for a uniform distribution of parallactic angles and \rachel{the position angle of the LSST camera} over all visits.

Additive shear systematics, such as those induced by errors in modeling the point-spread function (PSF) and errors arising from the CCD charge transfer, often have a coherent spatial pattern in single exposures. This type of systematic can potentially be mitigated and averaged down in coadded images, depending on the details of the dithering and observing strategy

In \cite{2020MNRAS.499.1140A}, we developed a physically-motivated \husniNew{analysis} related to the impact of observing strategy WL shear systematics, then used our findings with it to design a simpler proxy metric.  Here, we describe the physically-motivated \husniNew{analysis in four steps}: (a) We select a large number (e.g., 100,000) of random points at which the PSF is to be sampled, distributed uniformly in the WFD area of each survey, with cuts based on the co-added depth and dust extinction as explained in \autoref{sec:static_metrics}. (b) We create a toy model for the PSF modeling errors as a function of position in each exposure as a radial error in the outer 20\% of field of view (and no error in the inner 80\%), and for modeling CCD charge transfer errors and the brighter-fatter effect, we use a horizontal (CCD readout direction) error over the stars in the entire field of view. This model is motivated by observed spatial patterns in PSF model errors in ongoing surveys \citep{bosche2018,jarvis2016}. (c) To approximate the effect of coaddition, we average down the modeling errors across exposures via their second moments, since the coaddition process is linear in the image intensity and therefore in the (unweighted) moments. (d) We propagate the systematic errors for the \rachelNew{PSF in the} coadd\rachelNew{ed image} into the bias on the cosmic shear using the $\rho$-statistics formulation \citep{rowe2010,jarvis2016}. 

To create a proxy metric \husniNew{that connects more directly with survey parameters and is more practical to run for every simulation}, we note that given 
a \husniNew{chosen} dithering pattern (e.g., a random translational dither per visit with random rotational dithering at every filter change), the reduction in systematic errors is directly related to the number of visits that are used in \rachelNew{WL} analysis. This is due to the fact that more visits leads to a better sampling of a rotationally uniform distribution around the coherent direction associated with the additive shear systematic. We, therefore, use the 
average number of
$r$, $i$, and $z$-band visits for a large number of  objects -- for practical reasons, picked as the centers of cells in a HEALPix grid \citep{healpix}. The higher this number, the better a survey strategy performs. \husni{Increasing the number of visits need not be done at fixed exposure time, so it is not necessarily the case that increasing the number of visits requires a decrease in survey area; rather, there are a variety of area-depth tradeoffs possible for scenarios with increased numbers of visits.  Even a decrease to 20\,s exposures in some bands can be impactful for this metric.} Note that the bands to be used for \rachelNew{WL} shear estimation have not been decided yet; however, $riz$ are likely to dominate due to their higher S/N for \rachelNew{WL}-selected samples, which is why we choose to focus on them here.

\husni{Due to uncertainties in the level of detector effects and other sources of additive shear systematics, and in the performance of instrument signature removal methods (which  determine how sensor effects may contaminate the PSF estimated from bright stars), this metric has an arbitrary normalization and can only differentiate between the relative improvement between different strategies. Existence of on-sky LSSTCam data will provide a direct estimate of the level of additive shear systematics that need to be mitigated via observing strategy.  Therefore, the impact of this metric cannot currently be directly compared with that of the 3x2pt FoM defined in Section~\ref{sec:static_fom}. There is a potential trade-off between improving on 3x2pt FoM and mitigating WL shear systematics, as the 3x2pt FoM prefers an increase in area, while WL shear systematics are mitigated with a larger number of well-dithered visits; the relative importance of these metrics for science will likely only be clear at the time of commissioning. Strategies that increase the usable area for \rachelNew{WL} and decrease exposure time can lead to improvement in both metrics simultaneously.}


\subsection{Large-Scale Structure}
\authormacro{H. Awan, E. Gawiser}
\label{sec:lss}

Large-scale structure (LSS) constrains cosmological parameters via observations of galaxy clustering. LSS is a more localized tracer of the matter distribution, rather than an integral along a line of sight like \rachelNew{WL}. As a result, the constraining power of LSS is more sensitive to bias, scatter, and catastrophic errors in photometric redshift estimation, as these  
determine how much the clustering signal is degraded by projection along the line-of-sight 
\citep{2018MNRAS.477.3892C}.
Artificial modulations in the observed galaxy number density caused by depth variations and observing conditions (e.g., sky brightness, 
seeing, clouds) provide key systematic errors in measuring galaxy clustering \citep{Awan+2016}.  
Additionally, Galactic dust impacts the brightness and color of each galaxy 
\citep[e.g.,][]{2017AJ....153...88L}, and correcting for these effects becomes more difficult in regions with high levels of Galactic dust reddening.  

\begin{figure}
\centering
\includegraphics[width=1\linewidth]{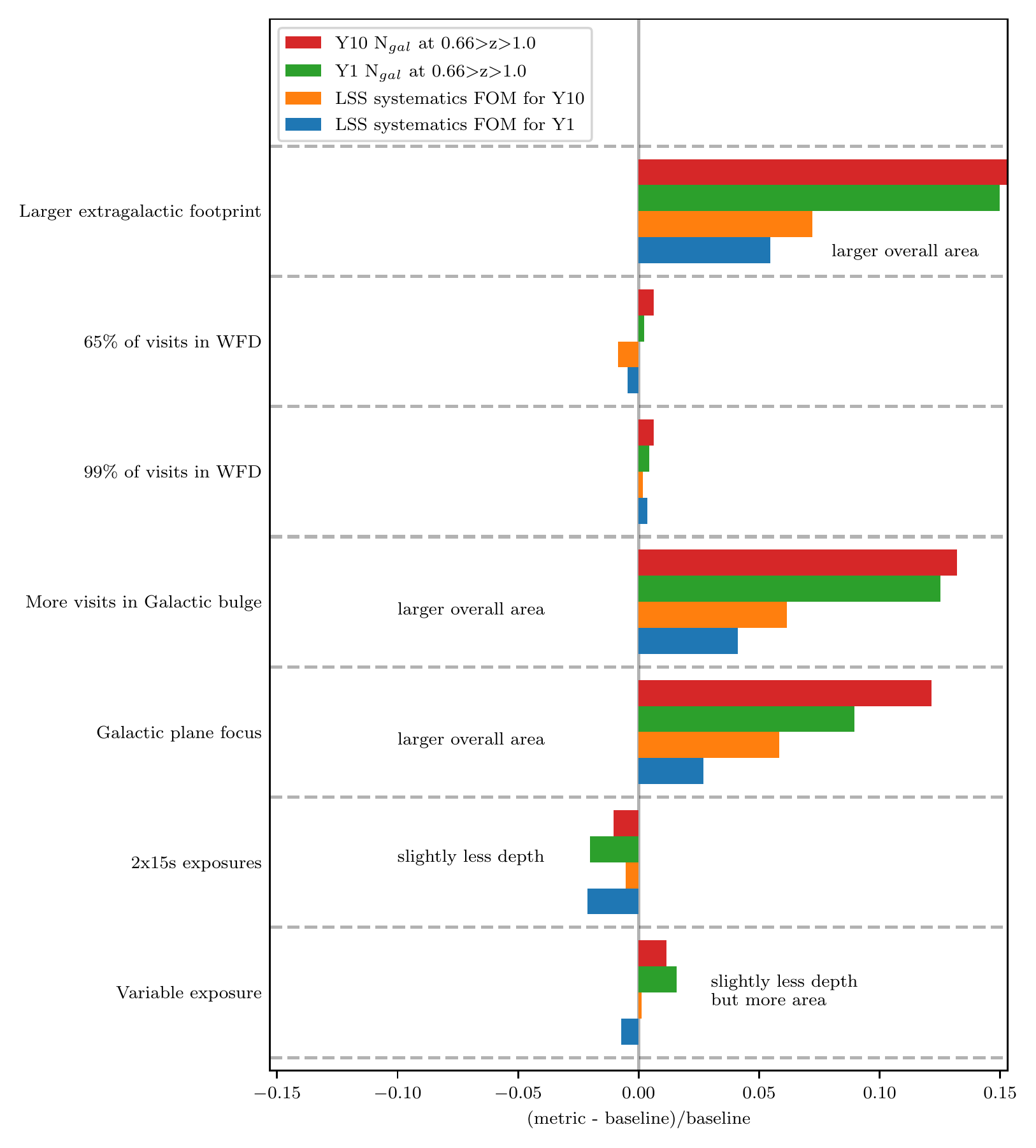}
\caption{Large-scale structure metrics as a function of selected observing strategies. \ml{\autoref{tab:sims_short} contains the exact simulation names corresponding to the short names used here.} As noted in the text, all of the strategies generate galaxy samples for which shot noise is small, making even 10\% changes in galaxy number unimportant for cosmological constraints and placing more emphasis on the LSS systematics FoM and the WL+LSS 3x2pt FoM of \autoref{fig:wl}.\label{fig:lss}}
\end{figure}

For the large-scale structure probe, we introduce 2 new metrics in detail below.  A summary of the results of these metrics can be seen in \autoref{fig:lss}.
\begin{enumerate}
\item Y1/Y10 $N_{\mathrm{gal}}$ at $0.66<z<1.0$	(\autoref{sec:lss_Ngal}): Estimated number of galaxies at $0.66<z<1.0$ based on Y1/Y10 $i$-band coadded depth in the effective survey area.	
\item LSS systematics FoM for Y1/Y10 (\autoref{sec:systematics})): LSS systematics diagnostic FoM for Y1/Y10, comparing the uncertainty added by Y1\humnaNew{/Y10} survey non-uniformity vs. that achieved for the baseline strategy using the Y10 gold sample (as defined in \autoref{sec:static_metrics}).
\end{enumerate}

\subsubsection{Galaxy Counts}
\label{sec:lss_Ngal}
In order to get cosmological constraints from $n$-point statistics, e.g., 2-point correlation functions or the 2-point power spectra, LSST will offer an unprecedentedly large and deep galaxy sample, allowing us to carry out analyses in a regime where statistical uncertainties will be subdominant to systematic ones. To estimate the number of galaxies, we follow \citet{Awan+2016} and propagate the simulated 5$\sigma$ coadded depth to the number of galaxies using light-cone mock catalogs. 
 
Using the \ttt{MAF} object from \citet{Awan+2016}, we create a new \ttt{MAF} object, \ttt{depthLimitedNumGalMetric}\footnote{ \url{https://github.com/humnaawan/sims_maf_contrib/blob/master/mafContrib/lssmetrics/depthLimitedNumGalMetric.py}}, that calculates the number of galaxies in the extragalactic footprint (as defined in~\autoref{sec:static_metrics}).

While we find small variations in the total galaxy counts across the observing strategies, 
these variations are not critical for LSS
since all strategies lead to samples comprising billions of galaxies that easily beat the shot noise limit and hence offer similar contributions to the 3x2pt FoM in 
\autoref{sec:static_fom}.

\subsubsection{Systematics Introduced by the Observing Strategy\label{sec:systematics}}

Spatial fluctuations in galaxy counts represent large-scale structure and hence are of interest for dark energy science.  
As discussed in \citet{Awan+2016}, 
artificial structure induced by the observing strategy leads to systematic uncertainties for LSS studies. Specifically, while 
a
systematic bias induced in the measured large-scale structure can be corrected, the uncertainty in our knowledge of this bias leads to uncertainties that affect our measurements. In order to quantify the effectiveness of each cadence in minimizing the uncertainties in the artificial structure that is induced by the observing strategy, we 
update 
the 
LSS \humna{FoM}
given 
in Equation 9.4 in the COSEP. Specifically, we have:
\begin{equation}
	\mathrm{\humna{LSS\ FoM}} \equiv
	\sqrt{
	\frac{
		\sum\limits_\ell{
		    \left( {\sqrt{ \frac{2}{f_{\mathrm{sky, Y10baseline}} (2\ell + 1)} }
									C_{\ell, \mathrm{LSS, Y10baseline}}
				    } \right)
				}^2
		}{
		\sum\limits_\ell
		    \left[{\left( { \sqrt{\frac{2}{f_{\mathrm{sky}} (2\ell + 1)}}
									\left\{{C_{\ell, \rm{LSS}} + \frac{1}{\bar{\eta}}} \right\}
						} \right ) ^2
					+ \sigma_{C_{\ell,\mathrm{OS}}}^2
				}\right]
		}
		}
\label{eq: LSS FoM COSEP modified}
\end{equation}
where the 
numerator differs from that in the COSEP as we now compare the uncertainty for each year vs. that achieved for the baseline strategy using the Y10 gold sample\humna{; briefly $f_\mathrm{{sky}}$ is the fraction of the sky used for analysis, while $C_\ell$ denotes the angular power spectrum and $\bar{\eta}$ is the surface number density of the galaxies in units of steradians$^{-1}$; we refer the reader to COSEP for further details}.  The first two terms in the denominator represent the standard sample variance and shot noise, and their combination adds to the  final term giving variance caused by the observing strategy \humna{(OS)}; the final term is calculated, as in \citet{Awan+2016}, as the standard deviation of $C_{\ell, \mathrm{OS}}$ across the $ugri$ bands, to model uncertainties due to detecting galaxy catalogs in different bands.  Note that this FoM approaches 1 if the observing strategy and shot noise contributions are negligible and the statistical power matches the Y10 baseline strategy.  It can be greater than 1 by Y10 for an observing strategy that covers more area than the baseline, but in that case, this improvement will duplicate that seen in the static science FoM.  Improvements in survey uniformity, however, will affect this LSS FoM and not the static science FoM.  

For $f_{\rm{fsky, baseline}}$, we use the Y10 sky coverage from the baseline for a given FBS simulation (i.e., we use \ttt{baseline\_v1.5\_10yrs} for v1.5 sims; \humna{\ttt{baseline\_nexp2\_v1.6\_10yrs} for v1.6 sims that implement the $2\times15$\,s exposure (identified by the \ttt{nexp2} tag) and  \ttt{baseline\_nexp1\_v1.6\_10yrs} for the rest};  \humna{\ttt{baseline\_nexp2\_v1.7\_10yrs}} for v1.7 sims); the footprint is the extragalactic footprint (as defined in~\autoref{sec:static_metrics}), designed to achieve the target Y10 gold sample of galaxies.

\subsection{General Conclusions from Weak Lensing and Large-Scale Structure}

\humna{For both \rachelNew{WL} and large-scale structure,} statistical constraining power 
\humna{and}
observational systematics are both impacted by choices in observing strategy.  \humna{As seen in}  \rachel{\autoref{fig:wl}}, 
changes in observing strategy that lead to more visits in $r$, $i$, or $z$ bands are preferred for the WL systematics metric.
\EG{The 3x2pt FoM benefits from any of the following:  larger survey area at low dust extinction, greater median depth, or  improved photometric redshifts.  While the latter two both favor depth versus area, within the variations available from simulated surveys, the 3x2pt FoM shows a greater improvement for simulations that maximize the area at low dust extinction.} 
\humna{The LSS metrics in \autoref{fig:lss} follow the general trend of the 3x2pt DETF FoM in favoring a larger effective survey area 
despite the corresponding modest loss of median depth.
\humna{The LSS FoM metric prefers both increased area and greater survey uniformity; the latter} responds favorably to nightly translational dithers (as shown in \citealt{Awan+2016}, COSEP), which have now been implemented as a default in the FBS simulations. } 
\humna{On a 
higher 
level, 
we find that the statistical power for combined WL and LSS prefers more area, as do the observational systematics for LSS, while observational systematics for WL prefer more visits. In the end,}  \EG{for the majority of the static science metrics explored in this section, the gain from larger area is greater than that from more visits.}

\begin{sloppypar}
\humna{To illustrate the tension specifically, f}or example, \ttt{footprint\_bluer\_footprintv1.5\_10yrs} and \ttt{wfd\_depth\_scale0.65\_noddf\_v1.5\_10yrs} both reduce exposure in the \husniNew{extragalactic} area (or in bands that are used for WL shear estimation) resulting in worse performance for both \humna{the 3x2pt FoM} and the systematics metric; while the opposite is true for \ttt{wfd\_depth\_scale0.99\_noddf\_v1.5\_10yrs}. For \ttt{bulges\_bs\_v1.5\_10yrs} and \ttt{footprint\_newAv1.5\_10yrs}, we see a trade-off between the two metrics, due to the fact that these simulations generally increase the area of the survey while decreasing the average number of visits in this area. The \humna{FBS} simulation \ttt{footprint\_big\_sky\_dustv1.5\_10yrs} is beneficial to the \humna{3x2pt FoM} due to the increase in area without harming the WL systematics metric due to reducing the area that is effectively ignored by the metric.
\end{sloppypar}

\section{Transient Science}
\label{sec:transients}
\subsection{Supernovae}
\authormacro{Ph. Gris, N. Regnault}

As of today, the Hubble diagram of type Ia supernovae (\sne) contains of the order of $\mathcal{O} (10^3)$ supernovae (SN) (\citealt{Betoule_2014,Scolnic_2018}). LSST will discover an unprecedented number of \sne~ -- $\mathcal{O} (10^6)$. The key requirements to turn a significant fraction of these discoveries ($\sim$10\%) into distance indicators useful for cosmology are 
(1) a regular sampling of the \sne~ light curve in several rest-frame bands,
(2) a relative photometric calibration (band-to-band) at the $10^{-3}$ level, (3) a good understanding of the \sne~ astrophysical environment, (4) a good estimate of the survey selection function, and (5) a precise measurement of the redshift host galaxy based on LSST photo-$z$ estimators. The first point is crucial. It determines how well we can extract the light-curve observables used by current and future standardization techniques (stretch, rest-frame color(s), rise time). It also determines how well photometric identification techniques are going to perform, as live spectroscopic follow-up will only be possible for 10\% of \sne.

\par
The average quality of \sne~ light curves depends primarily on the observing strategy through five key facets: (1) a high observing cadence (2 to 3 days between visits) delivers well sampled light curves, which is 
key to distance determination and photometric identification; (2) a regular cadence allows 
minimizing the number of large gaps ($>$10 days) between visits, which 
degrades the 
determination of luminosity distances, and potentially result in
rejecting large batches of light curves of poor quality; (3) a filter allocation ensuring the use at least 3 bands (rest-frame) to select high-quality supernovae; (4) the season length determines the number of \sne~ with observations before and after peak; due to time dilation maximizing season length is particularly important in the DDFs; (5) finally, the integrated signal-to-noise ratio over the \sne~ full light curve determines the  contribution of measurement noise to the distance measurement. It is a function of the visit depth and the number of visits in a given band.

\PhG{All the studies presented in this section on light-curve simulations of \sne. We have used the SALT2 model (\citealt{Guy_2007,Guy_2010}) where a type Ia supernova is described by five parameters: \snx, the normalization of the spectral energy distribution (SED) sequence; \snstretch, the stretch; \sncolor, the color; \daymax, the day of maximum luninosity; and $z$, the redshift. The time-distribution and the photometric errors of the light-curve points are estimated from observing conditions (cadence, 5$\sigma$-depth, season length) given by the scheduler.  
We consider two types of \sne~ defined by (\snstretch,\sncolor) parameters to estimate the metrics: (intrinsically) faint supernovae, defined by  (\snstretch,\sncolor)= (-2.0,0.2), and medium supernovae, defined by  (\snstretch,\sncolor)= (0.0,0.0). (\zfaint, \nsnfaint) gives an assessment of the size and depth of the redshift limited sample (i.e. the sample of supernovae usable for cosmology) with selection function having minimal dependence on understanding the noise properties.
(\zmed, \nsnmed) gives an assessment of the size and depth of the sample of \sne~with precise distances. We will get higher statistics with the medium sample, but need also a better understanding of noise to determine the selection function.

All the metrics described below are estimated from a sample of well-measured \sne~that passed the following light curve (LC) requirements: visits with SNR$>$ 10 in at least 3 bands;  5 visits before and 10 visits after peak, within [-10;+30] days (rest-frame); $\sigma_C<$ 0.04 where $\sigma_C$ is the SALT2 color uncertainty; all observations satisfying 
380~nm~$~<~\bar{\lambda}_{obs}/(1+z)~<$ 700 nm.
}

\begin{figure}
    \centering
    \includegraphics[width=1\columnwidth]{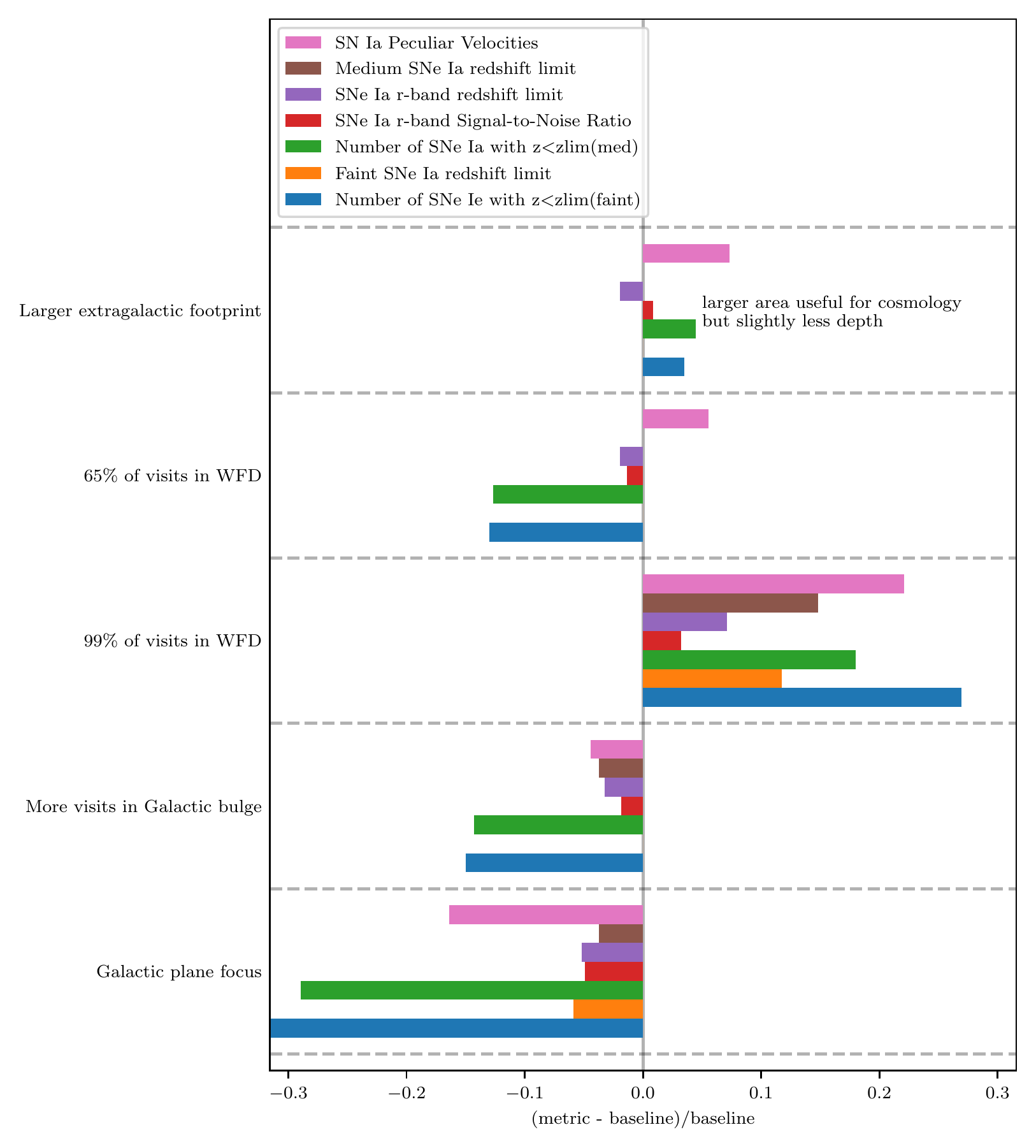}
    \caption{Supernova metrics as a function of selected observing strategies. \ml{\autoref{tab:sims_short} contains the exact simulation names corresponding to the short names used here. As described in \autoref{sec:nsn}, \zlim\ corresponds to the redshift beyond which supernovae no longer pass light curve requirements. ``med'' and ``faint'' refer to a sample of typical and faint supernovae respectively.}}
    \label{fig:sn}
\end{figure}

For the \sne~probe, we introduce \PhG{7} new metrics.  A summary of the results from the metrics can be seen in \autoref{fig:sn}.

\PhG{
\begin{enumerate}
\item Faint \sne~redshift limit - Redshift limit corresponding to a complete \sne~sample (z$_{lim}^{faint}$) (\autoref{sec:nsn})
\item  Medium \sne~redshift limit - Redshift limit corresponding to a complete \sne~sample (z$_{lim}^{medium}$) (\autoref{sec:nsn})
\item Number of \sne~with z$\leq$zlim(faint)	- Number of well-sampled \sne~with z$\leq$z$_{lim}^{faint}$ (\autoref{sec:nsn})
\item Number of  \sne~with z$\leq$ zlim(med) - 	Number of well-sampled \sne~with z$\leq$z$_{lim}^{medium}$ (\autoref{sec:nsn})
\item \sne~$r$-band Signal-to-Noise Ratio - Fraction of faint \sne~with a $r$-band Signal-to-Noise Ratio (SNR) higher than a reference SNR corresponding to a regular cadence (\autoref{sec:rate_zlim}). 
\item \sne~$r$-band redshift limit	- $r$-band redshift limit of faint \sne~(\autoref{sec:rate_zlim})
\item Peculiar velocities - \sne~host galaxy velocities (\autoref{sec:peculiar})

\end{enumerate}
}

\subsubsection{Number of well-measured type Ia supernovae/Survey completeness} \label{sec:nsn}
\vspace*{0.1cm}

We use, as our primary metric, the size and depth of a subset of well-sampled \sne~using the redshift limit \zlim~and the number of well-sampled supernovae \nsnlim below this redshift. \zlim~corresponds to the redshift beyond which supernovae no longer pass light curve requirements. 
\par
The WFD footprint is quite large (at least 18000 deg$^2$) forbidding the use of full (time-consuming) LC simulations. We have opted for a slightly different approach.
The celestial sphere is pixellized in HEALPIX superpixels ($N_\mathrm{side}$=64, which corresponds to 0.8 deg$^2$ per pixel).  The directions (i.e. (RA,Dec) positions)/ healpixel affected by a Galactic extinction $E(B-V)$ larger than 0.25 are masked (to minimize reddening effects) and not included in our assessment. We consider only the $griz$ observations which are the ones
that matter to derive SN luminosity distances.
  
We process observing strategies using a simple model of the LSST focal plane and estimate:
\begin{eqnarray}
  z_{\mathrm{lim}} & = & \mathrm{max}\left(z | \mathrm{LC(z)\ fulfill\ requirements}\right) \\
  N_{z<z_{\mathrm{lim}}} &= & \delta\Omega_{\mathrm{pix}} \int_0^{z_\mathrm{lim}} \frac{\Delta T_{\mathrm{step}}}{1+z}\ {\mathcal{R}}(z)\ dV(z)
\end{eqnarray}
where $\delta\Omega_{\mathrm{pix}}$ is the solid angle subtended by one pixel; $dV$ is the (differential) co-moving volume; $\Delta T_{\mathrm{step}}$ is the time interval for supernovae simulations (in observer frame days), that is only supernovae with a peak luminosity during this time range are simulated; $\mathcal{R}(z)$ is the SN~Ia volumetric rate \citep{2012AJ....144...59P}.  We also compute the average cadence (in day$^{-1}$), i.e. the number of $g, r,i$ or $z$ visits in a fiducial restframe interval. The quantities above are determined for each pixel and each night (identified by its Modified Julian Date MJD) and may be 
used to build full sky maps
giving, as a function of the position on the sky (1) the density of supernovae, (2) the median maximum redshift (over the observed area), and (3) the median cadence.  \par
This metric is the most precise to assess observing strategies, but also the most intricate to implement. A lot of effort has been put to design algorithms combining speed, reliability, and accuracy. The codebase is accessible through Github\footnote{\url{https://github.com/LSST-nonproject/sims\_maf\_contrib/blob/master/science/Transients/SN\_NSN\_zlim.ipynb}} in the Metric Analysis Framework.\par
\ml{We note that while we expect the quality cuts described here will ensure accurate classification of \sne~and separate them from other classes of transients, we do not yet have a transient metric to ensure this is the case and leave this important step to future work (see \autoref{sec:future} for a detailed discussion).}

\subsubsection{$SNR_{\mathrm{rate}}$ and redshift limit}
\label{sec:rate_zlim}
\vspace*{0.1cm}
\vspace*{0.1cm}

While the metric in \autoref{sec:nsn} is our most accurate metric, we developed two proxy metrics 
that have also been incorporated in MAF where straightforward, fast and easy-to-run metrics are preferred. These two metrics are quite simple (they do not require the use of a light curve fitter) and just need templates of supernova light curves as input. They are estimated for each band thus providing tools for further comparison of observing strategy performance. They are sensitive to two key points of observing strategies: the median cadence, and the inter-night gap variations. The codebase is accessible through Github\footnote{\url{https://github.com/LSST-nonproject/sims\_maf\_contrib/blob/master/science/Transients/SNSNR.ipynb}}.\par

These two metrics rely on Signal-to-Noise Ratio (SNR) of the light curves (per band) that may be written as:
\begin{equation} \label{eq:snr}
SNR_{b} = \sqrt{\sum_{i}{\left({{f_i^{b}}\over{\sigma_i^{b}}}\right)^2}}
\end{equation}
where $b$ is the band, $f^{b}$ are the fluxes and $\sigma^{b}$ the flux errors (summation over LC points). In the background-dominated regime, flux errors may be expressed as a function of the $5-\sigma$ limiting flux of each visit $f_{i|5}$. We may rewrite \autoref{eq:snr} by defining $\delta_b$ as the number of observations per bin in time $\Delta T$ and per band $b$:
\begin{equation} \label{eq:snrb}
f_{|5}<\delta_b>^{-1/2} = \frac{5 \sqrt{\Delta T} \sqrt{\sum_{i}  (f_i^b)^2}}{SNR_{b}}
\end{equation}

We describe below two metrics that may be extracted from \autoref{eq:snr} and \autoref{eq:snrb}: the $SNR_{\mathrm{rate}}$ and the redshift limit metrics. \par
$SNR_{b}$ (\autoref{eq:snr}) is the result of the 
combination of observing strategy features (5$\sigma$-depth and cadence) and supernovae \PhG{parameters.} 
We fix some of these parameters (\PhG{we considered faint supernovae with $z$=0.3, where the sample is not affected by the Malmquist bias}) so as to estimate
$SNR_{b}(t)$ for a supernova with \daymax=t-10. We also evaluate $SNR_{b}^{regular}(t)$ using the same supernova parameters but opting for median values for 5-$\sigma$ depth and cadence. The $SNR_{rate}$ metric is then defined as the fraction of time (in a season) when the requirement $SNR_{b}(t) \geq SNR_{b}^{regular}(t)$  is fullfilled.
\par
The two above-mentioned contributions to $SNR_{b}$ are clearly visible in \autoref{eq:snrb}, with observing conditions on the left side (5-$\sigma$ limiting flux times cadence), and flux (supernovae properties) on the right side. We fix some of the supernova parameters (\PhG{faint supernovae with \daymax=0.})
and we use median values of 5-$\sigma$ depth and cadences to estimate redshift values defining the second metric, dubbed as \zlim. 
We used \autoref{eq:snrb} with the following SNR (ANDed) requirements:  
$SNR_g > 30, SNR_r > 40, SNR_i > 30, SNR_z > 20$. Combining these selections is equivalent to requesting $\sigma_c \leq$ 0.04 and will ensure observation of well-measured \sne. 

\newpage
\subsubsection{Peculiar Velocities}
\authormacro{A.~G.~Kim, S.~{Gontcho A Gontcho}, N.~Regnault}
\label{sec:peculiar}
\PhG{The goal of the peculiar velocities metric is to study modified gravity through its effects on the overdensities and velocities of type Ia supernova host galaxies.} Gravitational models
are efficiently parameterized by the growth index, $\gamma$, which
influences the linear growth rate as $f=\Omega_M^\gamma$ \PhG{where $\Omega_M=\frac{\Omega_{M_0}}{\Omega_{M_0} + (1-\Omega_{M_0})a^3}$ and $\Omega_{M_0}$ is the mass density today. The parameter dependence enters through $fD$, where $D$ is  the spatially-independent ``growth factor'' in the linear evolution of density perturbations and
$f \equiv \frac{d\ln{D}}{d\ln{a}}$ is the linear growth rate where $a$ is the scale factor  \citep{2006PhRvD..73l3526H,2011ApJ...741...67D}. Two
surveys with the same fractional precision in fD will have different precision in $\gamma$, with the one at lower redshift providing the tighter constraint. We thus use the uncertainty in the growth index, $\sigma_\gamma$, as the peculiar velocity metric.}

For a parameterized survey and redshift-independent
$\gamma$,
$\sigma_\gamma$ is
bounded in
\citet{Kim2019} by using the 
Fisher matrix
\begin{align}
F_{ij} 
& = \frac{\Omega}{8\pi^2} \int_{r_{\rm min}}^{r_{\rm max}}  \int_{k_{\rm min}}^{k_{\rm max}}  \int_{-1}^{1} r^2 k^2 \text{Tr}\left[ C^{-1} \frac{\partial C}{\partial \lambda_i} C^{-1}
\frac{\partial C}{\partial \lambda_j} \right] d\mu\,dk\,dr
\label{fisher:eqn}
\end{align}
where $\Omega$ is the sky coverage of the survey, $r_{max}$ ($r_{min}$) are the comoving distances corresponding to the upper (lower) redshift limits of each redshift bin, and we set $k_{max}$ = 0.2h Mpc$^{-1}$ and $k_{min} = 2\pi/r_{max}$. $\mu$ is the cosine of the angle $\phi$ between the k-vector and the observer's line-of-sight.
The covariance matrix $C$  is defined by:
\begin{equation}
C(k,\mu)  =
  \begin{bmatrix}
   P_{\delta \delta}(k,\mu) + \frac{1}{n} &
   P_{v\delta}(k,\mu)  \\
   P_{v\delta}(k,\mu)  &
  P_{vv}(k,\mu) + \frac{\sigma^2}{n}
   \end{bmatrix}
\label{cov:eq}
\end{equation}
and the parameters considered are $\lambda \in \{\gamma,bD, \Omega_{M_0}\}$.

The \sne~host-galaxy radial peculiar velocity power spectrum is $P_{vv}\propto (fD\mu)^2$, the count overdensity
power spectrum is $P_{\delta \delta }\propto (bD + fD\mu^2)^2$, the overdensity-velocity cross-correlation is $P_{vg}
\propto  (bD + fD\mu^2)fD$, where $b$ is the galaxy bias and $\mu\equiv \cos{(\hat{k} \cdot \hat{r})}$ where $\hat{r}$ is the direction of
the line of sight and $n$ is $\epsilon\phi t$ where $\epsilon$ is the sample selection efficiency, $\phi$ is the observer-frame SN Ia rate and $t$ is the duration of the survey.  While the $bD$ term does contain information on $\gamma$, its constraining power is not used here.
The variance in $\gamma$ is $\left(F^{-1}\right)_{\gamma\gamma}$.  Our figure of
merit is the inverse variance, so that a larger value
corresponds to a more precise measurement and hence a better
survey strategy.

The 
parameters in \autoref{fisher:eqn} that are
primarily affected by survey strategy are the survey solid angle $\Omega$ and the \PhG{number
density $n$ of well-measured \sne}.  The other parameters related to the follow-up strategy of these SN discoveries are the survey depth $r_{\rm max}$
and the intrinsic velocity dispersion $\sigma$, which is related to the intrinsic \PhG{magnitude dispersion of well-measured \sne}.  The estimate
of $\sigma_\gamma$ is sensitive to both the sample variance $P_{vv}$
and shot noise $\frac{\sigma^2}{n}$ in the range of proposed
surveys,
meaning that its accurate determination cannot be taken
in either the sample- or shot-noise limit. 
A follow-up strategy must also be specified for the calculation of $\sigma_\gamma$ since Rubin/LSST will not generate all the information needed for this measurement, e.g. redshift, SN classification.  Here, we adopt a
maximum survey redshift of $z=0.2$ and follow-up that gives
0.08~mag magnitude dispersion per SN.  The minimum redshift
is $z=0.01$, number densities
are based on 65\% of the \sne~rates of \citet{2010ApJ...713.1026D},
and $k_{\rm max}=0.1h$\,Mpc$^{-1}$, $b=1.2$, $\Lambda$CDM
cosmology with $\Omega_{M_0}=0.3$, and overdensity power spectra
for the given cosmology as calculated by CAMB \citep{Lewis:2002ah}.

\PhG{The code used for the calculations are available in Github\footnote{\url{https://github.com/LSSTDESC/SNPeculiarVelocity/blob/master/doc/src/partials.py}}.}

\subsubsection{General Conclusions from Supernovae}

Collecting a {\it large} sample of {\it well-measured} type Ia supernovae is a prerequisite to measure cosmological parameters with high accuracy. Our analysis (\autoref{sec:nsn} to \autoref{sec:peculiar} and \autoref{fig:sn}) has shown that the key parameter to achieve this goal is the effective cadence delivered by the survey. For the WFD survey, a regular cadence in the $g$, $r$ and $i$ bands is essential to (1) secure a high-efficiency photometric identification of the detected \sne~and (2) secure precise standardized \sne~distances, by optimizing the integrated Signal-to-Noise Ratio along the \sne~light curves. Gaps of more than $\sim$ 10 days in the cadence have a harmful impact on the size and depth of the \sne~sample. The cadence of observations is by far the most important parameter for \sne~science, before observing conditions: on the basis of studies conducted, it is preferable to have pointings with sub-optimal observing conditions rather than no observation at all.

Three main sources of gaps have been identified: telescope down time (clouds, maintenance), filter allocation, and scanning strategy (ie the criteria used to move from one pointing to another). While we are aware that it is difficult to minimize the impact of down time, there is still room for improvement on filter allocation and scanning strategy.  Significant efforts have been made to make sure that nightly revisits of the same field are performed in different bands. Relaxing the veto on bluer bands around full moon, or increasing the density of visits (ie the number of visits per square degree) during a night of observation (by decreasing the observed area for instance) will help to achieve an optimal cadence for supernovae of 2 to 3 days in the $g$, $r$ and $i$ bands.

\subsection{Strong Lensing}
\label{sec:sl}

\authormacro{S.~Huber, S.~H.~Suyu, Ph.~Gris}

The Hubble constant $H_0$ is one of the key parameters to describe the
universe. Current observations of the CMB (cosmic microwave
background) assuming a flat $\Lambda$CDM cosmology and the standard
model of particle physics yield $H_0 = 67.36 \pm 0.54 \, {\rm km\,s^{-1}\,Mpc^{-1}}$
\suyuNew{\citep{PlanckVI:2020}}, \huberNew{which is in tension with} $H_0 =
74.03 \pm 1.42 \, {\rm km\,s^{-1}\,Mpc^{-1}}$ from the local \suyuNew{Cepheid} distance ladder
\citep{Riess:2016jrr,Riess:2018byc,Riess:2019cxk}\suyuNew{, although is statistically consistent with the $H_0$ derived using the tip of the red giant branch in the distance ladder by \citet{Freedman:2019}}. In order to verify or refute this
$4.4 \sigma$ tension, further independent methods are needed. 

One such method is lensing time delay cosmography which can determine
$H_0$ in a single step. The basic idea is to measure the time delays
between multiple images of a strongly lensed variable source
\citep{Refsdal:1964}. This time delay, in combination with mass
profile reconstruction of the lens and line-of-sight mass structure,
yields directly a ``time-delay distance'' that is inversely
proportional to the Hubble constant ($t \propto D_{\Delta t} \propto
H_0^{-1}$). Applying this method to six lensed quasar systems, the
H0LiCOW collaboration \citep{Suyu:2016qxx} together with the
COSMOGRAIL collaboration
\citep[e.g.][]{Eigenbrod:2005ie,2013Tewes,2017Courbin} measured $H_0 =
73.3^{+1.7}_{-1.8} \,{\rm km\,s^{-1}\,Mpc^{-1}}$ \citep{Wong:2019kwg} in flat
$\Lambda$CDM, which is in agreement with the
local distance ladder \huberNew{but} higher than CMB measurements.  Including the distance measurement to another lensed quasar system from the STRIDES collaboration \citep{Shajib:2020}, the newly formed TDCOSMO organisation has further investigated potential residual systematic effects \citep[e.g.,][]{Millon:2020, Gilman:2020, Birrer:2020}.  Another
promising approach goes back to the initial idea of
\cite{Refsdal:1964} using lensed supernovae (LSNe) instead of quasars
for time-delay cosmography \citep[e.g.,][]{2020ApJ...898...87G, Mortsell:2020, Suyu:2020}. In terms of discovering strong lens systems from the static LSST images for cosmological studies, having $g$-band observations with comparable seeing as in the $r$- and $i$-band would facilitate the detection of strong lens systems \citep{Verma:2019}.

In this section, we investigate the prospects of using
LSST for measuring time delays of both lensed supernovae and lensed
quasars.  In particular, we focus on the number of lens systems that we 
would detect for the various observing strategies as our metrics. 
From the investigation of LSNe by \huberNew{galaxies, we define} a metric
for the number of LSNe Ia with good time delay measurement.
For lensed quasars, we have additional metrics defining how well we can 
measure the time-delay distances.

\begin{figure}
\centering
\includegraphics[width=1\columnwidth]{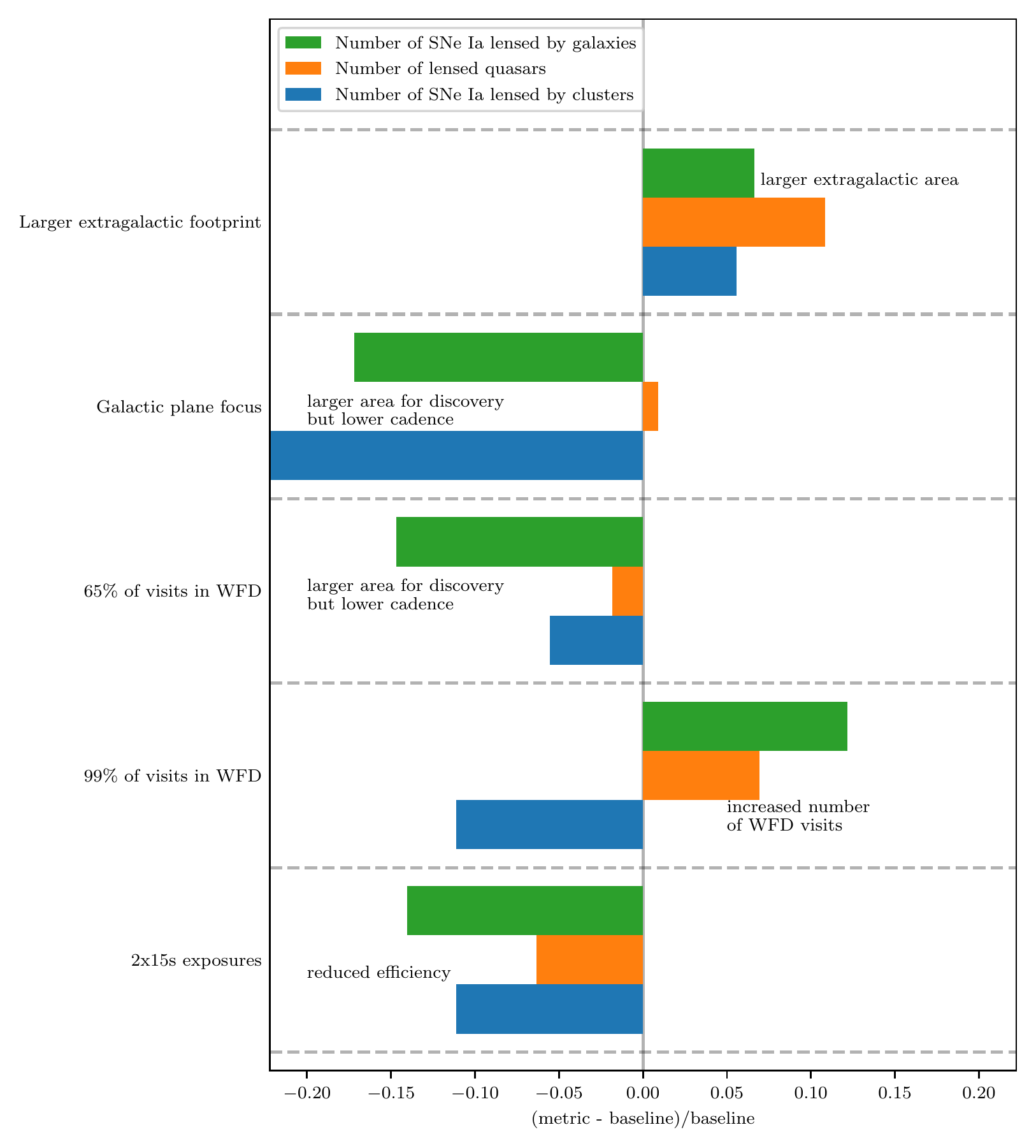}
\caption{Strong lensing metrics as a function of selected observing strategies. \ml{\autoref{tab:sims_short} contains the exact simulation names corresponding to the short names used here.} In general, a long cumulative season length and high sampling frequency at sufficient depth are preferred for strong lensing in terms of maximizing the number of lens systems, particularly SNe and quasars lensed by galaxies (blue and orange).  For the specific case of LSNe Ia in the five cluster fields, the number of LSNe Ia is sensitive to number of visits to the cluster fields, the image depth, the gap of these observations, and the filter used. For the observing strategies plotted here, the depth and the mean gap between the observations of the considered clusters remain roughly the same, while the main difference is the number of observations (lower than the baseline), which \huberNew{negatively affects} the number of expected cluster-lensed SNe~Ia. Furthermore, the number of observations to the clusters in WFD~65\% strategy is higher than the WFD~99\% one, which is also reflected in the expected cluster-lensed SNe~Ia.}
\label{fig:sl}
\end{figure}

For the Strong Lensing (SL) probe, we introduce 4 new metrics.  A summary of the results from these metrics can be seen in \autoref{fig:sl}
\begin{enumerate}
    \item Number of SNe Ia lensed by galaxies -	Number of SNe Ia strongly lensed by galaxies with accurate and precise time delays between the multiple SN images \suyu{(\autoref{sec:sl:slsn-gal})}
    \item Number of SNe Ia lensed by clusters - Number of strongly lensed SNe Ia in the multiply-imaged galaxies behind well-studied galaxy clusters \suyu{(\autoref{sec:sl_clusters})}
    \item Number of lensed quasars	- Number of strongly lensed quasars with accurate and precise time delays between the multiple quasar images \suyu{(\autoref{sec:sl_quasars})}

\end{enumerate}

\subsubsection{Number of supernovae lensed by galaxies}
\label{sec:sl:slsn-gal}
\authormacro{S. Huber, S. H. Suyu, Ph. Gris}

For constraining cosmological parameters with supernovae lensed by
galaxies as well as possible, ideally we would like to maximize the
number of accurate and precise time-delay distance measurements.
Currently there are only two known lensed SN systems, namely SN
``Refsdal'' \citep{Kelly:2015xvu,Kelly:2015vjq} and iPTF16geu
\citep{Goobar:2016uuf}, but LSST will play a key role \huberNew{in detecting} many
more LSNe \citep{Oguri:2010,Goldstein:2017bny, Wojtak:2019}.  A
measurement of a time-delay distance from a strongly lensed SN system
requires (1) the detection of the system, (2) the measurement of time
delays between the multiple SN images from their observed light
curves, and (3) the lens mass modeling of the system to infer the
distance from the time delays.  LSST's \huberNew{observing} strategies affect both
(1) and (2), and the uncertainties in the time delays from (2) enter
directly into the uncertainties on the time-delay
distances. Therefore, we use as \huberNew{a} metric the number of lensed supernovae
systems that could yield time-delay measurements with precision better
than 5\% and accuracy better than 1\%, in order to achieve $H_0$
measurement that has better than 1\% accuracy from a sample of lensed
SNe.  We refer to time delays that satisfy these requirements as having
``good'' delays.

\suyuNew{\cite{Huber:2019}} have presented a detailed study to obtain the number of lensed SNe, and \suyuNew{we} \huberNew{summarize the results here}.  To simulate
LSST observations of a lensed SN system, \suyuNew{\cite{Huber:2019}} have used 202 mock LSNe Ia
from the OM10 catalog \citep{Oguri:2010}, and produced synthetic light
curves for the mock SN images via ARTIS (Applied Radiative
Transfer In Supernovae) \citep{Kromer:2009ce} for the spherically symmetric 
SN Ia W7 model \citep{1984:Nomoto} in combination with
magnification maps from GERLUMPH \citep{Vernardos:2015wta} to include
the effect of microlensing similar to \cite{Goldstein:2017bny}. \suyuNew{\cite{Huber:2019} have
then simulated} data points for the light curves, following the
observation pattern from different \huberNew{observing} strategies from the \mlNew{FBS} scheduler\footnote{\url{https://cadence-hackathon.readthedocs.io/en/latest/current_runs.html}}  and uncertainties are calculated
according to The LSST Science Book. To
measure the time delay from the simulated observation, \suyuNew{\cite{Huber:2019}} have used the free
knot spline optimizer from PyCS (Python Curve Shifting)
\citep{2013:Tewesb,Bonvin:2015jia}. For each mock lens system from
OM10, \suyuNew{there are} 100 random starting configurations, where a starting
configuration corresponds to a random position in the microlensing
map, a random field on the sky where the system is located and a random
time of explosion in the observing seasons such that the detection
requirement from OM10 is fulfilled.  For each starting configuration,
\suyuNew{there are} 1000 noise realisations.  Through \suyuNew{these} realistic
simulations, \suyuNew{\cite{Huber:2019}} have then quantified the precision and accuracy of the
measured time delays for each mock system.  Taking into account the
different areas and cumulative seasons lengths of the different observing strategies, 
\suyuNew{\cite{Huber:2019}} have then estimated the number of lensed SNe that would be detected with LSST and
would have ``good'' delays.

\suyuNew{\cite{Huber:2019} have considered} two scenarios: (1) LSST data only for both detection and
time-delay measurements, and (2) LSST data for detection with
\huberNew{additional} follow-up observations with a more rapid cadence than LSST. 
\suyuNew{Follow-up observations are assumed to take place} every second night in three filters \huberNew{($\textit{g, r,}$ and \textit{i})} going to a \huberNew{baseline-like} 5$\sigma$ \huberNew{point-source} depth of 24.6, 24.2, and 23.7, respectively.  For our science
case of measuring time delays from as many lensed SNe as possible \citep{Huber:2019}, it
would be more effective to use LSST as a discovering machine with
additional follow-up, instead of relying on LSST completely for the
delay measurements. \suyuNew{Based on the investigations of \cite{Huber:2019}}, long
cumulative seasonal lengths and a more
frequent sampling are important \huberNew{to increase the number of LSNe Ia with well measured time delays}. \huberNew{Specifically, we request 10 seasons} with a season length of 170 days or longer. Rolling
cadences are clearly disfavored, because their shortened cumulative
season lengths (only 5 instead of 10 seasons for two declination bands) lead to overall a more negative impact on the number of
LSNe Ia with delays, compared to the gain from the more rapid sampling
frequency. To improve the
sampling, single visits per night, \huberNew{meaning that a given LSST field will be only observed once a night,} would be helpful. Since this will
make the science case of fast-moving transients impossible, we suggest
doing one revisit within a night but in a different band than the
first visit. Further improvements are the replacement of $2\times15$\,s exposure by $1\times30$\,s for an increased
efficiency and redistributing some of the visits in $y$-band to $g$, $r$, $i$
and $z$.

We note that \cite{Goldstein:2018bue} performed detailed simulations
of the LSN population using a completely independent technique and
pipeline, and reached similar conclusions to the ones presented here:
rolling cadences are strongly disfavored, and wide-area, long-season
surveys with well sampled light curves are optimal.

To evaluate further \huberNew{observing} strategies, we have defined a metric based on \suyuNew{the investigations of \citet{Huber:2019}}. The number of LSNe Ia with well measured time delays using LSST and follow-up observations for a given \huberNew{observing} strategy can be approximated as 
\begin{eqnarray}
    N_\mathrm{LSNeIa, \, good \, \,delay} \approx \nonumber
    45.7 \frac{\Omega_\mathrm{WFD}}{\SI{20000}{\square\deg}} \frac{t_\mathrm{csl}}{2.5 \, \mathrm{yr}} \cdot \frac{1}{2.15}  \mathrm{exp}(-0.37 t_\mathrm{gap})\\
    \label{eq:Metric SNe Ia lensed by galaxy}
\end{eqnarray}

The first part of the metric (separated by a dot from the second part) is the rescaling of the value predicted 
in OM10 \citep{Oguri:2010} by taking into account the survey area of the WFD \huberNew{($\Omega_\mathrm{WFD}$)} and the 
mean of the cumulative season length \huberNew{($t_\mathrm{csl}$; summed over all season lengths)} for a given \huberNew{observing}
strategy. The second part contains a fit based on the numbers of LSNe Ia with well measured time delay, presented in \cite{Huber:2019}, in comparison \huberNew{with} the total number of LSNe Ia which will be detected 
(first part of \autoref{eq:Metric SNe Ia lensed by galaxy}). The fit function depends on the median inter-night gap between any filters $t_\mathrm{gap}$ and is measured in days, which is an important parameter because we assume a detection of the LSNe Ia in \cite{Huber:2019} after the third data point exceeds the 5$\sigma$ \huberNew{point-source} depth in any filter. The inter-night gap \huberNew{($t_\mathrm{gap}$)}, survey area $\Omega_\mathrm{WFD}$, and cumulative season length $t_\mathrm{csl}$ can be 
calculated via MAF\footnote{\url{https://me.lsst.eu/gris/}} and a Python script \footnote{\url{https://github.com/shuber891/LSST-metric-for-LSNe-Ia/}} where we only take observations into account, \huberNew{with a 5$\sigma$ point-source depth greater than 22.7, 24.1, 23.7, 23.1, 22.2, and 21.4 for the filters $u$, $g$, $r$, $i$, $z$, and $y$. These cuts are motivated by \cite{Huber:2019} and are important to restrict visits to the WFD, where the metric from Equation \ref{eq:Metric SNe Ia lensed by galaxy} is valid}  The results are summarized in \autoref{fig:sl}, where we see that $1\times30$\,s exposures are favored over $2\times15$\,s exposures and more time on the WFD is preferred. Our conclusions including results from \cite{Huber:2019} are summarized in \autoref{sl: General Conclusions from Strong Lensing}.

\subsubsection{Number of supernovae lensed by galaxy clusters}
\authormacro{T.~Petrushevska}
\label{sec:sl_clusters}
Here, we focus on prospects of observing LSNe which are strongly lensed by galaxy clusters which have well-studied lens models. High-z galaxies that appear as multiple images in the cluster field can host SN explosions. The first discovery of this kind was SN Refsdal, which was classified as a core-collapse (CC) explosion \citep{Kelly2015, Kelly:2015xvu}.  Several teams predicted the re-appearance of SN Refsdal almost a year later, which allowed to test their lens models \citep[e.g.,][]{2016ApJ...822...78G, Kelly:2015xvu}. By measuring the time delays of SN Refsdal and having a high-quality strong lensing model of the galaxy cluster, it was shown that it is possible to measure \emph{H$_0$} with 6\% total uncertainty \citep{Grillo:2018ume,2020ApJ...898...87G}. Dedicated ground-based searches for lensed SNe behind galaxy clusters have been performed using near-infrared instruments at the Very Large Telescope \citep{2009A&A...507...61S, Second, Petrushevska2016, Petrushevska2018a}. Most notably, they reported the discovery of one of the most distant CC~SN ever found, at redshift $z = 1.703$ with a lensing magnification factor of $4.3 \pm 0.3$ \citep{2011ApJ...742L...7A}. Furthermore,  thanks to the power of the lensing cluster, it was possible to estimate the volumetric CC~SN rates for $0.4\leq z< 2.9$, and compare it with the predictions from cosmic star formation history \citep{Petrushevska2016}.  Knowing the absolute brightness of SNe~Ia allows estimating the absolute magnification of SNe~Ia, therefore to break the so-called mass-sheet degeneracy of gravitational lenses \citep{2001ApJ...556L..71H}. Thus, LSNe~Ia could be used to put constraints on the lensing potential, if the background cosmology is assumed to be known \citep[see e.g.,][]{PAtel2014,Nordin2014, Rodney2015}. 

As a metric we use the expected number of LSNe in the selected galaxy cluster fields in 10 years of LSST. For the details regarding the methods in this section, we refer to \cite{Petrushevska_2020} based on the work in 
\citep{Petrushevska2016, Petrushevska2018a, Petrushevska2018b}. Here, we present a short summary. We consider the six Hubble Frontier Fields clusters \citep{Lotz2017} and Abell 1689, given in \autoref{HFF_clusters}. These clusters have been extensively studied, and given the good quality data, well-constrained magnification maps and time delays can be obtained from the lensing models \citep{Petrushevska2016, Petrushevska2018a, Petrushevska2018b}. We consider the multiply imaged galaxies in the cluster fields that have a spectroscopic redshift.  Given the redshift range of the multiply-imaged galaxies considered here (see \autoref{HFF_clusters}), the most important bands are $i$, $z$ and $y$ (see Figure 2 in \citet{Petrushevska_2020}). The observability of a SN in the multiply-imaged galaxies is sensitive to  the redshift, star formation rate and magnification of the galaxy, but also on the observing strategy parameters such as depth, separation between the two consecutive observations and the filter. \tanjap{The expected number of LSNe~Ia in the five cluster fields is relatively low, mainly for two reasons. First, we have only considered 268 images of the background galaxies in the five cluster fields. Second, given the redshift range of $0.73<z<5.75$ (see \autoref{HFF_clusters}), the ground-based Rubin Observatory filter set is not optimal for detecting SNe in these galaxies. However, thanks to the magnification from the galaxy clusters, LSST is sensitive of detecting LSNe~Is to very high redshifts ($0.73 < z < 1.95$). We note that the resulting expected number of LSNe in the selected galaxy cluster fields are a lower limit, since we have only considered few clusters and the multiply-imaged galaxies with spectroscopic redshift. Beyond the clusters that we have considered here, LSST will observe $\sim 70$ galaxy clusters with Einstein radii larger than $\theta_E >20 "$ that have $\sim1000$ multiply-imaged background galaxies (The LSST Science Book).  As the expectations of strongly lensed SNe in cluster fields depend on several factors, including the star formation rate and the stellar mass of the host galaxy, it is not straightforward to make a reliable prediction and we leave  the expectation in all clusters visible to LSST for a future study. }

The conclusions of all the metrics presented in the Section 5.2 are presented in \autoref{sl: General Conclusions from Strong Lensing}. The conclusions for the galaxy-lensed SNe and cluster-lensed SNe are similar in general, with small differences (see \autoref{fig:sl}). Since here we are focusing on particular cluster fields, what matters more is the number of visits to the cluster fields rather than having generally large area. Furthermore, in order to optimize the sensitivity to high-redshift SNe with multiple images in galaxy cluster fields, deeper images in the reddest bands ($i$, $z$ and $y$) are preferred. This can be obtained by co-adding images from visits closely separated in time. As mentioned in the previous section, the LSST will serve to detect the LSNe, but additional follow-up by other photometric and spectroscopic instruments will be needed to securely measure the time delays.

\begin{table}[htpb]
	\begin{center}
		\caption{The galaxy clusters considered in this work\label{HFF_clusters}.  The number of unique galaxies behind the cluster is given in column 2 and the number of their multiple images of these galaxies in column 3. The redshift range  of these galaxies is given in column 4.}
		\begin{tabular}{lccc}
			\hline
			Cluster & $N_{systems}$ & $N_{images}$ & $z_{min}-z_{max}$ \\ 
			\hline
Abell 1689  & 18 & 51 & $ 1.15 - 3.4 $ \\ 
Abell 370  & 21 & 67 & $ 0.73 - 5.75 $ \\ 
Abell 2744  & 12 & 40 & $ 1.03 - 3.98 $ \\ 
Abell S1063  & 14 & 42 & $ 1.03 - 3.71 $ \\ 
MACS J0416.1-2403  & 23 & 68 & $ 1.01 - 3.87 $ \\ 
			\hline
							Total & 88 & 268 \\
							\hline
		\end{tabular}
	\end{center}
\end{table}

\subsubsection{Number of ``golden" lensed quasars}
\label{sec:sl_quasars}

The goal of this section\footnote{Summarized and updated version of
\citep[Cosmology chapter of][]{LSSTScienceCollaboration2017}} is to
evaluate the precision we can achieve in measuring time delays in
strongly lensed active galactic nuclei (AGN), and as such, the precision on the measurement
of the Hubble constant from all systems with measured time delays.

Anticipating that the time-delay accuracy would depend on night-to-night
cadence, season length, and campaign length, we carried out a large
scale simulation and measurement program that coarsely sampled these
schedule properties. In \cite{Liao2015}, we simulated 5 different
light curve datasets, each containing 1000 lenses, and presented them to
the strong lensing community in a ``Time Delay Challenge'' (TDC). These 5
challenge ``runs'' differed by their schedule properties. \anguita{Entries to the challenge consisted of samples of measured time delays, the quality of which the challenge team then measured via three primary diagnostic metrics: time
delay accuracy ($|A|$), time delay precision ($P$), and useable sample fraction ($f$). The accuracy of
a sample was defined to be the mean fractional offset between the estimated and true time delays
within the sample. The precision of a sample was defined to be the mean reported fractional
uncertainty within the sample.}

Focusing on the best challenge
submissions made by the community, we derived a simple power law model
for the variation of each of the time-delay accuracy, time-delay
precision, and useable sample fraction, with the schedule properties
cadence (cad), season length (sea) and campaign length (camp). They are
given by the following equations:

\begin{eqnarray}
|A|_{\rm model} &\approx 0.06\% \left(\frac{\rm cad} {\rm 3 days}  \right)^{0.0}
\left(\frac{\rm sea}  {\rm 4 months}\right)^{-1.0}
\left(\frac{\rm camp}{\rm 5 years} \right)^{-1.1}  \\
P_{\rm model} &\approx 4.0\% \left(\frac{\rm cad} {\rm 3 days}  \right)^{ 0.7}
\left(\frac{\rm sea}  {\rm 4 months}\right)^{-0.3}
\left(\frac{\rm camp}{\rm 5 years} \right)^{-0.6}  \\
f_{\rm model} &\approx 30\% \left(\frac{\rm cad} {\rm 3 days}  \right)^{-0.4}
\left(\frac{\rm sea}  {\rm 4 months}\right)^{ 0.8}
\left(\frac{\rm camp}{\rm 5 years} \right)^{-0.2} 
\end{eqnarray}

All three of these diagnostic metrics would, in an ideal world, be
optimized: this could be achieved by decreasing the night-to-night
cadence (to better sample the light curves), extending the observing
season length (to maximize the chances of capturing a strong variation
and its echo), and extending the campaign length (to increase the number
of effective time delay measurements).

As accuracy and precision in time-delay measurements (assuming identical lens ``modeling uncertainty'') is roughly proportional to the statistical uncertainty on the Hubble constant. Our analysis thus consists of selecting only the sky survey area that allows time-delay measurements with accuracies of $<1\%$ and precision $<5\%$.  This high accuracy and precision area can be used
to define a ``gold sample'' of lenses. The TDC useable fraction averaged over this area gives us
the approximate size of this sample: we simply re-scale the 400 lenses
predicted by \cite{Liao2015} by this fraction over the 30\% found
in TDC.  Note that naturally there is a strong dependence with footprint: assuming the single visit depth is deep enough to detect the typical lensed quasar image, the number of lenses will scale linearly with the survey area. The uncertainty on the Hubble constant, will then finally scale as one over the square root of the number of lenses. While these numbers are approximate, the ratios between
different observing and analysis strategies provide a useful indication of relative merit.

Our calculations are performed using the full 10 years of LSST operations with observations in all bands contributing equally (i.e. monochromatic intrinsic variability). However, it is important to note that there is a rather large caveat: Even when AGN variability can show almost negligible difference between bands close in wavelength, the difference can be important between the bluest and reddest LSST bands. As such, the numerical values have to considered as an optimistic lower limit but, as mentioned before, the ratios between observing strategies are an indication of relative merit.

\subsubsection{General Conclusions from Strong Lensing}
\label{sl: General Conclusions from Strong Lensing}

For our science case, a long cumulative season length and improved sampling frequency (cadence) is important. The cumulative season length provided by the baseline cadence is sufficient, but rolling cadences are disfavored as pointed out by \cite{Huber:2019}, because of the reduced cumulative season length. In terms of the cadence, \cite{Huber:2019} showed that, revisits of the same field within a single night should be done in different filters. Further improvements could be achieved by doing single snaps instead of two snaps, as shown in \autoref{fig:sl}. We also note that increasing the overall area would naturally yield an improvement in the number of ``golden'' lensed quasars, however, in this case it comes at the cost of having more visits in Galactic plane and having fewer WFD visits which would be detrimental for the number of lensed SNe with accurate and precise time delays.

\subsection{Kilonovae}

\authormacro{C. N. Setzer, H.V. Peiris, R. Biswas}
Within the next decade kilonovae (kNe) will mature as a probe of cosmological physics. Gravitational-wave (GW) observatories will begin to run nearly continuously and, with expected upgrades, become ever more sensitive to the mergers of binary neutron stars (BNS) that produce kNe. Detecting the electromagnetic (EM) counterpart to these sources, i.e., the kNe, will enable improved constraints on the accelerating expansion of the universe, via measurements of the Hubble constant \citep{Holz2005, Nissanke2013}, complementary to other probes of cosmic expansion \citep{Nissanke2010, Chen2018, Mortlock2018}. This is predicated on the detection of EM and gravitational waves from the same BNS mergers. The LSST will be able to detect kNe at distances beyond the range that future GW observatories operating in the 2020s will be sensitive \citep{Scolnic2018b,Setzer2019}. 

Combined with the large area and rapid cadence of LSST's WFD survey this offers the opportunity for the LSST to detect and identify kNe that will not be detected with any other instrument. However, GW detectors do not need to point in a given direction to make a measurement of a signal and are sensitive to signals from the entire sky. In principle, if GW detectors are operating coincidentally to a detection of a kN by the LSST, the signal of the BNS merger will be in the GW data, but possibly below the significance threshold used to claim a merger detection. The kN detection by the LSST can then be used as prior information to reverse-trigger a search through GW data for an accompanying merger signal \citep{Kelley2012}. With this information the population of sources with both EM and GW signals, i.e., standard sirens, used for studies of fundamental physics can be increased and studies can be made of the underlying BNS population. This may be critical for studying the selection effects of GW-detection of standard sirens.

\begin{figure}
\centering
\includegraphics[width=1\linewidth]{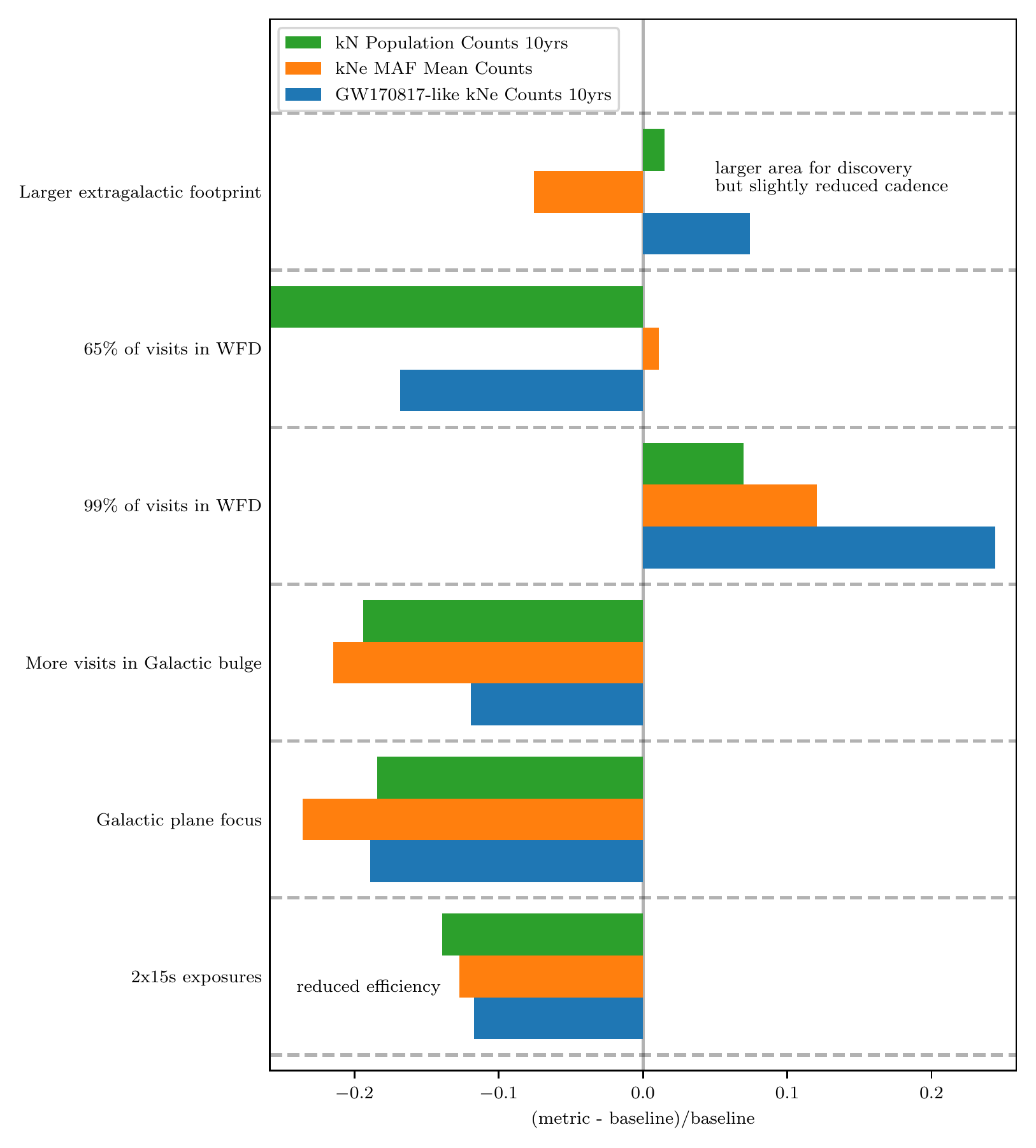}
\caption{Kilonova metrics as a function of selected observing strategies. \ml{\autoref{tab:sims_short} contains the exact simulation names corresponding to the short names used here.}}
\label{fig:kn}
\end{figure}

For the kilonovae probe, we introduce 3 new metrics.  A summary of the results from this section can be seen in \autoref{fig:kn}.
\begin{enumerate}
\item GW170817-like kNe Counts 10yrs - This metric represents the number of GW170817/AT2017gfo-like \citep{Abbott2017} kilonovae that are detected according to a set of criteria over the full 10 year survey.	

\item kNe MAF Mean Counts	- This metric represents the MAF implementation which evaluates the number of GW170817/AT2017gfo-like kilonovae that are detected on average per region of the sky assuming a kilonova is always ``going-off'' at a fixed redshift of 0.075.
\item kN Population Counts 10yrs - This metric represents the number of kilonovae drawn from a population that are detected according to a set of criteria over the full 10 year survey.
\end{enumerate}		

\subsubsection{Serendipitous Kilonova Detections}
\label{sec:kn_metrics}
To show the potential for this promising science case we use the number of detected kNe as a metric to compare survey strategies. \ml{We here focus only on serendipitous detection of kilonovae and not target-of-opportunity follow-up of gravitational wave triggers which has different requirements.} We introduce three metrics: the expected kilonova population counts after 10 years, the GW170817-like counts of kilonova after 10 years, and the MAF-Mean Counts of kilonovae.

To classify a kNe as being detected we use the criteria from \citep{Scolnic2018b} and used again by \citep{Setzer2019}. These criteria are the following:
\begin{itemize}
  \item Two \ml{LSST} alerts separated by $\geq 30$ minutes.
  \item Observations in at least two filters with $\mathrm{SNR} \geq 5$.
  \item Observations with $\mathrm{SNR} \geq 5$ separated by maximum of 25 days \ml{(i.e. no large gaps)}.
  \item A minimum of one observation of the same location within 20 days before the first $\mathrm{SNR} \geq 5$ observation.
  \item A minimum of one observation of the same location within 20 days after the last $\mathrm{SNR} \geq 5$ observation.
\end{itemize}

\ml{Several of these requirements were implemented to reject potential contaminants such as asteroids, AGN and superluminous supernovae \citep{Scolnic_2018, Setzer2019}. However, they are insufficient to perfectly separate kNe from other transient classes. While work is ongoing to implement a transient classification metric (as discussed in \autoref{sec:future}) and we anticipate such a metric to correlate with the same quantities that impact our kNe metric, we acknowledge that this is an important step that must be studied in future work.}

As was concluded by \citet{Setzer2019}, survey strategies which increase the cadence of observing a region of the sky in multiple filters, e.g., obtaining the nightly pair of observations in different filters, provide greater numbers of detected kNe. We direct the reader to \citet{Setzer2019} for a full discussion of the kN models and methodology used for this similar analysis. Since that work, the default setting for new observing strategies has been to implement the nightly pairs of observations in different filters. In this analysis of additional survey strategies, we find several more features which improve the numbers of detected kNe relative to the baseline strategy.
 
The first feature-change which benefits our science is when the number of visits in the WFD survey is increased. Whether this is achieved through the use of $1\times30$\,s exposures instead of $2\times15$\,s to increase survey efficiency or directly increasing the time-allocation of the WFD, more extra-galactic observations increases the number of detected kNe. This increase in the number of observations, given a fixed survey length, effectively increases the cadence of observations for any field.
 
This is expected, kNe are fast-evolving transients and their detection will be sensitive to the cadence of observations. However, we find that rolling observing strategies, which most directly increase the cadence of observations, do not notably improve the number of kN-detections over the baseline observing strategy. Kilonovae are not only fast-evolving, but they are also rare, and thus our metric is sensitive to the amount of cosmological volume, effectively sky-area, that is observed. Rolling-style observing strategies fundamentally decrease the observed sky-area active within an observing season by separating the sky into bands of declination. We find any improvement that might be expected from the increased cadence is, at best, negated by the substantial decrease in actively observed sky-area.
 
Conversely, increasing the total survey area with a fixed number of observations will decrease the cadence of observations. We find if only the observed sky-area is increased, there is not an improvement to our metric. However, if the increase in sky-area is accompanied by the addition of more observations, such as a reduction in total time allocated to mini-surveys, e.g., the removal of {\it uiy}-band observations outside the WFD survey, we see a modest increase in the number of detected kNe. We conclude increasing the footprint area of the survey is beneficial for increasing the number of detected kNe only if the number of observations per field is not significantly decreased. Furthermore, a larger extra-galactic sky area can be achieved by moving the observed sky-area to regions of low dust-extinction. We find, additionally, the proposed survey strategies which avoid the Milky-way by a dust or galactic-latitude cut are quite beneficial for detecting greater numbers of kNe.
 
Lastly, we note that while kNe are expected to be quite red transients, we do not find a redistribution of observations into redder filters to improve detections. Of the proposed distributions of observations by filter, the baseline filter-distribution performs best for both the population model of kNe and the model based on GW170817. \ml{This can be understood by considering the LSST bandpass efficiency, which is highest for the $g$-band and decreases for the redder bands (LSST SRD). The best filter distribution with which to detect kNe is hence dependent on both the underlying spectral energy distribution of kNe and the bandpass efficiencies of LSST. It appears that the current baseline filter allocation is close to optimal, at least for the chosen population model.}

While the full detection analysis
is in general computationally prohibitive for hundreds of proposed survey strategies, a simplified version of this is implemented into the Metric Analysis Framework (MAF) that is used by the LSST Project and is publicly available. The MAF implementation\footnote{ \url{https://github.com/LSST-nonproject/sims_maf_contrib/blob/master/mafContrib/GW170817DetMetric.py}}, labelled above descriptively as {\it kN MAF Mean Counts}, considers only GW170817-like kNe. Like other MAF transient metrics, it considers that these transients go off at a single, user-specified, cosmological redshift and occur one after another at each point on the sky. From this, observations are made of the light curves according to the chosen survey strategy and these are checked whether they pass certain detection criteria. Of the criteria listed above, we are limited to implementing only criteria two and three. From this we obtain a number of detected kNe; however, as this is not a proper cosmological transient distribution based on comoving volumetric rates this metric should not be used to forecast detected counts of kNe per survey, as was done in the full analysis. In this case, we instead compute the mean number of detections, i.e., the number of detected kNe averaged over the number of fields in the analyzed survey strategy, as our metric for comparison. The use of these numbers should only be in comparison to other survey strategies. Given the limitations in the MAF implementation, the MAF metric does not exactly emulate all metric results we see from the full analysis. However, the number of outliers is small and the overall trends are reproduced.

\subsubsection{General Conclusions from Kilonovae}

We conclude the most important observing strategy feature for improving the number of kilonovae detections is still obtaining a pair of observations in different filters within a single night. Second to this, increasing the number of the observations in the extragalactic WFD survey is very beneficial. Lastly, we find observing strategies that increase the observed sky-area in low dust-extinction regions, and do not substantially decrease the cadence of observations to achieve this larger area, are also preferable.

\section{Discussion}
\label{sec:discussion}
Combining the insights gained from our cosmology-related metrics is not a trivial task. In an ideal world, a full DETF FoM could be calculated for each simulation allowing the objective determination of the ``optimal'' observing strategy for cosmology. However, as we expect systematic effects to play a significant role in final cosmological constraints with LSST, we do not currently have a realistic enough FoM to make final decisions based on it alone. We make use of the 3x2pt FoM described in \autoref{sec:static_fom} to summarize the impact of observing strategy on the main static science probes, but also consider separately \rachelNew{WL} systematics and transient metrics. These simpler, more interpretable metrics can assist in gaining deeper insight and making general recommendations for observing strategy. We expect metrics such as the total number of well-measured supernovae and survey uniformity to correlate strongly with cosmological parameter constraining power once systematic effects are taken into account. In this section, we use select metrics to investigate and draw insight into various aspects of observing strategy.

\subsection{Footprint, area and depth}
\begin{figure*}[ht!]
\centering
\includegraphics[width=0.8\linewidth]{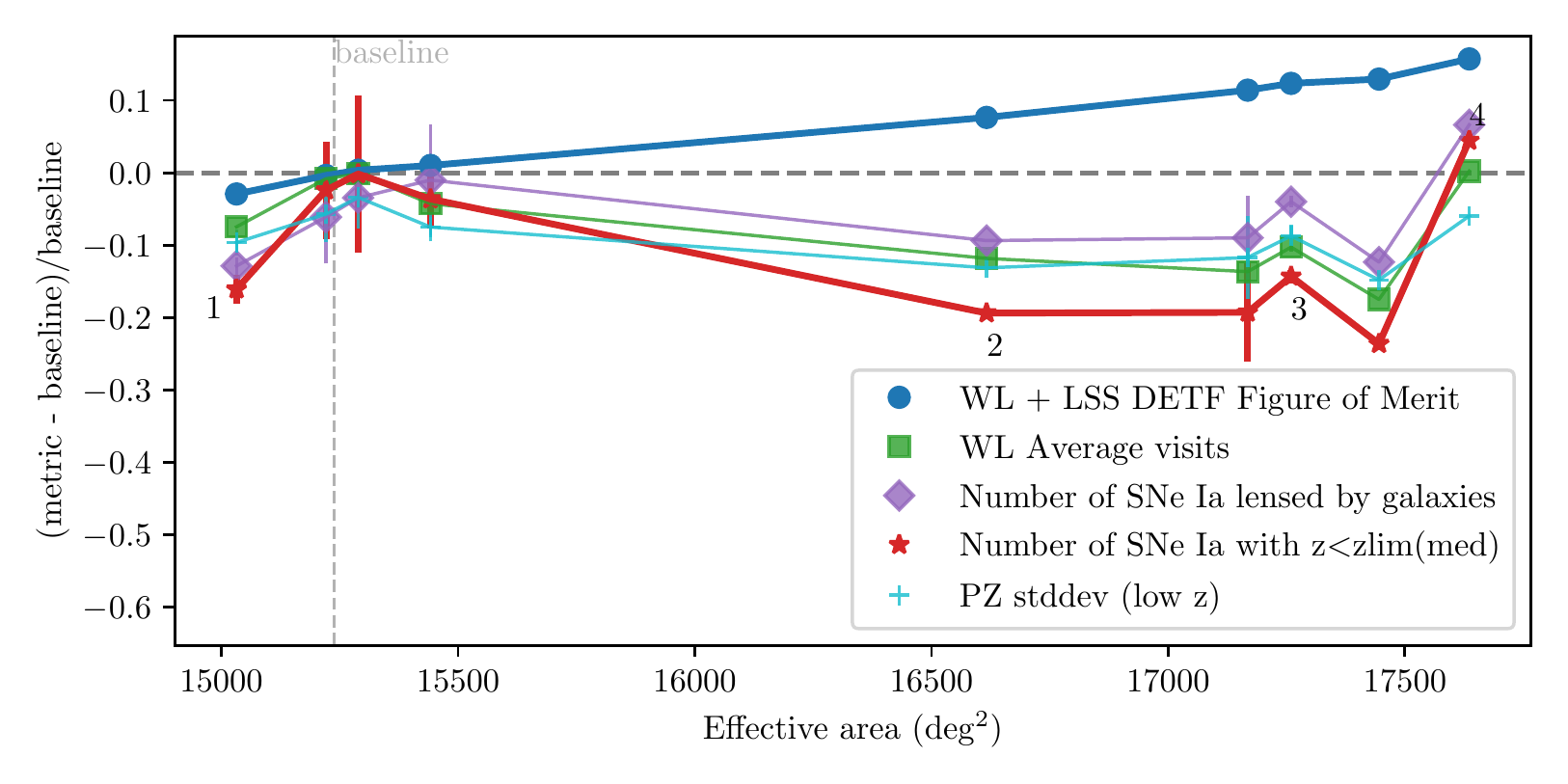}
\caption{Selected metrics, relative to their values at baseline, as a function of effective area (i.e., the area that meets the cuts described in \autoref{sec:static_metrics}) of different observing strategies. \ml{To improve readability, points that are nearby in the $x$-axis are binned with only the mean and error bar plotted for that bin.} It can be seen that the 3x2pt FoM \EG{metric simply prefers} more area while the situation is more complex for time-dependent metrics. \ml{We have highlighted specific simulations with numbered annotations. The simulation at 1 reduces the overall area available for cosmology by reducing visits in the crucial redder bands. Simulations 2 and 3 have larger area footprints, but degrade transient metrics and photometric redshifts due to a reduced number of visits in the extragalactic part of the WFD. Simulation 4 reverses this by completely prioritizing visits in areas of low dust extinction, resulting in both large area and a large number of visits and the best performance for all cosmology metrics. List of annotations: 1-\texttt{footprint\_bluer\_footprintv1.5\_10yrs}, 2-\texttt{footprint\_newBv1.5\_10yrs}, 3-\texttt{bulges\_bs\_v1.5\_10yrs}, 4-\texttt{footprint\_big\_sky\_dustv1.5\_10yrs}.}}
\label{fig:comparison_area}
\end{figure*}

In general, there is a trade-off between depth and area in any survey. \autoref{fig:comparison_area} shows several key metrics as a function of effective survey area (\mlNew{the survey area meeting the cuts described in \autoref{sec:static_metrics}}). The first thing to note is that the 3x2pt FoM has a simple linear relationship with area and the larger the survey, the better the 3x2pt FoM. Photometric redshifts however tend to prioritize depth over area. There is also a trade-off between \rachelNew{WL} shear systematics, which improve as more visits are taken (essentially ending up with more depth) and the 3x2pt FoM, which prefers larger area. To fully quantify this trade-off, the \rachelNew{WL} systematics would need to be included in the full DETF FoM pipeline. Each of the transient metrics has a more complicated relationship with area since they are impacted by other observing strategy choices such as cadence and filter distribution. 

Of particular interest are the observing strategies with the largest area. It can be seen that some strategies yield large area (around $16,500$ deg$^2$) but poor performance for the transient metrics and the \rachelNew{WL systematics} metric. This is because these strategies have a large footprint but prioritize visits in the Galactic bulge and plane, reducing cadence in the extragalactic area. The largest area simulation, which gives simultaneous high performance for the 3x2pt FoM and the number of supernovae, is the \verb!footprint_big_sky_dust! simulation which uses an extinction-based cut to define the WFD footprint. This allows both increased area and depth for extragalactic science. The number of visits in WFD and the other surveys are the same as baseline, \ml{however the number of visits in the Galactic plane is dramatically reduced}.

\begin{figure*}[ht!]
\centering
\includegraphics[width=0.8\linewidth]{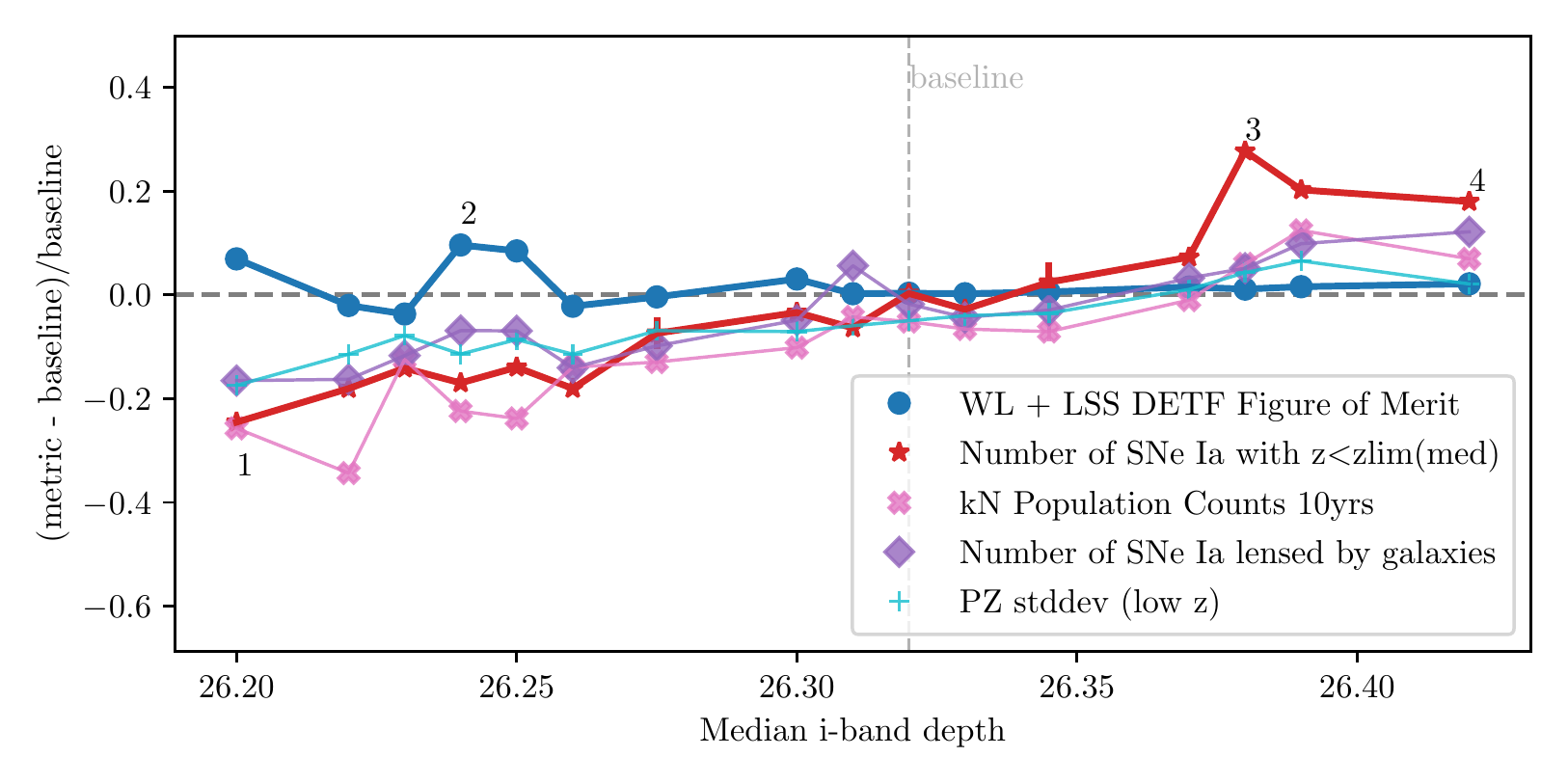}
\caption{Selected metrics, relative to their values at baseline, as a function of $i$-band median co-added depth of different observing strategies. \ml{To improve readability, points that are nearby in the $x$-axis are binned with only the mean and error bar plotted for that bin.} Here we note that the 3x2pt FoM metric is generally indifferent to depth, implying that larger area is more important (assuming the changes in depth remain small). Photometric redshifts improve with more depth and the transient science probes, particularly supernovae and kilonovae, are strongly affected since generally greater depth corresponds to increased cadence. \ml{We have highlighted specific simulations with numbered annotations. Simulation 1 has the poorest performance because it removes visits from the extragalactic part of the footprint. Simulation 2 is a large area footprint which improves the static science metrics, but still reduces overall cadence and depth. Simulations 3 and 4 artificially perform well for the transient metrics because they unrealistically ignore all mini-surveys and put all visits into WFD. List of annotations: 1-\texttt{footprint\_newAv1.5\_10yrs}, 2-\texttt{bulges\_cadence\_i\_heavy\_v1.5\_10yrs}, 3-\texttt{wfd\_depth\_scale0.99\_v1.5\_10yrs}, 4-\texttt{wfd\_depth\_scale0.99\_noddf\_v1.5\_10yrs}.}}
\label{fig:comparison_depth}
\end{figure*}

\autoref{fig:comparison_depth} shows a subset of metrics as a function of median co-added $5\sigma$ $i$-band depth. Here we see the lack of sensitivity of the 3x2pt FoM, as long as sufficient depth is achieved. On the other hand, two transient metrics, number of supernovae and number of kilonovae, improve by as much as 27\% for a deeper survey. This is really a consequence of an improved number of visits and thus increased cadence. However, most of the simulations that are significantly deeper than the baseline are not actually realistic, as they place artificially large amounts of survey time into WFD. While these experiments are useful to understand the behavior of the metric, they do not represent viable observing strategies. However, it is entirely possible to choose a footprint or strategy that reduces the overall depth of the survey, to the detriment of the transient probes and photometric redshifts.

Within the constraints of the observing strategy requirements from the LSST SRD, it appears to be quite challenging to achieve a depth much greater than the current baseline. The ideal approach from the standpoint of cosmology is to maintain a WFD footprint of 18,000 deg$^2$ but prioritize regions with low dust extinction, allowing for increased area while maintaining current depth.

\newpage
\subsection{A simple combined metric to analyze the depth-area trade-off}
\begin{figure*}
\centering
\includegraphics[width=0.8\linewidth]{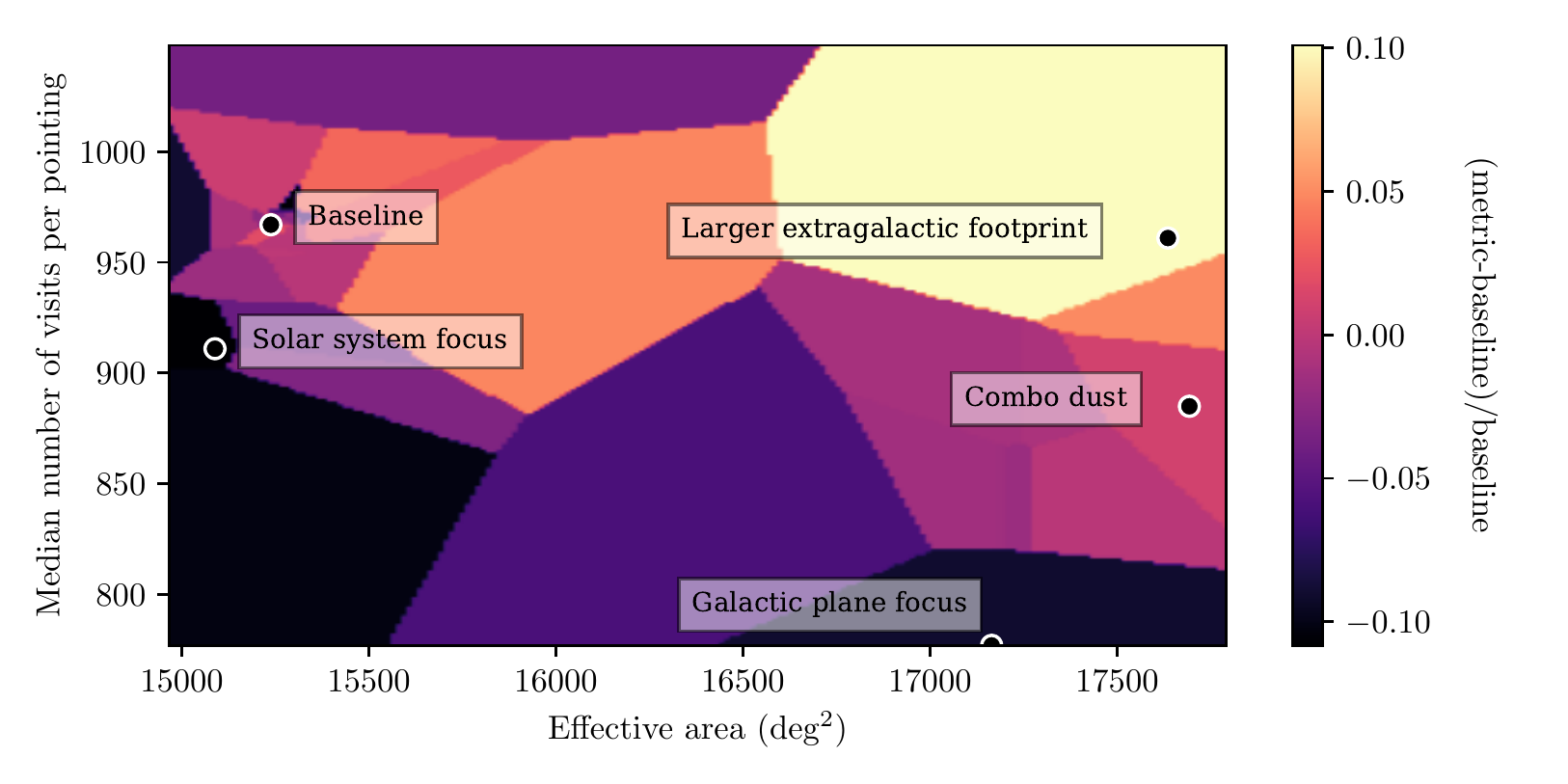}
\caption{A simple attempt to combine multiple metrics to produce a combined cosmology metric. We show the combined metric (using the usual (metric-baseline)/baseline)) as a function of effective area (i.e., the area that meets the cuts defined in \autoref{sec:static_metrics}) and median number of visits (which correlates with depth for most simulations). Although the produced map is complex, the trade-off between depth and area can be seen. Several specific strategies are highlighted, \ml{the simulation names of which can be found in \autoref{tab:sims_short}. Unlike for the rest of the figures which focus on simulations that more or less vary one aspect of observing strategy at a time, we have highlighted here strategies from FBS v1.6 which are proposed observing strategies that aim to satisfy multiple science goals. ``Solar system focus'' and ``Galactic plane focus'' both take observations away from WFD to prioritize other science cases. ``Combo dust'' is a proposed large area footprint which has more area but less depth than ``Baseline'', thus producing similar performance. ``Larger extragalactic footprint'' is a larger area footprint that is defined by dust extinction and gives improved cosmological constraints due to the area gained by avoiding extinction.} We caution the reader against assuming the performance in this metric would correspond to the exact percentage improvement/degradation in cosmological constraints. Only a full DETF FoM including systematics can indicate the exact numerical impact of observing strategy choice.}
\label{fig:mega_metric}
\end{figure*}

\mlNew{To better understand the trade-off between area and depth, we combine metrics in a very simple way} based on the DESC SRD. \ml{Because of the complexity involved in combining the systematic related metrics such as the \rachelNew{WL systematics} and photometric redshift metrics, we here only include the 3x2pt FoM and the number of well-measured supernovae \EG{metric}, in a 50--50 ratio \EG{after normalizing each metric to represent the ratio in the amount of information provided by a given simulation versus the baseline}. We note that both these metrics incorporate some (but not all) systematic effects.}  While we caution the reader that this is a simple approximation of the much more complex full DETF FoM that includes \ml{all} systematics, it can be helpful in gaining insight. The general trade-off between number of visits (and hence depth) and area can be seen and, while it is difficult to improve over the current baseline, many simulations are much worse. \mlNew{Our preferred footprint, indicated by ``Larger extragalactic footprint'' that implements a dust extinction cut, has the best overall performance for cosmology by increasing area without reducing depth.} One proposed simulation, indicated as ``Combo dust'' also has a larger area footprint defined by dust extinction, \mlNew{but includes a significant number of visits in the Galactic plane}. This simulation is roughly equivalent to baseline \mlNew{for this combined metric}, as can be seen in \autoref{fig:mega_metric}, because what is gained in area is lost in depth.  

It is important to note that key probes such as strong lensing and kilonovae are not included in the combined metric because their constraining power is not comparable to the other probes and they would be unfairly downweighted. Although they do not contribute as significantly to the \EG{overall} DETF FoM, these probes are still essential for other aspects of cosmology (unravelling the $H_0$ tension, for example). We thus consider them separately from this combined metric. 

Both kilonovae and strong lensing suffer if the area is increased so our recommendation is that any increases in area should maintain the depth in WFD \ml{as far as possible} by redefining the footprint, \ml{using rolling cadence} \mlNew{(discussed in \autoref{sec:future})} and \EG{re-optimizing the number of 
visits and filter distribution} 
in the Galactic plane and other \EG{mini-surveys for those specific science cases}.

\newpage
\subsection{Overlap with other surveys}
\label{sec:overlap}
\ml{Data from other surveys is key to enhancing the cosmological analysis with LSST data. Planned surveys using multi-object spectroscopy including the Dark Energy Spectroscopic Instrument \citep[DESI;][]{DESI} and the 4-metre Multi-Object Spectroscopic Telescope \citep[4MOST;][]{4MOST} will provide spectroscopic data for millions of objects. This can be used for photometric redshift calibration and training, cross correlation to improve \rachelNew{WL} and host galaxy redshift identification for transients \citep{Mandelbaum2019}. The Time-Domain Extragalactic Survey \citep[\tides;][]{TIDES} on 4MOST will provide live spectra for up to 30,000 transients, including supernovae. The Euclid telescope \citep{Euclid} will survey over 15,000 deg$^2$, obtaining photometric and spectroscopic data in the infrared bands, complementary to LSST and enabling better photometric redshift calibration and training, improvements to \rachelNew{WL} and improved de-blending of galaxies \citep{Capak2019}.

\autoref{tab:overlap} shows how the overlap with DESI, 4MOST and Euclid can be improved with a larger area footprint. Even better coverage, particularly for DESI, can be achieved by extending the footprint further North, as described in \autoref{sec:conclusions}.}

\begin{table}[ht!]
    \centering
    \caption{Area overlap for the 10-year LSST survey  \humnaNew{(extragalactic footprint) }with other surveys for two types of footprint: baseline and a larger area footprint with a boundary defined by a dust extinction cut (see \autoref{tab:sims_short} for the precise simulation names). The larger footprint provides nearly 2000 deg$^2$ additional overlap with DESI and over 1000 deg$^2$ with 4MOST and Euclid. Even greater gains can be made by extending the footprint further North.}
    \begin{tabular}{lccc}
        \hline
        \input{overlap.tbl}
    \end{tabular}
    
    \label{tab:overlap}
\end{table}

\ml{Finally, we also note that external data in the DDFs will be crucial for photometric redshift training and supernova spectroscopic follow-up (both live and later host galaxy redshift determination). It is particularly important and challenging to align DDF observations temporally with telescopes such as Euclid, meaning that the optimal strategy will likely be to prioritize certain DDFs in certain years. We leave it to future work to determine the optimal DDF strategy for cosmological measurements with LSST.}

\subsection{Filter distribution}
\begin{figure*}
\centering
\includegraphics[width=0.8\linewidth]{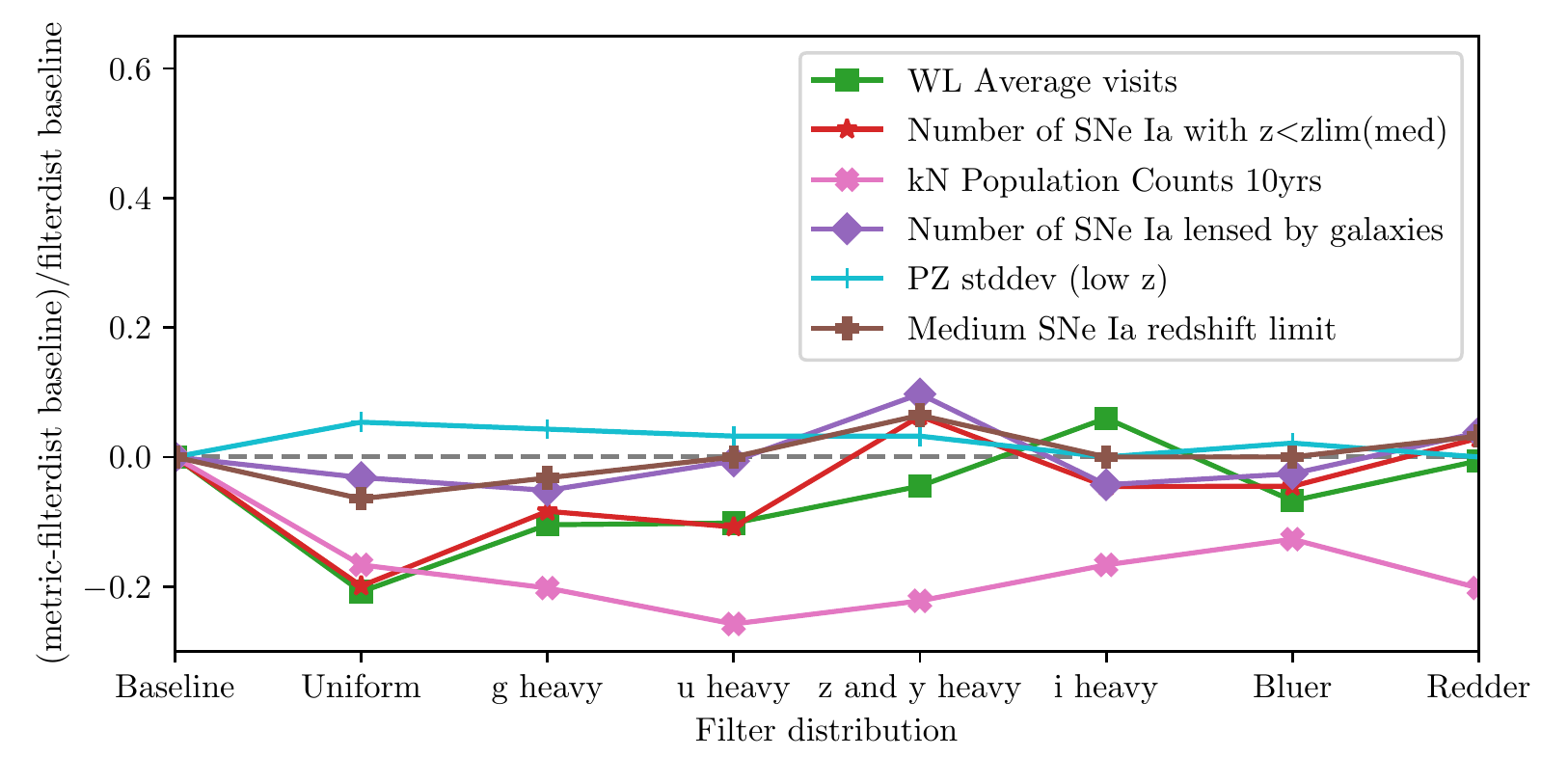}
\caption{Selected metrics, relative to their values at baseline, as a function of varying choices of filter distributions. Note that these simulations use a different footprint so all metrics are measured against the baseline \texttt{filter\_dist} simulation, and not the standard baseline. We do not show the 3x2pt FoM or LSS metrics because they do not vary significantly here. We find no compelling reason to vary the baseline filter distribution: different distributions improve different probes but at the cost of others.}
\label{fig:comparison_filter_dist}
\end{figure*}

While the LSST SRD enforces a minimum number of visits in each filter, there is still room to increase or decrease the number of visits (and hence the cadence and depth) in any particular band. This choice of filter distribution has a complex impact on our cosmology metrics, as seen in \autoref{fig:comparison_filter_dist}. Here we see some small tensions between probes: supernovae prefer a redder distribution \ml{while any change from baseline is detrimental to the number of kilonovae}. \ml{The choice of filter distribution is especially complex for transients, where the spectral energy distribution of the transient, the underlying efficiency of the LSST bandpasses and the cadence in each band all play a role. For example, kilonovae are likely intrinsically redder transients and yet our metric does not improve with a redder filter distribution. This could be because the LSST $g$-band single visit depth is much deeper than the redder bands, meaning the optimum for kNe detection is somewhere between the redder and the bluer distribution and ends up being close to baseline. The baseline filter allocation was heavily influenced by photometric redshifts so any radical departures from it could prove detrimental. Our photometric redshift metric actually improves slightly for some of the filter distribution simulations, but not significantly.} From our results, it is difficult to find a compelling reason to change the distribution from baseline. While the overall filter distribution may already be acceptable for cosmology, \emph{when} those filters are used can still have significant impact on transient metrics, as discussed in \autoref{sec:discussion_cadence}.

\newpage
\subsection{Visit pairs and exposure time}
\label{sec:visit_pairs_exposure}
\begin{figure*}[ht!]
\centering
\includegraphics[width=0.7\linewidth]{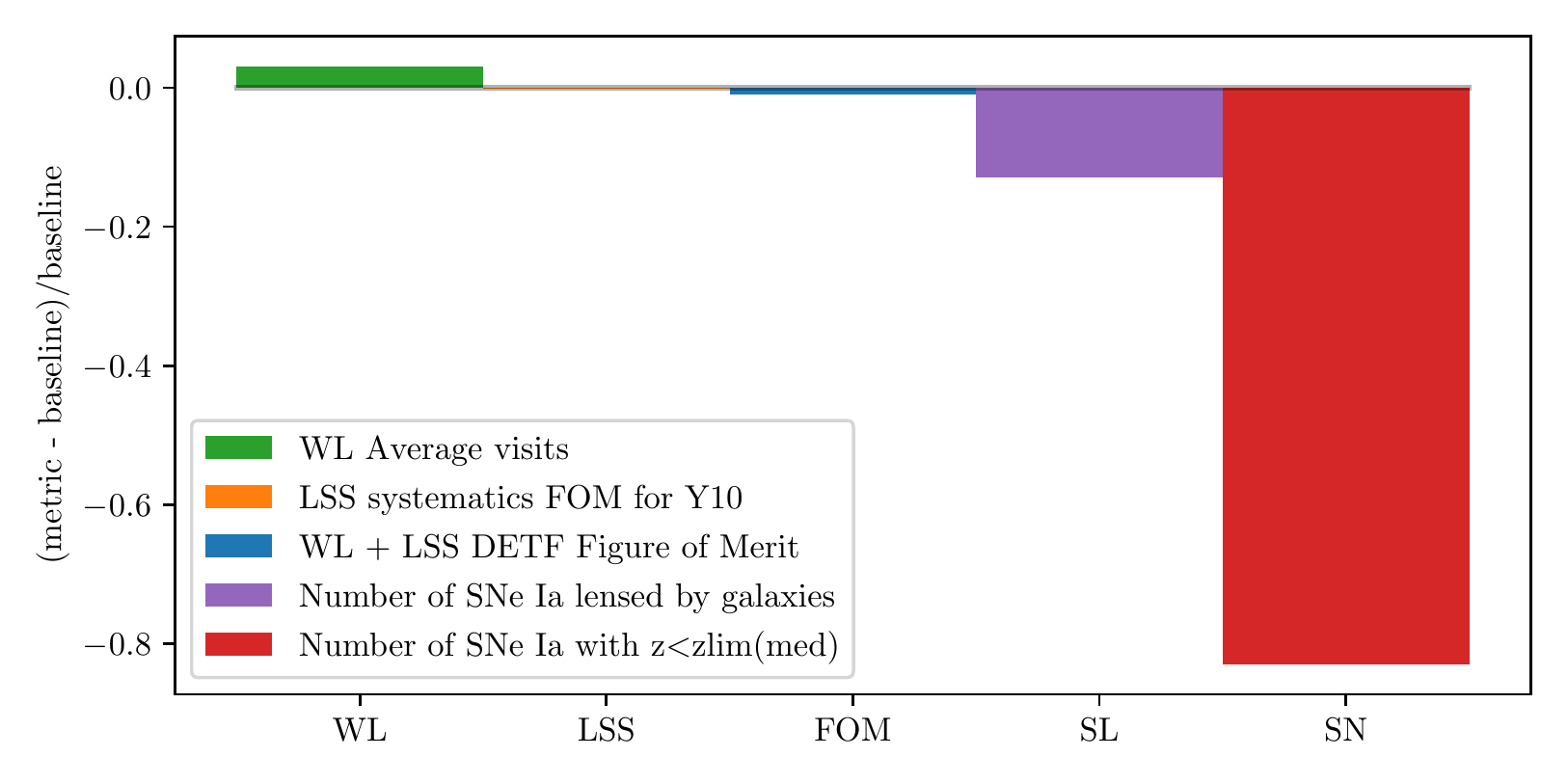}
\caption{Selected metrics, relative to their values at baseline, for the \texttt{baseline\_samefilt\_v1.5\_10yrs} simulation. This clearly shows that while taking visit pairs in the same filter does not impact static science much, it dramatically degrades the transient metrics.}
\label{fig:comparison_samefilt}
\end{figure*}

Cadence remains a critical choice when designing observing strategy for any survey that aims to detect transients. One of the most important changes to observing strategy simulations has been to ensure visit pairs are taken in different filters. \autoref{fig:comparison_samefilt} shows the dramatic decrease in the number of supernovae detected if this is not enforced. Although taking pairs in the same filter is more efficient, it severely degrades the overall cadence of the survey. \mlNew{Taking visit pairs in different filters has been the default in simulations since the 2018 white paper call.}

\begin{figure*}
\centering
\includegraphics[width=0.7\linewidth]{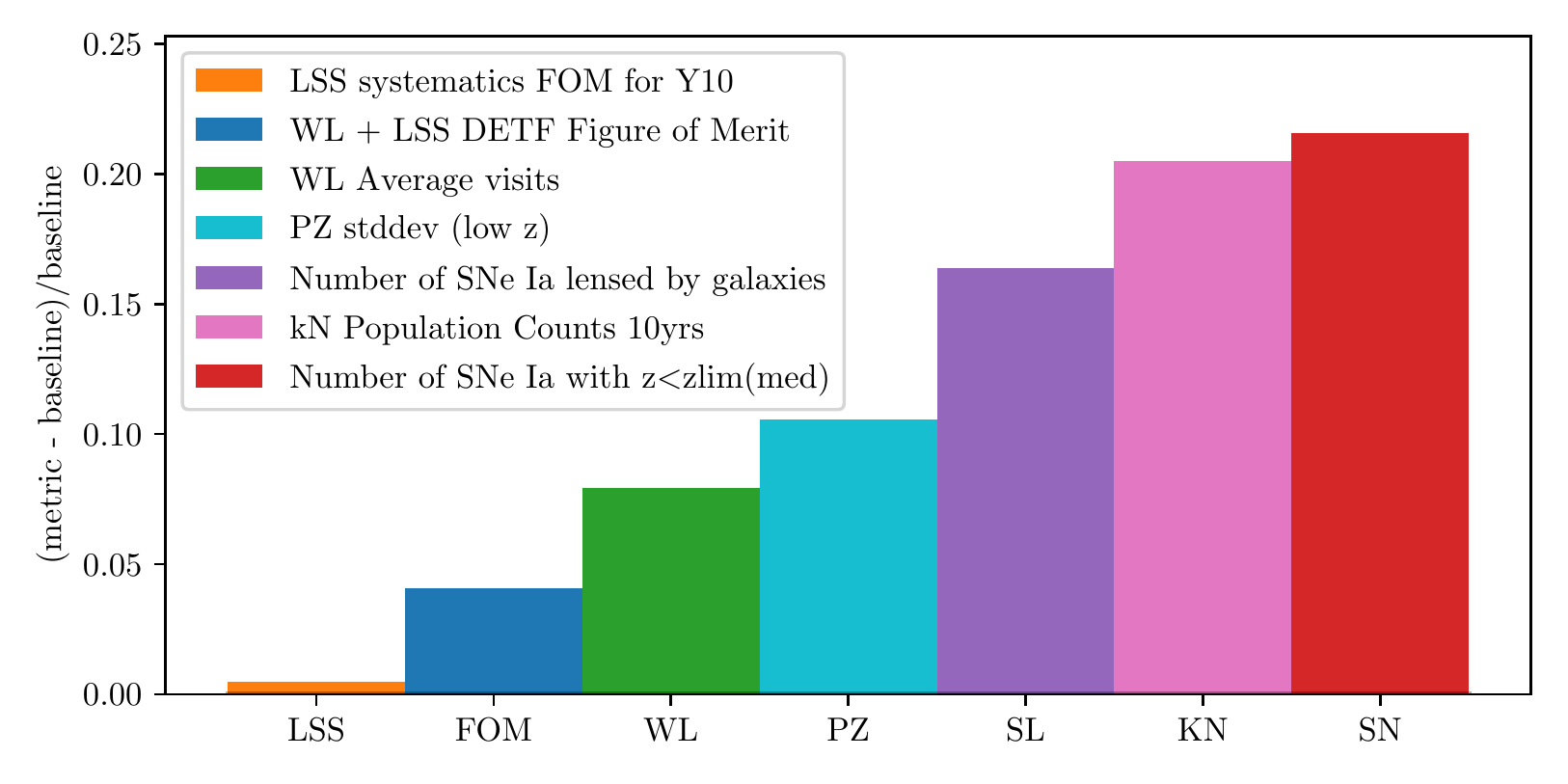}
\caption{Selected metrics, relative to their values at baseline, for the \texttt{baseline\_nexp1\_v1.7\_10yrs.db} simulation, as compared to the current v1.7 baseline called \texttt{baseline\_nexp2\_v1.7\_10yrs.db}. The only difference between these two simulations is the choice of whether an observation is taken in two 15\,s exposures (the current baseline) as opposed to a single 30\,s exposure. It is clear that the efficiency gained through a single exposure improves all metrics, some dramatically.}
\label{fig:comparison_nexp}
\end{figure*}

\ml{Another important decision to be made is whether the visits should be made in a single exposure ($1\times30$\,s) or separated into two exposures ($2\times15$\,s), which could help reject cosmic rays. While this decision will only be made during commissioning, \autoref{fig:comparison_nexp} shows how beneficial the $1\times30$\,s exposure would be to all major cosmology metrics.}

\newpage
\subsection{Cadence}
\label{sec:discussion_cadence}
\begin{figure*}[hb!]
\centering
\includegraphics[width=0.8\linewidth]{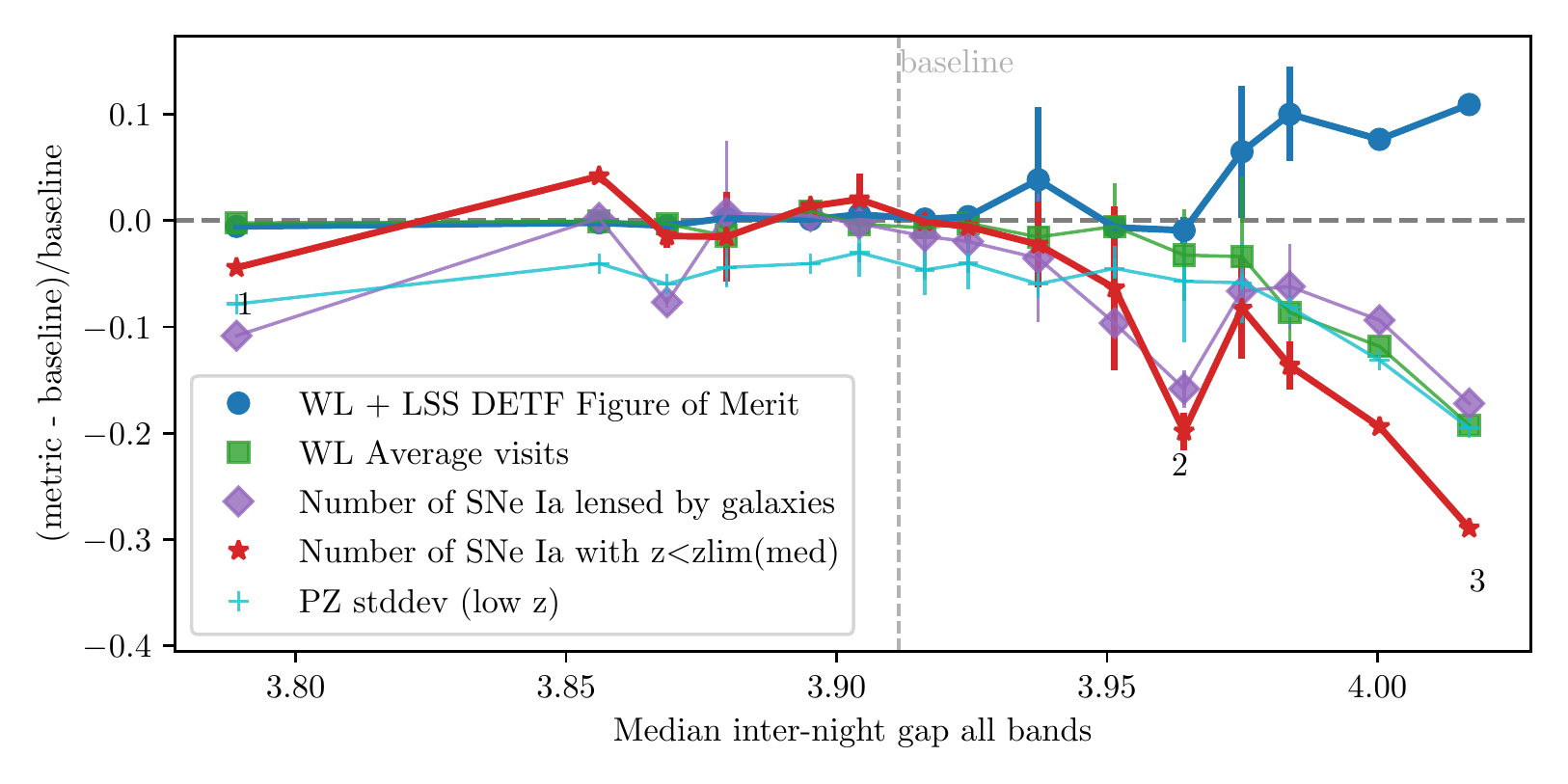}
\caption{Selected metrics, relative to their values at baseline, as a function of the median inter-night gap of different observing strategies. \ml{To improve readability, points that are nearby in the $x$-axis are binned with only the mean and error bar plotted for that bin.} The supernova metric prefers shorter gaps between observations but the situation is complicated by many other factors. Generally it is important the cadence is at least no worse than baseline. \ml{We have highlighted specific simulations with numbered annotations. Simulation 1 includes visits at high airmass, which is generally avoided, indicating that high airmass observations can improve cadence without too much degradation of metrics. Simulation 2 is when $2\times15$\,s exposures are used instead of $1\times30$\,s. Simulation 3 is a large area footprint but removes visits from the extragalactic area reducing cadence. List of annotations: 1-\texttt{dcr\_nham2\_ugr\_v1.5\_10yrs}, 2-\texttt{baseline\_2snaps\_v1.5\_10yrs}, 3-\texttt{footprint\_newAv1.5\_10yrs}.}}
\label{fig:comparison_cadence}
\end{figure*}

\autoref{fig:comparison_cadence} shows how different metrics behave as a function of the median inter-night gap. While the median inter-night gap is a good measure of overall cadence, it gives an incomplete picture as the tails of the distribution of inter-night gap can be quite broad. It is still useful however to quantify the average cadence of a simulation. The behavior of the metrics correlates fairly well with the usual area and depth trade-off: larger areas tend to produce worse cadence overall. An interesting point is that while there are only a few simulations that produce better cadence than baseline, there are many that can dramatically reduce the cadence which has a serious impact on the transient metrics and metrics that depend on depth.

\begin{figure*}
\centering
\includegraphics[width=0.8\linewidth]{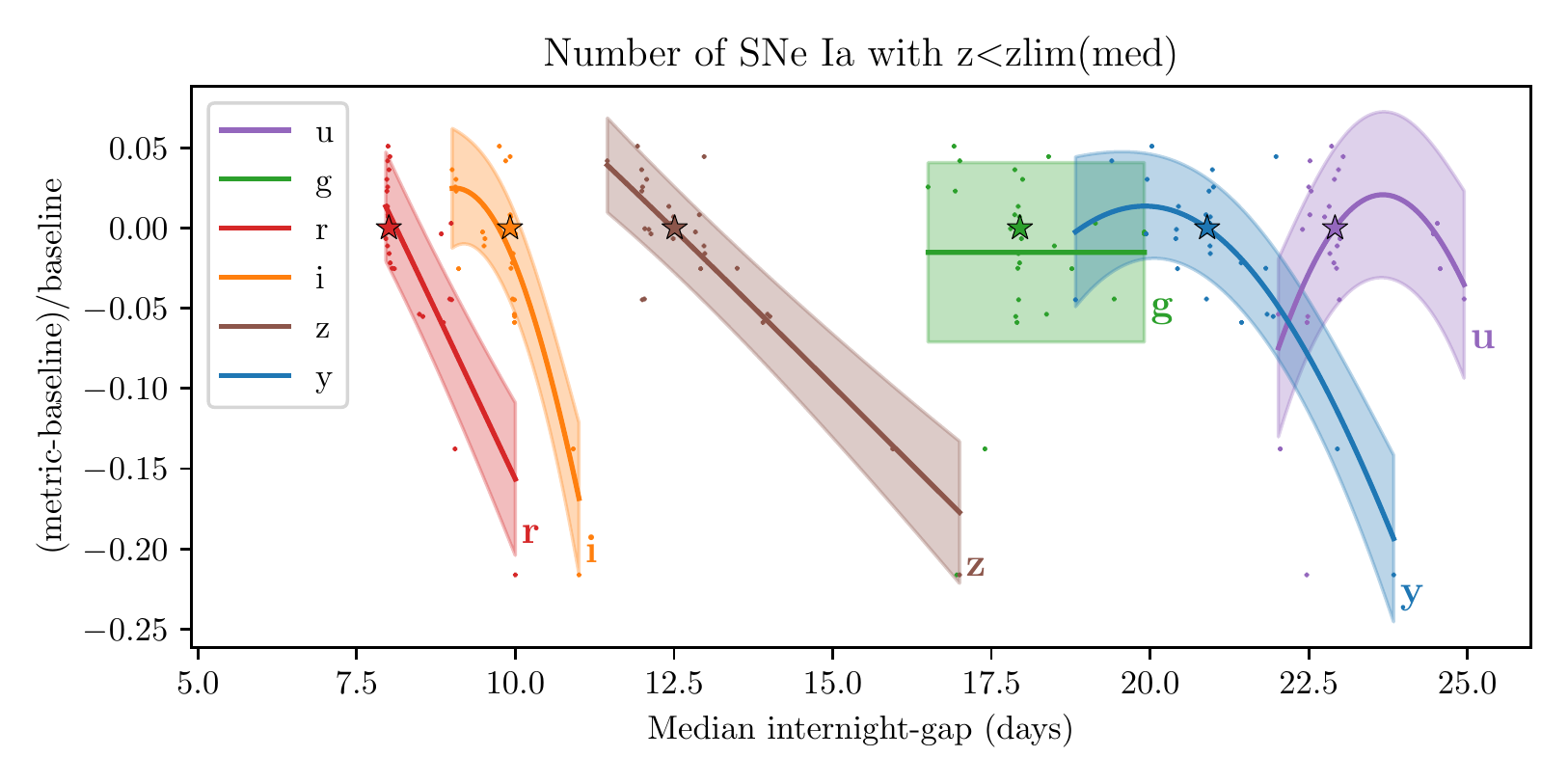}
\caption{The relative improvement over baseline for the number of supernovae as a function median inter-night gap in each filter. \ml{The baseline cadence is represented by star markers.} In general, the number of supernovae improve as the cadence improves, but note the relatively small dependence on $g$ and $u$ band, indicating the higher importance of good cadence in the redder bands. A Gaussian process has been fitted to the set of points in each filter to help guide the eye; the mean is plotted with a solid line and the standard deviation indicated with a coloured envelope. Also note that we have restricted this plot to simulations with similar number of visits in the WFD.}
\label{fig:internight_gap_filters}
\end{figure*}

\autoref{fig:internight_gap_filters} takes the cadence analysis a step further by investigating the impact of inter-night gap per filter on the number of supernovae detected. For fair comparison, only simulations with a similar number of visits in WFD are included. The average cadence varies significantly between the bands, mostly due to the non-uniform filter distribution but also due to decisions of what filters to use in different moon conditions. For supernova observations, the redder bands seem to be more important than the bluer bands. We find similar results for strongly lensed supernovae and kilonovae, although kilonovae have less dependence on $z$-band. 

\begin{figure*}
\centering
\includegraphics[width=0.8\linewidth]{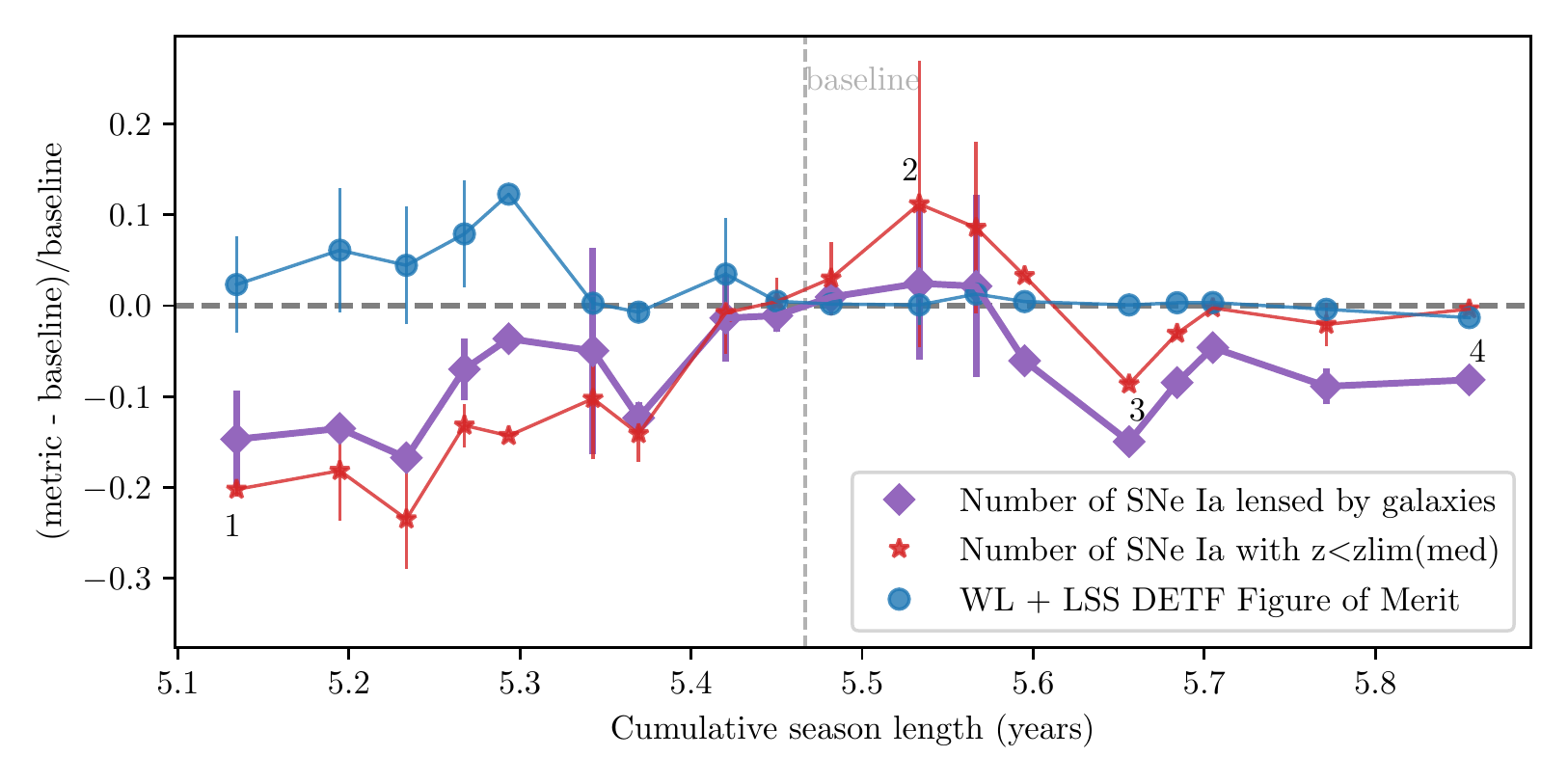}
\caption{Selected metrics, relative to their values at baseline, as a function of the cumulative season length (the total amount of time a field is observed) of different observing strategies. \ml{To improve readability, points that are nearby in the $x$-axis are binned with only the mean and error bar plotted for that bin.} Shorter season lengths degrade strong lensing and supernovae performance but as long as the season length does not fall much below it's value at baseline, these metrics do not seem to be strongly affected. \ml{We have highlighted specific simulations with numbered annotations. Simulation 1 is a larger area footprint which removes visits from the extragalactic area. Simulation 2 artificially improves cadence (and hence SN performance) by removing the mini surveys. Simulation 3 replaces some visits with short exposures that cannot be used for SN observations thus reducing signal-to-noise. And simulation 4 takes some observations at high airmass, which is usually avoided. This indicates that taking some number of high airmass observations is an effective way to increase season length without degrading metric performance significantly. List of annotations: 1-\texttt{footprint\_newBv1.5\_10yrs}, 2-\texttt{wfd\_depth\_scale0.99\_v1.5\_10yrs}, 3-\texttt{short\_exp\_5ns\_5expt\_v1.5\_10yrs}, 4-\texttt{dcr\_nham1\_ugri\_v1.5\_10yrs}}}
\label{fig:comparison_season}
\end{figure*}

\autoref{fig:comparison_season} shows selected metrics as a function of cumulative season length (how long in years a field is observed for on average). We find that as long as the season length remains comparable to baseline, most metrics are fairly insensitive to changes in season length.

\subsection{Weather and seeing}
\begin{figure*}[ht!]
\centering
\includegraphics[width=0.8\linewidth]{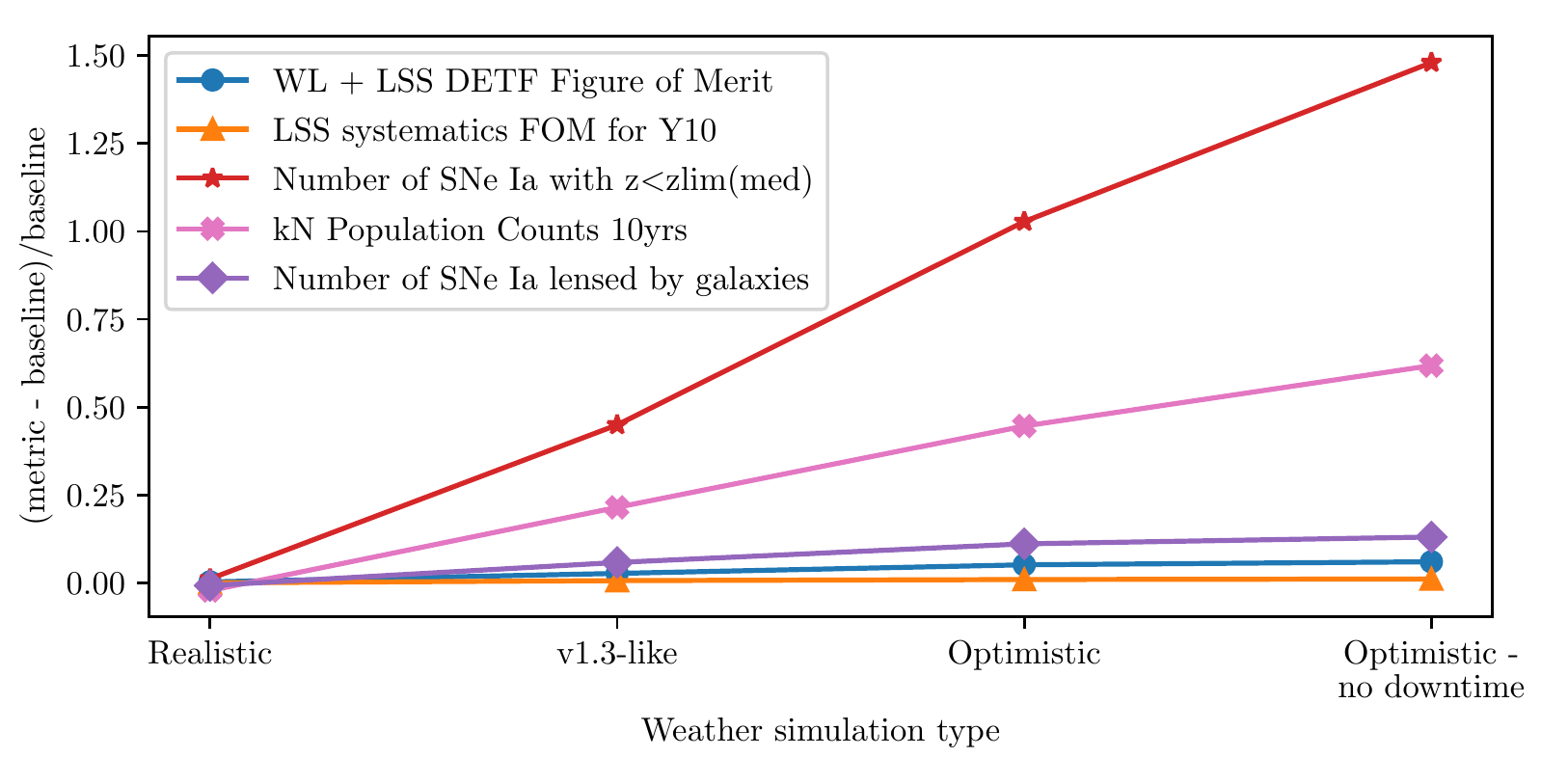}
\caption{Selected metrics, relative to their values at baseline, as a function of different weather simulations, \mlNew{for an older FBS version, v1.3}. The simulations make use of real weather data and differ on what cloud coverage would be required before \mlNew{observations are paused}: 30\% (realistic), 70\% (v1.3-like in other words similar to previous simulations), 120\% (optimistic, where the dome would never close due to bad weather) and no down time (optimistic-no downtime, where the telescope would always observe no matter what). While bad weather only has a small impact on static science metrics by slightly reducing the median depth, it has enormous impact on transient science metrics by as much as 150\%. This highlights the importance of including realistic weather in simulations to make accurate predictions of scientific returns.}
\label{fig:comparison_weather}
\end{figure*}

\autoref{fig:comparison_weather} shows observing strategies that simulated different weather scenarios, from realistic to impossibly optimistic. Naturally, metrics sensitive to cadence suffer immensely when weather conditions restrict observing time. Although these optimistic simulations are not realistic, they do indicate that much could be gained in transient science by ensuring fields missed by bad weather are observed as soon as possible to avoid large light curve gaps. 

\newpage
\section{Conclusions}
\label{sec:conclusions}

Because of its effect on a wide variety of science cases, optimization of LSST observing strategy is critical and has become an active area of research \mlNew{that has already resulted in significant improvements to the scheduler and baseline strategy}. In this paper, we have introduced metrics used within the LSST DESC to investigate the impact of observing strategy choices on cosmological measurements with LSST. While we continue to work towards our goal of a \mlNew{full DETF Figure of Merit combining all probes}, which includes systematics, the metrics introduced here are still valuable in gaining insights about observing strategy.

\subsection{LSST DESC recommendations for observing strategy}
\label{sec:obs_strat_factors}

Here we summarize our conclusions drawn from the metrics outlined in this paper, combined with those from the observing strategy white paper \citep{Lochner2018}:

\begin{itemize}

 \item {\bfseries Footprint and area:} The nominal area for the LSST WFD survey is 18,000 deg$^2$, but its area and definition can be changed. \cite{Lochner2018} and \cite{Olsen2018} both proposed to shift the WFD footprint 
 away from regions of high Galactic dust extinction to better benefit extragalactic science and rather have a dedicated Galactic plane survey that does not have the same requirements as extragalactic. \autoref{fig:comparison_area} shows that all metrics decrease with increased effective area, aside from the 3x2pt FoM, but that one simulation, \texttt{footprint\_big\_sky\_dust}, improves all metrics. This simulation proposes a large area footprint but uses a dust extinction cut to define the boundary. \ml{However, it should be noted that this footprint would degrade Galactic science and so, while it should be considered the ideal for cosmology, some compromise must be made to ensure the integrity of other science goals of LSST.} \ml{Area is not the only consideration of course;} there is a natural trade-off between depth and area, with depth (and correspondingly good cadence) being critical for mitigating systematic effects and for transient probes. \autoref{fig:mega_metric} shows a simple attempt to illustrate this trade-off. \ml{In \cite{Lochner2018}, we recommended a $18,000\ \rm{deg}^2$ footprint, that uses an E(B-V)\textless0.2 extinction cut, with declination limits of $-70<\rm{dec.}<12.5$, which is similar to \texttt{footprint\_big\_sky\_dust}. This footprint obtains a large area suitable for extragalactic science and, crucially, extends the footprint North to increase overlap with other surveys like DESI. Our work in this paper shows that such a footprint would ensure cosmology goals with LSST are met. Of course, dedicating $18,000\  \rm{deg}^2$ to the extragalactic area means visits must be moved from the WFD to the Galactic plane to support Galactic science goals, potentially reducing the cadence and impacting the transient probes and some systematics. This could be mitigated with a change in strategy such as implementing rolling cadence (see \autoref{sec:future}).}
  \item {\bfseries Survey uniformity:} Uniformity of depth across the WFD is a strict requirement for cosmology to avoid systematic effects being introduced by non-uniform measurements. We have a metric to measure uniformity (described in \autoref{sec:uniformity}) which finds that most strategies do not deviate significantly from baseline in terms of uniformity, indicating the SRD requirements are being met. However, it is worth pointing out that there is some flexibility at what points in the survey the observations need to be uniform.  \cite{Lochner2018} proposed to ensure observations be uniform by the data releases at the end of years 1, 3, 6, 10 to enable periodic cosmological analyses; however these are not yet finalized and can be changed. The important point is that any strategy should allow regular ``checkpoints'' where uniformity is achieved and analysis-ready data can be released. \ml{This should naturally be straightforward with most strategies but we note that rolling cadence (further discussed in \autoref{sec:future}) could affect when these uniform ``checkpoints'' are possible.} 
  
  \item {\bfseries{Dithering:}} The process of adding random rotational and translational offsets when repeating an observation in a field, known as dithering, is extremely important for cosmology systematics. Uniform coverage of 180 deg of camera rotation angle in every field is important for reduction of camera-based PSF systematics. All simulations used here implement translational and rotational dithers, which help maintain survey uniformity. While some investigation remains, we consider the current baseline dithering strategy to be largely effective. We note, however, that it is important that any translational dithers used in the DDFs should be as small as possible so that the depth of the fields is not compromised.
  
  \item {\bfseries Overlap with other surveys:} \ml{As discussed in \autoref{sec:overlap}, overlap with other surveys such as DESI, 4MOST:\tides\ and Euclid is critical for a number of cosmological probes. Spectroscopic data and infrared photometry can provide better calibration and training of photometric redshifts, improved \rachelNew{WL} constraints and more accurate de-blending of galaxies. Transient probes, in particular supernovae, will rely heavily on spectroscopy for host galaxy redshift identification and live spectroscopic follow-up to provide training sets for photometric classifiers. Improving overlap with other surveys further supports our recommendation for a larger footprint, particularly one which extends further North to improve overlap with DESI and Euclid.} 
  
 \item {\bfseries Exposure time:} The LSST SRD describes a visit to a field as a pair of 15\,s exposures ($2\times15$\,s). This is designed to robustly reject cosmic rays, but  \cite{Lochner2018} reported a potential gain of 7\% efficiency, as well as improved image quality, 
 by switching to a single exposure ($1\times30$\,s) and using 
 standard 
 cosmic-ray rejection techniques (e.g., \citealt{aihara2017}). \ml{Our results show that a single exposure can improve our metrics by as much as 20\% (see \autoref{fig:comparison_nexp}). The only advantage of short exposures is that they may mitigate saturation effects for very low redshift ($z< 0.05$) SNe. However, this advantage does not outweigh the severe degradation to the larger SNe sample.} While we strongly advocate for single 30\,s exposures, we acknowledge that the final decision about exposure times will only be taken after commissioning tests prove that cosmic rays and satellite trails can be rejected accurately.

 \item {\bfseries Repeated visits in a night:} A requirement in the SRD is that any given LSST field must be visited at least twice within a short period of time (15--40 minutes) to ensure accurate asteroid detection and assist with orbit characterization. There is a clear trade-off between the number of \emph{intra-night} visits and \emph{inter-night} visits, which can heavily impact cosmology with transient objects. Hence,  \cite{Lochner2018} proposed 
 that 
 repeat visit pairs be in different filters to improve transient characterisation. This has been adopted for all the simulations used in this paper and remains one of our strongest recommendations, as it significantly improves the transient metrics \ml{(\autoref{fig:comparison_samefilt})}.
 
 \item {\bfseries Cadence:} The return time with which each field in the WFD is observed is of critical importance to many science cases and is one of the most difficult factors to optimize. While the LSST SRD states that every field must be observed on average 825 times over the survey, it does not specify how those visits should be distributed and while each field is visited every 3 days \emph{on average}, there is a large spread in this distribution. \ml{\autoref{fig:comparison_cadence} shows the impact of the median internight gap on various metrics.} Significant gains can be made by changing how the filters are allocated and ensuring the $griz$ filters maintain regular cadence whatever the moon conditions. We find that the redder filters, $riz$, are particularly important for supernova characterisation and need to have high cadence (see \autoref{fig:internight_gap_filters}). $u$-band, while important for photometric redshifts, does not heavily impact our transient probes 
 and
 thus does not require high cadence. \ml{Cadence and filter allocation are expected to also be important for transient classification which, while not included as a metric, is discussed in \autoref{sec:future}.}

\end{itemize}

\subsection{Future work}
\label{sec:future}
While significant gains have been made in understanding the impact of observing strategy on cosmological constraints with LSST, there remain some unanswered questions and active areas of research:

\begin{itemize}
    \item{\bfseries Rolling cadence:}  
 This is a proposed strategy that can increase the cadence in a certain area of sky by prioritizing it for a period of time at the expense of other areas. This would then be repeated for the initially de-prioritized areas, allowing a pattern of rolling cadence. While rolling cadence is very promising, it has proved difficult to simulate. \ml{The LSST SRD proper motion requirements mean that the survey must have uniform coverage at its start and end, meaning a rolling cadence strategy must start and stop smoothly. It also has to roll for full observing seasons, constraining when a roll can start. Finally, rolling cadence strategies have many parameters to optimize including how large an area to roll at a time, for how long and whether to roll completely or allow the remaining part of the survey to still be observed but at a lower priority. All these complexities and constraints have made it difficult to find a rolling strategy that works well for transient science without impacting other science cases. Although it is difficult, we still consider it a promising option to obtain large sky area without reducing cadence, and are continuing to research this approach.} 
 \item{\bfseries Transient classification:} All our transient metrics have made the explicit assumption that transients detected with LSST can be perfectly classified, which is obviously not true. \ml{Not only does photometric transient classification impact the ability to constrain cosmology with SNe, kNe and strongly lensed SNe, it is also in turn affected by the quality of the photometric redshift estimation. Additionally, transient classification will likely be highly sensitive to observing strategy, making a transient classification metric critical. This metric is a challenge to develop as it would be affected by many aspects of observing strategy, would need to encompass a broad array of transient classes and could be somewhat dependent on the choice of classifier used. Because photometric transient classification for large surveys is still a young research field, not enough is known to build a reliable emulator to use as a fast metric. Thus, to evaluate a single observing strategy, sophisticated and expensive simulations must be run. An important step in this direction was taken by the \plasticc\ challenge \citep{plasticc2018}, which was a community challenge that released a set of LSST-like simulations incorporating a wide variety of transients and variables. Work is ongoing to understand the effect of observing strategy on existing transient classification pipelines \citep[e.g.,][]{Lochner2016, Moeller2019} using the \plasticc\ simulations themselves, and also by running new simulations based on some of the observing strategies discussed in this paper.}  
 \item{\bfseries A third visit:} There is a proposal to add an additional third visit some time after the visit pair, repeating one of the filters used in the original pair. \ml{The motivation is to improve early classification of transients, 
 which will affect all transient science and be of particular importance to spectroscopic follow-up. However, a third observation in the same band will reduce overall cadence and hence the number of well-measured supernovae, kilonovae and strongly lensed transients. As stated in the previous point, we do not yet have a reliable classification metric to evaluate this approach. Classification performance must be carefully weighed against overall cosmological constraints for a balance to be found. Considerations must also be made about the spectroscopic follow-up strategy to build training sets for classifiers. Thus, the question of whether or not to add a third visit is actually only a single component of an important and complex analysis that is ongoing.} 
 \item {\bfseries DDFs:} 
Cosmology with LSST relies heavily on the DDFs for both photometric redshift training and particularly to provide a deep sample of high redshift supernovae.
While this paper focused on the WFD, many of the metrics developed for the SN probe in \autoref{sec:nsn} can be applied to evaluate different DDF strategies. The optimal outcome for the DDFs is that they produce a well-measured SNIa sample at significantly higher redshifts than the WFD.  The cadence for a DDF survey will be on the scale of every $\sim2$ days rather than every $10-15$ days, which \mlNew{results in higher quality light curves than WFD}.  However, \mlNew{to improve the redshift range of supernovae detected in the DDFs,} we advocate here for two strategies:
1. Rolling deep fields such that in a given year, 1--2 fields can go significantly deeper than the other fields, and 2. A significantly skewed filter-allocation towards the redder bands.

\end{itemize}
The metrics developed in this paper have been invaluable in gaining insights about observing strategy and lay a useful foundation for ongoing work on this challenging optimization problem.

\section*{Acknowledgements}
\input{author_contributions}

\input{author_acknowledgements}
\input{desc_acknowledgements}
\bibliography{refs}

\appendix

\section{Metric descriptions and transformations}
\label{sec:metrics_appendix}

\begin{longtable}{lccc}
    \hline
    \input{metric_transforms.tbl}
    \caption{Metrics used in this paper. The first column describes the metric, the second indicates how the metric ($x$) is transformed such that it can be directly compared with other metrics and interpreted as ``larger is better''. The third column indicates the original metric value at baseline and the last column directs the reader to the section where the metric is described.}
    \label{tab:metric_transforms}
\end{longtable}

\section{List of observing strategy simulations}
\label{sec:simulations_appendix}
All simulations used in this paper, \mlNew{as well as those that have been explicitly excluded}. We have focused on the v1.5 FBS simulations but indicate where some have been excluded, usually due to bugs or a variation in the simulation that artificially changes a metric and is thus not directly comparable to the baseline. We highlight which figures each simulation is used in, noting particularly that Figure~\ref{fig:internight_gap_filters} is restricted to simulations with similar total number of visits. We do not indicate the list of simulations used in the figures in Sections~\ref{sec:static} and~\ref{sec:transients} since it is explicitly indicated there.
    \begin{longtable}{lll}
    \hline
    \input{simulations_table.tbl}
    \caption{List of all simulations used in this paper, \mlNew{which baseline simulation it is compared against} and which figures the simulation was used in (aside from early figures in the paper where the simulation is explicitly indicated).}
    \label{tab:simulations_full_list}
    \end{longtable}
\end{document}

%% file: authors.tex
\author{Michelle Lochner} 
\affiliation{Department of Physics and Astronomy, University of the Western Cape, Bellville, Cape Town, 7535, South Africa} 
\affiliation{South African Radio Astronomy Observatory (SARAO), The Park, Park Road, Pinelands, Cape Town 7405, South Africa} 

\author{Dan Scolnic} 
\affiliation{Department of Physics, Duke University, 120 Science Drive, Durham, NC, 27708, USA} 

\author{Husni Almoubayyed} 
\affiliation{McWilliams Center for Cosmology, Department of Physics, Carnegie Mellon University, Pittsburgh, PA 15213, USA} 

\author{Timo Anguita} 
\affiliation{Departamento de Ciencias Fisicas, Universidad Andres Bello, Fernandez Concha 700, Las Condes, Santiago, Chile} 
\affiliation{Millennium Institute of Astrophysics, Nuncio Monsenor Sotero Sanz 100, Of. 104, Providencia, Santiago, Chile } 

\author{Humna Awan} 
\affiliation{Leinweber Center for Theoretical Physics, Department of Physics, University of Michigan, Ann Arbor, MI 48109, USA} 
\affiliation{Department of Physics and Astronomy, Rutgers, the State University, Piscataway, NJ 08854, USA} 

\author{Eric Gawiser} 
\affiliation{Department of Physics and Astronomy, Rutgers, the State University, Piscataway, NJ 08854, USA} 

\author{Satya Gontcho A Gontcho} 
\affiliation{Department of Physics \& Astronomy, University of Rochester, 500 Wilson Blvd., Rochester, NY 14627} 

\author{Philippe Gris} 
\affiliation{Laboratoire de Physique de Clermont,IN2P3/CNRS, Universit\'e Clermont Auvergne, F-63000 Clermont-Ferrand, France} 

\author{Simon Huber} 
\affiliation{Max-Planck-Institut f{\"u}r Astrophysik, Karl-Schwarzschild-Str.~1, 85748 Garching, Germany} 
\affiliation{Physik-Department, Technische Universit\"at M\"unchen, James-Franck-Stra\ss{}e~1, 85748 Garching, Germany} 

\author{Saurabh W. Jha} 
\affiliation{Department of Physics and Astronomy, Rutgers, the State University, Piscataway, NJ 08854, USA} 

\author{R. Lynne  Jones} 
\affiliation{Department of Astronomy, University of Washington, Seattle, WA  98195, USA} 
\affiliation{DiRAC Institute, University of Washington, Seattle, WA  98195, USA} 

\author{Alex G. Kim} 
\affiliation{Lawrence Berkeley National Laboratory, 1 Cyclotron Rd., Berkeley, CA, 94720, USA} 

\author{Rachel Mandelbaum} 
\affiliation{McWilliams Center for Cosmology, Department of Physics, Carnegie Mellon University, Pittsburgh, PA 15213, USA} 

\author{Phil Marshall} 
\affiliation{Kavli Institute for Particle Astrophysics and Cosmology,
P.O. Box 20450, MS29,
Stanford,
CA 94309,
U.S.A.} 

\author{Tanja Petrushevska} 
\affiliation{Centre for Astrophysics and Cosmology, University of Nova Gorica, Vipavska 11c, 5270 Ajdov\v{s}\u{c}ina, Slovenia} 

\author{Nicolas Regnault} 
\affiliation{Laboratoire de Physique Nucleaire et de Hautes-Energies, Barre 12-22 1er etage, 4 place Jussieu, 75005 Paris, France} 

\author{Christian N. Setzer} 
\affiliation{The Oskar Klein Centre for Cosmoparticle Physics, Department of Physics, Stockholm University, AlbaNova, Stockholm, SE-106 91, Sweden} 

\author{Sherry H. Suyu} 
\affiliation{Max-Planck-Institut f{\"u}r Astrophysik, Karl-Schwarzschild-Str.~1, 85748 Garching, Germany} 
\affiliation{Physik-Department, Technische Universit\"at M\"unchen, James-Franck-Stra\ss{}e~1, 85748 Garching, Germany} 
\affiliation{Academia Sinica Institute of Astronomy and Astrophysics (ASIAA), 11F of ASMAB, No.1, Section 4, Roosevelt Road, Taipei 10617, Taiwan} 

\author{Peter Yoachim} 
\affiliation{Department of Astronomy, University of Washington, Seattle, WA  98195, USA} 

\author{Rahul  Biswas} 
\affiliation{The Oskar Klein Centre for Cosmoparticle Physics, Department of Physics, Stockholm University, AlbaNova, Stockholm, SE-106 91, Sweden} 

\author{Tristan Blaineau} 
\affiliation{Universit\'e Paris-Saclay, CNRS/IN2P3, IJCLab, 91405 Orsay, France} 

\author{Isobel Hook} 
\affiliation{Department of Physics, Lancaster University, Lancaster, Lancs LA1 4YB, U.K.} 

\author{Marc Moniez} 
\affiliation{Universit\'e Paris-Saclay, CNRS/IN2P3, IJCLab, 91405 Orsay, France} 

\author{Eric Neilsen} 
\affiliation{Fermi National Accelerator Laboratory, P.O. Box 500, Batavia, IL 60510, USA} 

\author{Hiranya Peiris} 
\affiliation{The Oskar Klein Centre for Cosmoparticle Physics, Department of Physics, Stockholm University, AlbaNova, Stockholm, SE-106 91, Sweden} 
\affiliation{Department of Physics \& Astronomy, University College London, Gower Street, London WC1E 6BT, UK} 

\author{Daniel Rothchild} 
\affiliation{Department of Electrical Engineering and Computer Sciences, UC Berkeley, 253 Cory Hall, Berkeley, CA 94720-1770} 

\author{Christopher Stubbs} 
\affiliation{Department of Physics, Harvard University, 17 Oxford St, Cambridge MA 02138} 
\affiliation{Department of Astronomy, Harvard University, 60 Garden St, Cambridge MA 02138} 

\author{the LSST Dark Energy Science Collaboration}

%% file: sims_short.tbl
Simulation name & Short name\\ 
\hline 
\texttt{baseline\_v1.5\_10yrs.db} & Baseline \\ 
\texttt{baseline\_2snapsv1.5\_10yrs.db} & 2x15s exposures \\ 
\texttt{bulges\_bs\_v1.5\_10yrs.db} & More visits in Galactic bulge \\ 
\texttt{footprint\_big\_sky\_dustv1.5\_10yrs.db} & Larger extragalactic footprint \\ 
\texttt{footprint\_bluer\_footprintv1.5\_10yrs.db} & Bluer filter distribution \\ 
\texttt{footprint\_newAv1.5\_10yrs.db} & Galactic plane focus \\ 
\texttt{var\_expt\_v1.5\_10yrs.db} & Variable exposure \\ 
\texttt{wfd\_depth\_scale0.65\_noddf\_v1.5\_10yrs.db} & 65\% of visits in WFD \\ 
\texttt{wfd\_depth\_scale0.99\_noddf\_v1.5\_10yrs.db} & 99\% of visits in WFD \\ 
\texttt{ss\_heavy\_v1.6\_10yrs.db} & Solar system focus \\ 
\texttt{combo\_dust\_v1.6\_10yrs.db} & Combo dust \\ 
\hline

%% file: overlap.tbl
Simulation short name & 4MOST overlap (deg$^2$) & DESI overlap (deg$^2$) & Euclid overlap (deg$^2$)\\ 
\hline 
Baseline & 14421.22 & 4178.63 & 8201.79\\ 
Larger extragalactic footprint & 14925.47 & 5951.48 & 9332.37\\ 
\hline 

%% file: author_contributions.tex
\noindent Author contributions are listed below.\\ 
\noindent \textbf{HAl:} Developed WL systematics metric, contributed MAF metric for 3x2pt Figure of Merit. Analysis and interpretation for weak lensing metric results. Writing and discussion for sections 4 and 6. \\
\textbf{TA:} Helped develop time delay lensed quasar metric, run metrics, analysis and discussion of Section 5.2.\\
\textbf{HAw:} Provided technical assistance with simulations. Developed MAF metrics for the extragalactic footprint and the number of galaxies. Provided numbers for basic static metrics (e.g depth statistics) and those for LSS. Helped with discussions, and writing sections 3 and 4.\\
\textbf{RB:} Contributed to software and metric development\\
\textbf{TB:} Contributed to long time-scale microlensing efficiency understanding\\
\textbf{EG:} Conceptualization, discussion, and writing in sections 3 and 4.  Edited sections 1-4 and 7.  \\
\textbf{SGG:} Section 5.1.3, metric for SNe PV\\
\textbf{PG:} Section 5.1: discussion, oversight and writing. Developed a set of supernovae metrics implemented in MAF. Processed all observing strategies. Developed real-time display (\url{https://me.lsst.eu/gris/OS_Videos/}) to visualize the progress of observing strategies on a daily basis to study cadence/gap effects. Implementation of the lensed SNe Ia by galaxies metric in MAF and processing of observing strategies (section 5.2.1).\\
\textbf{IH:} Section 5.1:discussion. Writing (review and editing) of Sections 1,2 and 5.\\
\textbf{SH:}  Section 5.2.1:metric, Section 5.2:validation, analysis, writing, editing, \\
\textbf{SWJ:} Conceptualization and development of observing strategies related to SN Ia (section 5.1) and rolling cadences. Editing of sections 1 and 2.\\
\textbf{RLJ:} Created simulations, technical assistance with simulations and metric interpretation. Worked with various groups to develop metrics and analysis.\\
\textbf{AGK:} Section 5.1.3, writing/metric-creation\\
\textbf{ML:} Conceptualization,  software, validation, formal analysis, investigation, data curation, writing - original draft, writing - review and editing, visualization, supervision\\
\textbf{RM:} Discussion and writing in sections 3 and 4\\
\textbf{PM:} Section 5.2 (designed and implemented metrics, wrote text)\\
\textbf{MM:} Microlensing discussion and writing in introduction\\
\textbf{EHN:} Improved seeing input to opsim simulations (section 2.5, 6.7), general descussions and recommendations\\
\textbf{HP:} Section 5.3: conceptualization, validation, writing - review, supervision, funding acquisition\\
\textbf{TP:} Metric development and writing of subsection 5.2.2\\
\textbf{NR:} Section 5.1: discussion and writing. Developed a set of metrics to evaluate the expected quality on the SN light curves, on the DDF and WFD surveys. Built a pipeline to evaluate the pre-white paper, white-paper and fbs1.3 - fbs1.7 cadences. \\
\textbf{DR:} Developed deterministic scheduler (alt sched) as a conceptual alternative to OpSim for comparison purposes\\
\textbf{DS:} Conceptualization,  writing - original draft, writing - review \& editing, visualization, supervision\\
\textbf{CNS:} Software, metrics, validation, analysis and writing of Section 5.3.\\
\textbf{CS:} Developed deterministic scheduler (alt sched) as a conceptual alternative to OpSim for comparison purposes\\
\textbf{SHS:} Coordinated strong lensing group; writing, validation, analysis and editing of Section 5.2.\\
\textbf{PY:} Created survey simulations. Section 2 writing. Helped with development of MAF metrics for Section 3, 5.\\

%% file: author_acknowledgements.tex
TA acknowledges support from Proyecto Fondecyt N 1190335 and the ANID-Millennium Science Initiative Program – ICN12\_009 awarded to the Millennium Institute of Astrophysics MAS. HAw also acknowledges support by the Rutgers Discovery Informatics Institute (RDI2) Fellowship of Excellence in Computational and Data Science, Rutgers University \& Bevier Dissertation Completion Fellowship, Leinweber Postdoctoral Research Fellowship, and DOE grant DE-SC009193. HAw also thanks the LSSTC Data Science Fellowship Program, which is funded by LSSTC, NSF Cybertraining Grant \#1829740, the Brinson Foundation, and the Moore Foundation, as participation in the program has benefited this work.  This material is based upon work supported by the U.S. Department of Energy, Office of Science, Office of High Energy Physics Cosmic Frontier Research pro- gram under Award Numbers DE-SC0011636 and DE- SC0010008, which supported EG and HAw.   SGG acknowledges support from the University of Rochester and the DOE Office of High Energy Physics under grant number DE-SC0008475.

SH and SHS thank the European Research Council (ERC) for support under the European Union’s Horizon 2020 research and innovation programme (LENSNOVA: grant agreement No. 771776)

         This material is based upon work supported by the U.S. Department of Energy, Office of Science, Office of High Energy Physics Cosmic Frontier Research program under Award Numbers DE-SC0011636, which supported SWJ. AGKs work is supported by the U.S. Department of Energy, Office of Science, Office of High Energy Physics, under contract No. DE-AC02-05CH11231. ML acknowledges support from the South African Radio Astronomy Observatory and the National Research Foundation (NRF) towards this research. Opinions expressed and conclusions arrived at, are those of the authors and are not necessarily to be attributed to the NRF. RM and H. Almoubayyed are supported by the Department of Energy grant DE-SC0010118. This document was prepared by DESC using the resources of the Fermi National Accelerator Laboratory (Fermilab), a U.S. Department of Energy, Office of Science, HEP User Facility. Fermilab is managed by Fermi Research Alliance, LLC (FRA), acting under Contract No. DE-AC02-07CH11359 CS, RB and HVP's work was partially supported the research environment grant “Gravitational Radiation and Elec- tromagnetic Astrophysical Transients (GREAT)” funded by the Swedish Research Council (VR) under Dnr 2016- 06012 and the research project grant ‘Understanding the Dynamic Universe’ funded by the Knut and Alice Wallenberg Foundation under Dnr KAW 2018.0067. TP acknowledges the financial support from the Slovenian Research Agency (grants I0-0033, P1-0031, J1-8136 and Z1-1853). SHS thanks the Max Planck Society for support through the Max Planck Research Group.


%% file: desc_acknowledgements.tex
The DESC acknowledges ongoing support from the Institut National de 
Physique Nucl\'eaire et de Physique des Particules in France; the 
Science \& Technology Facilities Council in the United Kingdom; and the
Department of Energy, the National Science Foundation, and the LSST 
Corporation in the United States.  DESC uses resources of the IN2P3 
Computing Center (CC-IN2P3--Lyon/Villeurbanne - France) funded by the 
Centre National de la Recherche Scientifique; the National Energy 
Research Scientific Computing Center, a DOE Office of Science User 
Facility supported by the Office of Science of the U.S.\ Department of
Energy under Contract No.\ DE-AC02-05CH11231; STFC DiRAC HPC Facilities, 
funded by UK BIS National E-infrastructure capital grants; and the UK 
particle physics grid, supported by the GridPP Collaboration.  This 
work was performed in part under DOE Contract DE-AC02-76SF00515. This paper has undergone internal review in the LSST Dark Energy Science Collaboration. The authors thank the internal reviewers, who were Ariel Goobar, Joshua Meyers and Samuel Schmidt. The authors also thank Jean-Charles Cuillande and Christopher Frohmaier for providing the footprint maps for Euclid and 4MOST respectively.

%% file: metric_transforms.tbl
Metric description & Transform & Value at baseline & Section in paper\\ 
\hline 
Y1 median $i$-band coadded depth & $1/\big(10^{-\frac{x}{2.5}}\big)^2$ & 25 & \ref{sec:static_metrics}\\ 
Y3 median $i$-band coadded depth & $1/\big(10^{-\frac{x}{2.5}}\big)^2$ & 26 & \ref{sec:static_metrics}\\ 
Y6 median $i$-band coadded depth & $1/\big(10^{-\frac{x}{2.5}}\big)^2$ & 26 & \ref{sec:static_metrics}\\ 
Y10 median $i$-band coadded depth & $1/\big(10^{-\frac{x}{2.5}}\big)^2$ & 26 & \ref{sec:static_metrics}\\ 
Y1 $i$-band coadded depth stddev & $1/\big(10^{-\frac{x}{2.5}}\big)^2$ & 0.16 & \ref{sec:static_metrics}\\ 
Y3 $i$-band coadded depth stddev & $1/\big(10^{-\frac{x}{2.5}}\big)^2$ & 0.13 & \ref{sec:static_metrics}\\ 
Y6 $i$-band coadded depth stddev & $1/\big(10^{-\frac{x}{2.5}}\big)^2$ & 0.12 & \ref{sec:static_metrics}\\ 
Y10 $i$-band coadded depth stddev & $1/\big(10^{-\frac{x}{2.5}}\big)^2$ & 0.12 & \ref{sec:static_metrics}\\ 
Y1 area for static science (deg$^2$) & $x$ & 1.5e+04 & \ref{sec:static_metrics}\\ 
Y3 area for static science (deg$^2$) & $x$ & 1.5e+04 & \ref{sec:static_metrics}\\ 
Y6 area for static science (deg$^2$) & $x$ & 1.5e+04 & \ref{sec:static_metrics}\\ 
Y10 area for static science (deg$^2$) & $x$ & 1.5e+04 & \ref{sec:static_metrics}\\ 
PZ outlier fraction (low z) & $1 - 2x$ & 0.0019 & \ref{sec:pz}\\ 
PZ stddev (low z) & $1/x^2$ & 0.0096 & \ref{sec:pz}\\ 
PZ outlier fraction (high z) & $1 - 2x$ & 0.044 & \ref{sec:pz}\\ 
PZ stddev (high z) & $1/x^2$ & 0.022 & \ref{sec:pz}\\ 
WL + LSS DETF Figure of Merit & $x$ & 36 & \ref{sec:static_fom}\\ 
WL Average visits & $x$ & 6.1e+02 & \ref{sec:wl}\\ 
LSS systematics FOM for Y1 & $x$ & 0.94 & \ref{sec:systematics}\\ 
LSS systematics FOM for Y10 & $x$ & 0.98 & \ref{sec:systematics}\\ 
Y1 $N_{gal}$ at 0.66\textless z\textless 1.0 & $x$ & 4.5e+08 & \ref{sec:lss_Ngal}\\ 
Y10 $N_{gal}$ at 0.66\textless z\textless 1.0 & $x$ & 9.5e+08 & \ref{sec:lss_Ngal}\\ 
Number of SNe Ie with z$\textless$zlim(faint) & $x$ & 4.5e+04 & \ref{sec:nsn}\\ 
Faint SNe Ia redshift limit & $x$ & 0.17 & \ref{sec:nsn}\\ 
Number of  SNe Ia with z$\textless$zlim(med) & $x$ & 1.6e+05 & \ref{sec:nsn}\\ 
SNe Ia r-band Signal-to-Noise Ratio & $x$ & 0.59 & \ref{sec:rate_zlim}\\ 
SNe Ia r-band redshift limit & $x$ & 0.31 & \ref{sec:rate_zlim}\\ 
Medium SNe Ia redshift limit & $x$ & 0.27 & \ref{sec:nsn}\\ 
SN Ia Peculiar Velocities & $x$ & 1.4 & \ref{sec:peculiar}\\ 
Number of SNe Ia lensed by galaxies & $x$ & 8.7 & \ref{sec:sl:slsn-gal}\\ 
Number of lensed quasars & $x$ & 3.3e+02 & \ref{sec:sl_quasars}\\ 
Number of SNe Ia lensed by clusters & $x$ & 0.18 & \ref{sec:sl_clusters}\\ 
GW170817-like kNe Counts 10yrs & $x$ & 97 & \ref{sec:kn_metrics}\\ 
kNe MAF Mean Counts & $x$ & 34 & \ref{sec:kn_metrics}\\ 
kN Population Counts 10yrs & $x$ & 40 & \ref{sec:kn_metrics}\\ 
\hline

%% file: simulations_table.tbl
Simulation name & Comparison baseline & Included in which figures\\ 
\hline
\texttt{weather\_0.3\_v1.4\_10yrs} & \texttt{weather\_0.3\_v1.4\_10yrs} & \ref{fig:comparison_weather} \\ 
\texttt{weather\_0.7\_v1.4\_10yrs} & \texttt{weather\_0.3\_v1.4\_10yrs} & \ref{fig:comparison_weather} \\ 
\texttt{weather\_1.2\_ndt\_v1.4\_10yrs} & \texttt{weather\_0.3\_v1.4\_10yrs} & \ref{fig:comparison_weather} \\ 
\texttt{weather\_1.2\_v1.4\_10yrs} & \texttt{weather\_0.3\_v1.4\_10yrs} & \ref{fig:comparison_weather} \\ 
\texttt{agnddf\_v1.5\_10yrs} & \texttt{baseline\_v1.5\_10yrs} & \ref{fig:comparison_area}, \ref{fig:comparison_cadence}, \ref{fig:comparison_depth}, \ref{fig:comparison_season}, \ref{fig:internight_gap_filters}, \ref{fig:mega_metric} \\ 
\texttt{alt\_dust\_v1.5\_10yrs} & \texttt{baseline\_v1.5\_10yrs} & Excluded \\ 
\texttt{alt\_roll\_mod2\_dust\_sdf\_0.20\_v1.5\_10yrs} & \texttt{baseline\_v1.5\_10yrs} & Excluded \\ 
\texttt{baseline\_2snaps\_v1.5\_10yrs} & \texttt{baseline\_v1.5\_10yrs} & \ref{fig:comparison_area}, \ref{fig:comparison_cadence}, \ref{fig:comparison_depth}, \ref{fig:comparison_season}, \ref{fig:mega_metric} \\ 
\texttt{baseline\_samefilt\_v1.5\_10yrs} & \texttt{baseline\_v1.5\_10yrs} & \ref{fig:comparison_samefilt} \\ 
\texttt{baseline\_v1.5\_10yrs} & \texttt{baseline\_v1.5\_10yrs} & \ref{fig:comparison_area}, \ref{fig:comparison_cadence}, \ref{fig:comparison_depth}, \ref{fig:comparison_nexp}, \ref{fig:comparison_samefilt}, \ref{fig:comparison_season}, \ref{fig:comparison_weather}, \ref{fig:internight_gap_filters}, \ref{fig:mega_metric} \\ 
\texttt{bulges\_bs\_v1.5\_10yrs} & \texttt{baseline\_v1.5\_10yrs} & \ref{fig:comparison_area}, \ref{fig:comparison_cadence}, \ref{fig:comparison_depth}, \ref{fig:comparison_season}, \ref{fig:mega_metric} \\ 
\texttt{bulges\_bulge\_wfd\_v1.5\_10yrs} & \texttt{baseline\_v1.5\_10yrs} & \ref{fig:comparison_area}, \ref{fig:comparison_cadence}, \ref{fig:comparison_depth}, \ref{fig:comparison_season}, \ref{fig:mega_metric} \\ 
\texttt{bulges\_cadence\_bs\_v1.5\_10yrs} & \texttt{baseline\_v1.5\_10yrs} & \ref{fig:comparison_area}, \ref{fig:comparison_cadence}, \ref{fig:comparison_depth}, \ref{fig:comparison_season}, \ref{fig:mega_metric} \\ 
\texttt{bulges\_cadence\_bulge\_wfd\_v1.5\_10yrs} & \texttt{baseline\_v1.5\_10yrs} & \ref{fig:comparison_area}, \ref{fig:comparison_cadence}, \ref{fig:comparison_depth}, \ref{fig:comparison_season}, \ref{fig:mega_metric} \\ 
\texttt{bulges\_cadence\_i\_heavy\_v1.5\_10yrs} & \texttt{baseline\_v1.5\_10yrs} & \ref{fig:comparison_area}, \ref{fig:comparison_cadence}, \ref{fig:comparison_depth}, \ref{fig:comparison_season}, \ref{fig:mega_metric} \\ 
\texttt{bulges\_i\_heavy\_v1.5\_10yrs} & \texttt{baseline\_v1.5\_10yrs} & \ref{fig:comparison_area}, \ref{fig:comparison_cadence}, \ref{fig:comparison_depth}, \ref{fig:comparison_season}, \ref{fig:mega_metric} \\ 
\texttt{dcr\_nham1\_ug\_v1.5\_10yrs} & \texttt{baseline\_v1.5\_10yrs} & \ref{fig:comparison_area}, \ref{fig:comparison_cadence}, \ref{fig:comparison_depth}, \ref{fig:comparison_season}, \ref{fig:internight_gap_filters}, \ref{fig:mega_metric} \\ 
\texttt{dcr\_nham1\_ugr\_v1.5\_10yrs} & \texttt{baseline\_v1.5\_10yrs} & \ref{fig:comparison_area}, \ref{fig:comparison_cadence}, \ref{fig:comparison_depth}, \ref{fig:comparison_season}, \ref{fig:internight_gap_filters}, \ref{fig:mega_metric} \\ 
\texttt{dcr\_nham1\_ugri\_v1.5\_10yrs} & \texttt{baseline\_v1.5\_10yrs} & \ref{fig:comparison_area}, \ref{fig:comparison_cadence}, \ref{fig:comparison_depth}, \ref{fig:comparison_season}, \ref{fig:internight_gap_filters}, \ref{fig:mega_metric} \\ 
\texttt{dcr\_nham2\_ug\_v1.5\_10yrs} & \texttt{baseline\_v1.5\_10yrs} & \ref{fig:comparison_area}, \ref{fig:comparison_cadence}, \ref{fig:comparison_depth}, \ref{fig:comparison_season}, \ref{fig:internight_gap_filters}, \ref{fig:mega_metric} \\ 
\texttt{dcr\_nham2\_ugr\_v1.5\_10yrs} & \texttt{baseline\_v1.5\_10yrs} & \ref{fig:comparison_area}, \ref{fig:comparison_cadence}, \ref{fig:comparison_depth}, \ref{fig:comparison_season}, \ref{fig:internight_gap_filters}, \ref{fig:mega_metric} \\ 
\texttt{dcr\_nham2\_ugri\_v1.5\_10yrs} & \texttt{baseline\_v1.5\_10yrs} & Excluded \\ 
\texttt{descddf\_v1.5\_10yrs} & \texttt{baseline\_v1.5\_10yrs} & \ref{fig:comparison_area}, \ref{fig:comparison_cadence}, \ref{fig:comparison_depth}, \ref{fig:comparison_season}, \ref{fig:internight_gap_filters}, \ref{fig:mega_metric} \\ 
\texttt{filterdist\_indx1\_v1.5\_10yrs} & \texttt{baseline\_v1.5\_10yrs} & \ref{fig:comparison_filter_dist} \\ 
\texttt{filterdist\_indx2\_v1.5\_10yrs} & \texttt{baseline\_v1.5\_10yrs} & \ref{fig:comparison_filter_dist} \\ 
\texttt{filterdist\_indx3\_v1.5\_10yrs} & \texttt{baseline\_v1.5\_10yrs} & \ref{fig:comparison_filter_dist} \\ 
\texttt{filterdist\_indx4\_v1.5\_10yrs} & \texttt{baseline\_v1.5\_10yrs} & \ref{fig:comparison_filter_dist} \\ 
\texttt{filterdist\_indx5\_v1.5\_10yrs} & \texttt{baseline\_v1.5\_10yrs} & \ref{fig:comparison_filter_dist} \\ 
\texttt{filterdist\_indx6\_v1.5\_10yrs} & \texttt{baseline\_v1.5\_10yrs} & \ref{fig:comparison_filter_dist} \\ 
\texttt{filterdist\_indx7\_v1.5\_10yrs} & \texttt{baseline\_v1.5\_10yrs} & \ref{fig:comparison_filter_dist} \\ 
\texttt{filterdist\_indx8\_v1.5\_10yrs} & \texttt{baseline\_v1.5\_10yrs} & \ref{fig:comparison_filter_dist} \\ 
\texttt{footprint\_add\_mag\_cloudsv1.5\_10yrs} & \texttt{baseline\_v1.5\_10yrs} & \ref{fig:comparison_area}, \ref{fig:comparison_cadence}, \ref{fig:comparison_depth}, \ref{fig:comparison_season}, \ref{fig:internight_gap_filters}, \ref{fig:mega_metric} \\ 
\texttt{footprint\_big\_sky\_dustv1.5\_10yrs} & \texttt{baseline\_v1.5\_10yrs} & \ref{fig:comparison_area}, \ref{fig:comparison_cadence}, \ref{fig:comparison_depth}, \ref{fig:comparison_season}, \ref{fig:internight_gap_filters}, \ref{fig:mega_metric} \\ 
\texttt{footprint\_big\_sky\_nouiyv1.5\_10yrs} & \texttt{baseline\_v1.5\_10yrs} & Excluded \\ 
\texttt{footprint\_big\_skyv1.5\_10yrs} & \texttt{baseline\_v1.5\_10yrs} & Excluded \\ 
\texttt{footprint\_big\_wfdv1.5\_10yrs} & \texttt{baseline\_v1.5\_10yrs} & \ref{fig:comparison_area}, \ref{fig:comparison_depth}, \ref{fig:comparison_season}, \ref{fig:mega_metric} \\ 
\texttt{footprint\_bluer\_footprintv1.5\_10yrs} & \texttt{baseline\_v1.5\_10yrs} & \ref{fig:comparison_area}, \ref{fig:comparison_cadence}, \ref{fig:comparison_depth}, \ref{fig:comparison_season}, \ref{fig:mega_metric} \\ 
\texttt{footprint\_gp\_smoothv1.5\_10yrs} & \texttt{baseline\_v1.5\_10yrs} & \ref{fig:comparison_area}, \ref{fig:comparison_cadence}, \ref{fig:comparison_depth}, \ref{fig:comparison_season}, \ref{fig:internight_gap_filters}, \ref{fig:mega_metric} \\ 
\texttt{footprint\_newAv1.5\_10yrs} & \texttt{baseline\_v1.5\_10yrs} & \ref{fig:comparison_area}, \ref{fig:comparison_cadence}, \ref{fig:comparison_depth}, \ref{fig:comparison_season}, \ref{fig:mega_metric} \\ 
\texttt{footprint\_newBv1.5\_10yrs} & \texttt{baseline\_v1.5\_10yrs} & \ref{fig:comparison_area}, \ref{fig:comparison_cadence}, \ref{fig:comparison_depth}, \ref{fig:comparison_season}, \ref{fig:mega_metric} \\ 
\texttt{footprint\_no\_gp\_northv1.5\_10yrs} & \texttt{baseline\_v1.5\_10yrs} & \ref{fig:comparison_area}, \ref{fig:comparison_cadence}, \ref{fig:comparison_depth}, \ref{fig:comparison_season}, \ref{fig:internight_gap_filters}, \ref{fig:mega_metric} \\ 
\texttt{footprint\_standard\_goalsv1.5\_10yrs} & \texttt{baseline\_v1.5\_10yrs} & \ref{fig:comparison_area}, \ref{fig:comparison_cadence}, \ref{fig:comparison_depth}, \ref{fig:comparison_season}, \ref{fig:internight_gap_filters}, \ref{fig:mega_metric} \\ 
\texttt{footprint\_stuck\_rollingv1.5\_10yrs} & \texttt{baseline\_v1.5\_10yrs} & Excluded \\ 
\texttt{goodseeing\_gi\_v1.5\_10yrs} & \texttt{baseline\_v1.5\_10yrs} & \ref{fig:comparison_area}, \ref{fig:comparison_cadence}, \ref{fig:comparison_depth}, \ref{fig:comparison_season}, \ref{fig:internight_gap_filters}, \ref{fig:mega_metric} \\ 
\texttt{goodseeing\_gri\_v1.5\_10yrs} & \texttt{baseline\_v1.5\_10yrs} & \ref{fig:comparison_area}, \ref{fig:comparison_cadence}, \ref{fig:comparison_depth}, \ref{fig:comparison_season}, \ref{fig:internight_gap_filters}, \ref{fig:mega_metric} \\ 
\texttt{goodseeing\_griz\_v1.5\_10yrs} & \texttt{baseline\_v1.5\_10yrs} & \ref{fig:comparison_area}, \ref{fig:comparison_cadence}, \ref{fig:comparison_depth}, \ref{fig:comparison_season}, \ref{fig:internight_gap_filters}, \ref{fig:mega_metric} \\ 
\texttt{goodseeing\_gz\_v1.5\_10yrs} & \texttt{baseline\_v1.5\_10yrs} & \ref{fig:comparison_area}, \ref{fig:comparison_cadence}, \ref{fig:comparison_depth}, \ref{fig:comparison_season}, \ref{fig:internight_gap_filters}, \ref{fig:mega_metric} \\ 
\texttt{goodseeing\_i\_v1.5\_10yrs} & \texttt{baseline\_v1.5\_10yrs} & \ref{fig:comparison_area}, \ref{fig:comparison_cadence}, \ref{fig:comparison_depth}, \ref{fig:comparison_season}, \ref{fig:internight_gap_filters}, \ref{fig:mega_metric} \\ 
\texttt{greedy\_footprint\_v1.5\_10yrs} & \texttt{baseline\_v1.5\_10yrs} & \ref{fig:comparison_area}, \ref{fig:comparison_cadence}, \ref{fig:comparison_depth}, \ref{fig:comparison_season}, \ref{fig:internight_gap_filters}, \ref{fig:mega_metric} \\ 
\texttt{roll\_mod2\_dust\_sdf\_0.20\_v1.5\_10yrs} & \texttt{baseline\_v1.5\_10yrs} & Excluded \\ 
\texttt{rolling\_mod2\_sdf\_0.10\_v1.5\_10yrs} & \texttt{baseline\_v1.5\_10yrs} & Excluded \\ 
\texttt{rolling\_mod2\_sdf\_0.20\_v1.5\_10yrs} & \texttt{baseline\_v1.5\_10yrs} & Excluded \\ 
\texttt{rolling\_mod3\_sdf\_0.10\_v1.5\_10yrs} & \texttt{baseline\_v1.5\_10yrs} & Excluded \\ 
\texttt{rolling\_mod3\_sdf\_0.20\_v1.5\_10yrs} & \texttt{baseline\_v1.5\_10yrs} & Excluded \\ 
\texttt{rolling\_mod6\_sdf\_0.10\_v1.5\_10yrs} & \texttt{baseline\_v1.5\_10yrs} & Excluded \\ 
\texttt{rolling\_mod6\_sdf\_0.20\_v1.5\_10yrs} & \texttt{baseline\_v1.5\_10yrs} & Excluded \\ 
\texttt{short\_exp\_2ns\_1expt\_v1.5\_10yrs} & \texttt{baseline\_v1.5\_10yrs} & \ref{fig:comparison_area}, \ref{fig:comparison_cadence}, \ref{fig:comparison_depth}, \ref{fig:comparison_season}, \ref{fig:mega_metric} \\ 
\texttt{short\_exp\_2ns\_5expt\_v1.5\_10yrs} & \texttt{baseline\_v1.5\_10yrs} & \ref{fig:comparison_area}, \ref{fig:comparison_cadence}, \ref{fig:comparison_depth}, \ref{fig:comparison_season}, \ref{fig:mega_metric} \\ 
\texttt{short\_exp\_5ns\_1expt\_v1.5\_10yrs} & \texttt{baseline\_v1.5\_10yrs} & \ref{fig:comparison_area}, \ref{fig:comparison_cadence}, \ref{fig:comparison_depth}, \ref{fig:comparison_season}, \ref{fig:mega_metric} \\ 
\texttt{short\_exp\_5ns\_5expt\_v1.5\_10yrs} & \texttt{baseline\_v1.5\_10yrs} & \ref{fig:comparison_area}, \ref{fig:comparison_cadence}, \ref{fig:comparison_depth}, \ref{fig:comparison_season}, \ref{fig:mega_metric} \\ 
\texttt{spiders\_v1.5\_10yrs} & \texttt{baseline\_v1.5\_10yrs} & \ref{fig:comparison_area}, \ref{fig:comparison_cadence}, \ref{fig:comparison_depth}, \ref{fig:comparison_season}, \ref{fig:internight_gap_filters}, \ref{fig:mega_metric} \\ 
\texttt{third\_obs\_pt120v1.5\_10yrs} & \texttt{baseline\_v1.5\_10yrs} & \ref{fig:comparison_area}, \ref{fig:comparison_cadence}, \ref{fig:comparison_depth}, \ref{fig:comparison_season}, \ref{fig:internight_gap_filters}, \ref{fig:mega_metric} \\ 
\texttt{third\_obs\_pt15v1.5\_10yrs} & \texttt{baseline\_v1.5\_10yrs} & \ref{fig:comparison_area}, \ref{fig:comparison_cadence}, \ref{fig:comparison_depth}, \ref{fig:comparison_season}, \ref{fig:internight_gap_filters}, \ref{fig:mega_metric} \\ 
\texttt{third\_obs\_pt30v1.5\_10yrs} & \texttt{baseline\_v1.5\_10yrs} & \ref{fig:comparison_area}, \ref{fig:comparison_cadence}, \ref{fig:comparison_depth}, \ref{fig:comparison_season}, \ref{fig:internight_gap_filters}, \ref{fig:mega_metric} \\ 
\texttt{third\_obs\_pt45v1.5\_10yrs} & \texttt{baseline\_v1.5\_10yrs} & \ref{fig:comparison_area}, \ref{fig:comparison_cadence}, \ref{fig:comparison_depth}, \ref{fig:comparison_season}, \ref{fig:internight_gap_filters}, \ref{fig:mega_metric} \\ 
\texttt{third\_obs\_pt60v1.5\_10yrs} & \texttt{baseline\_v1.5\_10yrs} & \ref{fig:comparison_area}, \ref{fig:comparison_cadence}, \ref{fig:comparison_depth}, \ref{fig:comparison_season}, \ref{fig:internight_gap_filters}, \ref{fig:mega_metric} \\ 
\texttt{third\_obs\_pt90v1.5\_10yrs} & \texttt{baseline\_v1.5\_10yrs} & \ref{fig:comparison_area}, \ref{fig:comparison_cadence}, \ref{fig:comparison_depth}, \ref{fig:comparison_season}, \ref{fig:internight_gap_filters}, \ref{fig:mega_metric} \\ 
\texttt{twilight\_neo\_mod1\_v1.5\_10yrs} & \texttt{baseline\_v1.5\_10yrs} & Excluded \\ 
\texttt{twilight\_neo\_mod2\_v1.5\_10yrs} & \texttt{baseline\_v1.5\_10yrs} & Excluded \\ 
\texttt{twilight\_neo\_mod3\_v1.5\_10yrs} & \texttt{baseline\_v1.5\_10yrs} & Excluded \\ 
\texttt{twilight\_neo\_mod4\_v1.5\_10yrs} & \texttt{baseline\_v1.5\_10yrs} & Excluded \\ 
\texttt{u60\_v1.5\_10yrs} & \texttt{baseline\_v1.5\_10yrs} & Excluded \\ 
\texttt{var\_expt\_v1.5\_10yrs} & \texttt{baseline\_v1.5\_10yrs} & \ref{fig:comparison_area}, \ref{fig:comparison_cadence}, \ref{fig:comparison_depth}, \ref{fig:comparison_season}, \ref{fig:mega_metric} \\ 
\texttt{wfd\_depth\_scale0.65\_noddf\_v1.5\_10yrs} & \texttt{baseline\_v1.5\_10yrs} & \ref{fig:comparison_area}, \ref{fig:comparison_depth}, \ref{fig:comparison_season}, \ref{fig:mega_metric} \\ 
\texttt{wfd\_depth\_scale0.65\_v1.5\_10yrs} & \texttt{baseline\_v1.5\_10yrs} & \ref{fig:comparison_area}, \ref{fig:comparison_depth}, \ref{fig:comparison_season}, \ref{fig:mega_metric} \\ 
\texttt{wfd\_depth\_scale0.70\_noddf\_v1.5\_10yrs} & \texttt{baseline\_v1.5\_10yrs} & \ref{fig:comparison_area}, \ref{fig:comparison_depth}, \ref{fig:comparison_season}, \ref{fig:mega_metric} \\ 
\texttt{wfd\_depth\_scale0.70\_v1.5\_10yrs} & \texttt{baseline\_v1.5\_10yrs} & \ref{fig:comparison_area}, \ref{fig:comparison_depth}, \ref{fig:comparison_season}, \ref{fig:mega_metric} \\ 
\texttt{wfd\_depth\_scale0.75\_noddf\_v1.5\_10yrs} & \texttt{baseline\_v1.5\_10yrs} & \ref{fig:comparison_area}, \ref{fig:comparison_depth}, \ref{fig:comparison_season}, \ref{fig:mega_metric} \\ 
\texttt{wfd\_depth\_scale0.75\_v1.5\_10yrs} & \texttt{baseline\_v1.5\_10yrs} & \ref{fig:comparison_area}, \ref{fig:comparison_depth}, \ref{fig:comparison_season}, \ref{fig:mega_metric} \\ 
\texttt{wfd\_depth\_scale0.80\_noddf\_v1.5\_10yrs} & \texttt{baseline\_v1.5\_10yrs} & \ref{fig:comparison_area}, \ref{fig:comparison_depth}, \ref{fig:comparison_season}, \ref{fig:internight_gap_filters}, \ref{fig:mega_metric} \\ 
\texttt{wfd\_depth\_scale0.80\_v1.5\_10yrs} & \texttt{baseline\_v1.5\_10yrs} & \ref{fig:comparison_area}, \ref{fig:comparison_depth}, \ref{fig:comparison_season}, \ref{fig:mega_metric} \\ 
\texttt{wfd\_depth\_scale0.85\_noddf\_v1.5\_10yrs} & \texttt{baseline\_v1.5\_10yrs} & \ref{fig:comparison_area}, \ref{fig:comparison_depth}, \ref{fig:comparison_season}, \ref{fig:internight_gap_filters}, \ref{fig:mega_metric} \\ 
\texttt{wfd\_depth\_scale0.85\_v1.5\_10yrs} & \texttt{baseline\_v1.5\_10yrs} & \ref{fig:comparison_area}, \ref{fig:comparison_depth}, \ref{fig:comparison_season}, \ref{fig:internight_gap_filters}, \ref{fig:mega_metric} \\ 
\texttt{wfd\_depth\_scale0.90\_noddf\_v1.5\_10yrs} & \texttt{baseline\_v1.5\_10yrs} & \ref{fig:comparison_area}, \ref{fig:comparison_depth}, \ref{fig:comparison_season}, \ref{fig:mega_metric} \\ 
\texttt{wfd\_depth\_scale0.90\_v1.5\_10yrs} & \texttt{baseline\_v1.5\_10yrs} & \ref{fig:comparison_area}, \ref{fig:comparison_cadence}, \ref{fig:comparison_depth}, \ref{fig:comparison_season}, \ref{fig:internight_gap_filters}, \ref{fig:mega_metric} \\ 
\texttt{wfd\_depth\_scale0.95\_noddf\_v1.5\_10yrs} & \texttt{baseline\_v1.5\_10yrs} & \ref{fig:comparison_area}, \ref{fig:comparison_depth}, \ref{fig:comparison_season}, \ref{fig:mega_metric} \\ 
\texttt{wfd\_depth\_scale0.95\_v1.5\_10yrs} & \texttt{baseline\_v1.5\_10yrs} & \ref{fig:comparison_area}, \ref{fig:comparison_depth}, \ref{fig:comparison_season}, \ref{fig:mega_metric} \\ 
\texttt{wfd\_depth\_scale0.99\_noddf\_v1.5\_10yrs} & \texttt{baseline\_v1.5\_10yrs} & \ref{fig:comparison_area}, \ref{fig:comparison_depth}, \ref{fig:comparison_season}, \ref{fig:mega_metric} \\ 
\texttt{wfd\_depth\_scale0.99\_v1.5\_10yrs} & \texttt{baseline\_v1.5\_10yrs} & \ref{fig:comparison_area}, \ref{fig:comparison_depth}, \ref{fig:comparison_season}, \ref{fig:mega_metric} \\ 
\texttt{barebones\_v1.6\_10yrs} & \texttt{baseline\_nexp1\_v1.6\_10yrs} & \ref{fig:mega_metric} \\ 
\texttt{baseline\_nexp1\_v1.6\_10yrs} & \texttt{baseline\_nexp1\_v1.6\_10yrs} & \ref{fig:mega_metric} \\ 
\texttt{combo\_dust\_v1.6\_10yrs} & \texttt{baseline\_nexp1\_v1.6\_10yrs} & \ref{fig:mega_metric} \\ 
\texttt{ddf\_heavy\_v1.6\_10yrs} & \texttt{baseline\_nexp1\_v1.6\_10yrs} & \ref{fig:mega_metric} \\ 
\texttt{dm\_heavy\_v1.6\_10yrs} & \texttt{baseline\_nexp1\_v1.6\_10yrs} & \ref{fig:mega_metric} \\ 
\texttt{even\_filters\_alt\_g\_v1.6\_10yrs} & \texttt{baseline\_nexp1\_v1.6\_10yrs} & \ref{fig:mega_metric} \\ 
\texttt{even\_filters\_altv1.6\_10yrs} & \texttt{baseline\_nexp1\_v1.6\_10yrs} & \ref{fig:mega_metric} \\ 
\texttt{even\_filters\_g\_v1.6\_10yrs} & \texttt{baseline\_nexp1\_v1.6\_10yrs} & \ref{fig:mega_metric} \\ 
\texttt{even\_filtersv1.6\_10yrs} & \texttt{baseline\_nexp1\_v1.6\_10yrs} & \ref{fig:mega_metric} \\ 
\texttt{mw\_heavy\_v1.6\_10yrs} & \texttt{baseline\_nexp1\_v1.6\_10yrs} & \ref{fig:mega_metric} \\ 
\texttt{rolling\_exgal\_mod2\_dust\_sdf\_0.80\_v1.6\_10yrs} & \texttt{baseline\_nexp1\_v1.6\_10yrs} & \ref{fig:mega_metric} \\ 
\texttt{rolling\_fpo\_2nslice0.8\_v1.6\_10yrs} & \texttt{baseline\_nexp1\_v1.6\_10yrs} & Excluded \\ 
\texttt{rolling\_fpo\_2nslice0.9\_v1.6\_10yrs} & \texttt{baseline\_nexp1\_v1.6\_10yrs} & Excluded \\ 
\texttt{rolling\_fpo\_2nslice1.0\_v1.6\_10yrs} & \texttt{baseline\_nexp1\_v1.6\_10yrs} & Excluded \\ 
\texttt{rolling\_fpo\_3nslice0.8\_v1.6\_10yrs} & \texttt{baseline\_nexp1\_v1.6\_10yrs} & Excluded \\ 
\texttt{rolling\_fpo\_3nslice0.9\_v1.6\_10yrs} & \texttt{baseline\_nexp1\_v1.6\_10yrs} & Excluded \\ 
\texttt{rolling\_fpo\_3nslice1.0\_v1.6\_10yrs} & \texttt{baseline\_nexp1\_v1.6\_10yrs} & Excluded \\ 
\texttt{rolling\_fpo\_6nslice0.8\_v1.6\_10yrs} & \texttt{baseline\_nexp1\_v1.6\_10yrs} & Excluded \\ 
\texttt{rolling\_fpo\_6nslice0.9\_v1.6\_10yrs} & \texttt{baseline\_nexp1\_v1.6\_10yrs} & Excluded \\ 
\texttt{rolling\_fpo\_6nslice1.0\_v1.6\_10yrs} & \texttt{baseline\_nexp1\_v1.6\_10yrs} & Excluded \\ 
\texttt{ss\_heavy\_v1.6\_10yrs} & \texttt{baseline\_nexp1\_v1.6\_10yrs} & \ref{fig:mega_metric} \\ 
\texttt{baseline\_nexp1\_v1.7\_10yrs} & \texttt{baseline\_nexp2\_v1.7\_10yrs} & \ref{fig:comparison_nexp} \\ 
\texttt{baseline\_nexp2\_v1.7\_10yrs} & \texttt{baseline\_nexp2\_v1.7\_10yrs} & \ref{fig:comparison_nexp} \\ 
\hline